\documentclass[3p]{elsarticle}

\usepackage{lineno,hyperref}
\modulolinenumbers[5]

\journal{Physics Reports}

\usepackage{graphicx, color, xcolor}
\usepackage{amsmath,amsfonts,amssymb}
\usepackage{slashed}
\usepackage{wasysym}
\usepackage{tikz}
\usepackage{wrapfig,framed,caption}
\usepackage{subfig}

\numberwithin{equation}{section}

\newcommand{\figref}[1]{Fig.~\protect\ref{#1}}

\newcommand{\fx}{\mathbf{x}}

\newcommand{\Ic}{I_{\text{c}}}
\newcommand{\ms}{S}
\newcommand{\mi}{I}
\newcommand{\mr}{R}
\newcommand{\rin}{\gamma}
\newcommand{\rhe}{\epsilon}
\newcommand{\rsu}{\zeta}
\newcommand{\rgg}{\lambda}
\newcommand{\rgb}{\xi}

\newcommand{\Dg}{D_{\rhe}}
\newcommand{\Dz}{D_{\rsu}}
\newcommand{\fy}{\mathbf{y}}

\usepackage{tcolorbox}

\newcommand{\beq}{\begin{equation}}
\newcommand{\eeq}{\end{equation}}
\newcommand{\bea}{\begin{eqnarray}}
\newcommand{\eea}{\end{eqnarray}}
\newcommand{\be}{\begin{equation}}
\newcommand{\ee}{\end{equation}}

\newcommand{\msl}[1]{n^S_{#1}}
\newcommand{\mil}[1]{n^I_{#1}}
\newcommand{\mrl}[1]{n^R_{#1}}

\definecolor{cerulean}{rgb}{0.0, 0.48, 0.65}

\newtcolorbox{summary}[2][]{colbacktitle=gray!40!white, colback=gray!10!white,coltitle=black, title={#2},fonttitle=\bfseries,#1}

\begin{document}

\begin{frontmatter}

\title{The field theoretical ABC of epidemic dynamics}

\author{Giacomo Cacciapaglia, Corentin Cot, Stefan Hohenegger, Shahram  Vatani}
\address{Institut de Physique des 2 Infinis (IP2I), CNRS/IN2P3, UMR5822, 69622 Villeurbanne, France\\
Universit\' e de Lyon, Universit\' e Claude Bernard Lyon 1, 69001 Lyon, France}

\author{Michele Della Morte}
\address{IMADA \& CP$^3$-Origins. Univ. of Southern Denmark, Campusvej 55, DK-5230 Odense, Denmark}

\author{Francesco Sannino\corref{mycorrespondingauthor}}
\cortext[mycorrespondingauthor]{Corresponding author}
\address{Scuola Superiore Meridionale, Largo S. Marcellino, 10, 80138 Napoli NA, Italy  \\ CP$^3$-Origins and D-IAS, Univ. of Southern Denmark,  Campusvej 55, DK-5230 Odense, Denmark\\
Dipartimento di Fisica, E. Pancini, Univ. di Napoli, Federico II and INFN sezione di Napoli \\  Complesso Universitario di Monte S. Angelo Edificio 6, via Cintia, 80126 Napoli, Italy}





\begin{abstract}
Infectious diseases are a threat for human health with tremendous impact on our society at large. They are events that recur with a frequency that is growing  with the exponential increase in the world population and growth of the human ecological footprint. The latter causes a frequent spillover of transmissible diseases from wildlife to humans. The recent COVID-19 pandemic, caused by the SARS-CoV-2, is the latest example of a highly infectious disease that, since late 2019, is ravaging the globe with a huge toll in terms of human lives and socio-economic impact.  It is therefore imperative to develop efficient mathematical models, able to substantially curb the damages of a pandemic by unveiling disease spreading dynamics and symmetries. This will help inform (non)-pharmaceutical prevention strategies. It is for the reasons above that  we decided to write this report. It goes at the heart of mathematical modelling of  infectious disease diffusion by simultaneously investigating the underlying microscopic dynamics in terms of percolation models, effective description via compartmental models and the employment of temporal symmetries naturally encoded in the mathematical language of critical phenomena. Our report reviews these approaches and determines their common denominators, relevant for theoretical epidemiology and its link to important concepts in theoretical physics. We show that the different frameworks exhibit common features such as criticality and self-similarity under time rescaling. These features are naturally encoded within the unifying field theoretical approach. The latter leads to an efficient description of the time evolution of the disease via a framework in which (near) time-dilation invariance is explicitly realised.  As important test of the relevance of symmetries we show how to mathematically account for observed phenomena such as multi-wave dynamics.  Although we consider the COVID-19 pandemic as an explicit phenomenological application, the models presented here are of immediate relevance for different realms of scientific enquiry from medical applications to the understanding of human behaviour. Our review offers novel perspectives on how to model, capture, organise and understand epidemiological data and disease dynamics for modelling real-world phenomena, and helps devising public health and socio-economics strategies. 
\end{abstract}

\begin{keyword}
epidemiology \sep field theory
\MSC[2021] 92D30 
\end{keyword}

\end{frontmatter}



\tableofcontents
\newpage


\section{Introduction}
 
Infectious diseases that can efficiently spread across the human population and cause a pandemic have always been a threat to humanity. This menace has been growing with the increase in the population and the progressive destruction of the wild environment 
with its impact on wildlife. The last century has been affected by, at least, three major worldwide pandemics: the 1918 ``Spanish'' influenza of 1918-1920 \cite{1918influenza}, HIV/AIDS \cite{AIDS1,AIDS2} and the most recent COVID-19 that started at the end of 2019. Understanding in a mathematically consistent way the diffusion of a pandemic is of paramount importance in designing effective policies apt at curbing and limiting its diffusion and the impact on the life loss and economic damage.
In this report we will review some crucial aspects of the mathematical modelling, ranging from the microscopic mechanisms encoded in diffusion models, to approaches based on symmetries. In this discussion, the application of field theory and other concepts borrowed from theoretical physics will play a crucial role.

The dynamics of physical phenomena, from the fundamental laws of nature to quantum and ordinary matter phase transitions, even including protein behaviour, is well captured by effective descriptions in terms of fields and their interactions.
Given the enormous success of the field theoretical interpretation of physical phenomena, it is highly interesting to  review several main mathematical models employed to describe the diffusion of infectious diseases and show how the different approaches are related within the field theoretical framework. 
We will show that the  models exhibit common features, such as criticality and self-similarity under time rescaling. These features are naturally encoded within the unifying field theoretical approach. The latter yields an efficient description of the time evolution of the disease via a framework in which (near) time-dilation invariance is explicitly realised. The models are extended to account for observed phenomena such as multi-wave dynamics.  Because of the immediacy of the COVID-19 pandemic and the high quality data available, we use it as an explicit and relevant phenomenological test of the models and their effectiveness. It should be clear, however, that the methodologies presented here are relevant for any infectious disease, and can be extended to different realms of scientific enquiry, from medical applications to the understanding of human behaviour.

We will complete this introduction with a historical overview of the mathematical modelling applied to infectious diseases, the contemporary applications and the role of field theory concepts, before offering a summary of the main body of the review.

\subsection{Historical Overview} \label{Sect:HistoO}

The first application of mathematical modelling to an epidemiological process is the work of Daniel~Bernoulli \cite{Bernoulli} on the effectiveness of an inoculation against smallpox in 1760. A more systematic application of mathematical methods to study the spread of infectious diseases occurred after the work of Robert Koch and Louis Pasteur, which showed that such diseases are caused by living organisms, triggering the question on how they are passed on from one individual to another. A related point in this regards is how (and why) outbreaks and epidemics  end. As outlined in  \cite{Heesterbeek}, there are two prevalent hypotheses:
\begin{itemize}
\item \underline{Farr's hypothesis} (mostly based on the work of W.~Farr in 1866 \cite{Farr}): epidemics stop because the {\it potency} of the microorganisms decreases with every new individual that is infected.
\item \underline{Snow's hypothesis} (mostly based on the work of J.~Snow in 1853 \cite{Snow}): epidemics end due to a lack of sufficient available new individuals to infect (the disease runs out of {\it ``fuel"}).
\end{itemize}
In view of closer studies of actual data stemming from outbreaks of communicable diseases, Farr's hypothesis was gradually  dropped from the scientific discussion. Moreover, the focus of research shifted towards explaining  regularities  of observed \emph{epidemic curves}. A first discovery along these lines can be found in the work of W.H.~Hamer \cite{Hamer,HamerLect1,HamerLect2,HamerLect3}, who described the biennial period of measles outbreaks in London and implicitly \cite{Heesterbeek} introduced the concept of mass-action law into epidemiology. The latter was  firmly established in the pioneering works of Sir~R.~Ross~\cite{Ross1911,Ross1916,RossHudson1916II,RossHudson1916III} and A.G.~McKendrick \cite{McKendrick1912,McKendrick1914,McKendrick1926}. Specifically, in a model of discretised time (with time steps $\delta t$) such as in \cite{Ross1911}, the mass action law can be formulated as follows:
\begin{align}
\text{number of cases at } t+\delta t\propto (\text{number of cases at }t)\times (\text{susceptibles at }t)\,.\nonumber
\end{align}
In models with a continuous time variable (as in later works of Ross and notably McKendrick), the model can  be formulated in terms of differential equations, including additional contributions capturing the population dynamics due to recovery from the disease, birth, death, migration, \emph{etc.}. Credit for the so-called SIR model,  still  widely used today (and which we review in Section~\ref{Sect:BasicSIRmodel}), is given to the work by W.O.~Kermack and A.G.~McKendrick in 1927 \cite{Kermack:1927}. The basic idea behind models of this type is that the disease is passed on among individuals in the form of {\it happenings} or {\it collisions}, in analogy to how reactions work in chemistry.  This led to numerous more refined models, see for example  the reviews \cite{BaileyBook,Becker,Castillo,Dietz,Dietz2,HethcoteThousand,HethcoteLevin,HethcoteStechDriessche,Wickwire}, including the historical overview in \cite{DietzSchenzle} .

In the second half of the 20th century, progress in different disciplines influenced epidemiological investigations. 
It was understood that, to describe (and combat) large scale outbreaks such as HIV/AIDS,  
human behaviour  plays a crucial role in modulating the spreading of the virus (\emph{e.g.} \cite{HETHCOTE1976335,AndersonMay}). Thereby, mathematical modelling started going beyond models inspired by basic chemical reactions. The appearance of a large number of reviews and books on epidemiological modelling is testimony to the depth and complexity of the analysis as well as the interdisciplinary attention this topic has received, \emph{e.g.}~\cite{HETHCOTErev,AndersonBook,AndersonMayBook,AndersonMayNature,Bailey2,Bartlett,Becker2,BusenbergCooke,Capasso,CliffHaggett,DaleyGani,DiekmannHeesterbeek,Frauenthal,GabrielLefeverPicard,GrenfellDobson,HethcoteArk,HethcoteYorke2,IshamMedley,Kranz,Lauwerier,LudwigCooke,Mollison2,Nasell,ScottSmith,Vanderplank,Waltman}. Besides influences and contributions from pure mathematics, chemistry and social sciences, certain features and symmetries of epidemic curves are similar to those found in particular physical systems. This has lead to novel approaches rooted in the physics of critical phenomena and phase transitions, such as \emph{percolation models} \cite{Flory1941,Stockmayer1944} and their relation to \emph{(scale invariant) field theories}. As we shall review in Section~\ref{Sect:GeneralPercolation}, there are various types of percolation models. Here we shall define them simply as collections of points in a given space, where some of which can be linked pairwise. The sets of points that are linked to each other are termed clusters and the spread of such clusters (following certain pre-determined rules) can be used to model the spread of a disease within a given population. In particular, the transition from finite sized clusters to the percolation phase (where all points are linked together) is a phase transition.  This important feature allows percolation models  to be organised in terms of universality classes and even to put them in correspondence  to   other physical systems. This property is useful to determine  important physical quantities. The first attempts  appeared in \cite{BroadbentHammersley} and their relation to phase transitions was pointed out in \cite{Domb}, while excellent reviews on more complicated models can be found in \cite{FrischHammersley,Essam1,ShanteKirkpatrick,Essam2,Kirkpatrick1,Kirkpatrick2,Welsh,Wu,Stauffer,Essam}. Direct formulations of percolation models in terms of field theories follow the approach of M.~Doi and L.~Peliti \cite{Doi1,Doi2,Peliti}, which have been reviewed, for example, in \cite{Pruessner}. Further work in this direction (notably the work by J.~Cardy and P. Grassberger \cite{Cardy_1985} and its relation to models in particle physics \cite{reggeon1,reggeon2}) shall be reviewed in Section~\ref{Sect:GeneralPercolation}.

\subsection{Current approaches to epidemiology}

The aim of our work is to summarise, review and connect various current approaches to understand and model the time evolution of pandemics. From the brief historical analysis of the previous subsection it is clear that, over the course of almost a century, many different approaches have been developed. Classifying them is often difficult. From a mathematical perspective one can distinguish {\bf stochastic} and {\bf deterministic} approaches, based on how the basic fundamental (microscopic) processes of the transmission and development of the disease are modelled: all epidemiological models generally assume that new infected individuals can appear when an uninfected one (usually called a \emph{susceptible} individual) comes in contact with an \emph{infectious} individual such that the disease is passed on. After some time, infected individuals may turn non-infectious (at least temporarily) via recovering or dying from the disease or by some other means of \emph{removal} from the actively involved population. Mathematically speaking, these processes can be modelled in two different fashions:

\begin{itemize}
\item {\bf Stochastic approach:} all (microscopic) processes between individuals are of a probabilistic nature. For instance, the contact between a susceptible and an infectious individual has a certain probability to lead to an infection of the former; infected individuals have a certain probability of removal after a certain time; \emph{etc}. In these approaches, time is understood as a 
discrete variable and time-evolution is typically described in the form of differential-difference equations (called \emph{master equations}). The solutions depend on a set of probabilities (\emph{e.g.} the probability of a contact among individuals leading to an infection), geometric parameters (such as the number of 'neighbouring' individuals that a single infectious individual can potentially infect) as well as the initial conditions. Furthermore, in order to make predictions or to compare with deterministic approaches, some sort of averaging process is required.
\item {\bf Deterministic approach:} the time evolution of the number of susceptible, infectious and removed individuals is understood as a fully predictable process and is typically described through systems of coupled, ordinary differential equations in time (the latter is understood as a continuous variable). Solutions of these systems are therefore determined by certain parameters (such as infection and recovery rates) as well as initial conditions (\emph{e.g.} the number of infectious individuals at the outbreak of the disease).
\end{itemize}

\noindent
In this review, we prefer to think of this classification in a somewhat different (but equivalent) fashion, which (as we shall explain) is closer to the concept of (energy) \emph{scale} in particle physics. Indeed, we prefer to think of models as ranging from microscopic models, in which fundamental interactions (\emph{i.e.} at the level of individuals) are explicitly modelled, to more and more macroscopic approaches, in which the microscopic interactions have been (at least partially) included into the interactions of new, effective degrees of freedom. A basic overview, with concrete models, is given in Figure~\ref{Fig:OverviewModels}: models in the left part of the diagram 
(red box) incorporate many details of how the disease spreads at a microscopic level, \emph{i.e.} between single individuals. These models are mostly of a stochastic nature, using probabilistic means to simulate the spread of the disease. As we shall explain, many of them are inspired by chemical models, in which a random movement of molecules is considered, with collisions leading with a certain probability to a chemical reaction (and the creation of new molecules). The models further to the right of the diagram (blue box) are more macroscopic, in the sense that they no longer model individual interactions (\emph{i.e.} the spread from one person to the next), but rather describe the time evolution of the disease  in a larger population (\emph{e.g.} an entire country). While, historically, the oldest models that have been developed to describe the spread of an infectious disease are in this category, many of them can be obtained from more microscopic approaches (\emph{e.g.} percolation models) through a 'replacement' of the degrees of freedom of the latter by more macroscopic ones. This can happen, for example, via a \emph{mean field approximation} or via certain averaging procedures or by describing the spread of the disease through suitable flow equations. The resulting models are mostly of a deterministic nature, but can retain stochastic elements. 

Besides the explicit models and approaches listed in Figure~\ref{Fig:OverviewModels} (some of which we shall review in the main part of this article), there are also data- and computer-driven approaches~\cite{Breiman,Flaxman}. These generally use machine learning (also called statistical learning) tools to analyse existing data with the goal of finding patterns and predicting the future development of pandemics. On the one hand, these approaches use the large advances in computer technology (in particular the development of artificial intelligence). On the other hand, they are made viable in recent years due to the dramatical increase in the volume and quality of available data on the spread and development (\emph{e.g.} its genetic mutations) of diseases in a large population. This allows data-driven approaches to be applied at any level, ranging from analysing microscopic interactions (see \emph{e.g.} \cite{WiemkenPneumonia}) to more effective descriptions that only aim at predicting 'global' key statistics of epidemics~\cite{ComparisonApproach,BigData}. Since the current review is aimed at studying field theoretic tools in epidemiology, we shall not discuss these methods here. However, we point out a number of excellent review articles \cite{WiemkenKelley} in the literature. Another class of models we will not discuss utilises complex networks to include the effect of human behaviour \cite{WANG20151}.

Microscopic approaches on the left spectrum of Figure~\ref{Fig:OverviewModels} generally utilise first principles, however at the expense of a lack of symmetries (usually also entailing a large computational cost). Effective theories on the right side of the graph are, usually, less intuitive (since basic interactions of the disease enter into a less obvious manner). However, they incorporate  basic symmetries that appear in the solutions of the microscopic models -- in the sense of making them manifest -- typically also leading to more streamlined and less expensive computations. Here is an incomplete list of the symmetries at the base of these approaches:

\begin{figure}[t]
\begin{center}
\scalebox{1}{\parbox{16.8cm}{\begin{tikzpicture}
\draw[ultra thick,->] (0,0.25) -- (15.8,0.25);
\begin{scope}[yshift=-0.1cm]
\draw[fill=yellow!40!white] (0,-0.1) -- (2,-0.1) -- (2,-0.95) -- (0,-0.95) -- (0,-0.1);
\node at (1.6,-0.5) {\parbox{3cm}{microscopic\\ models}};
\end{scope}
\begin{scope}[xshift=13.6cm,yshift=-0.1cm]
\draw[fill=yellow!40!white] (0,-0.1) -- (2,-0.1) -- (2,-0.95) -- (0,-0.95) -- (0,-0.1);
\node at (1.6,-0.5) {\parbox{3cm}{macroscopic\\ models}};
\end{scope}
\begin{scope}[xshift=-0.6cm,yshift=0cm]
\draw[fill=white!90!red] (0.6,0.65) -- (0.6,3.1) -- (4.3,3.1) -- (4.3,0.65) -- (0.6,0.65);
\node at (1.975,2.9) {$\bullet$ lattice models};
\node at (2.35,2.5) {$\bullet$ percolation models };
\node at (1.975,2.1) {$\bullet$ random walks};
\node at (2.15,1.7) {$\bullet$ diffusion models};
\node at (2.35,1.3) {$\bullet$ (epidemic) field th.};
\node at (2.1,0.9) {$\bullet$ network models };
\end{scope}

\begin{scope}[xshift=11.25cm,yshift=-0.25cm]
\draw[fill=white!90!blue] (-0.1,1.85) -- (-0.1,3.1) -- (4.55,3.1) -- (4.55,1.85) -- (-0.1,1.85);
\node at (1.975,2.9) {$\bullet$ compartmental models};
\node at (2.225,2.5) {$\bullet$ epidemic Renormalisation};
\node at (0.9,2.1) {Group};
\end{scope}

\node[red] at (7.25,2.5) {effective description};
\draw[ultra thick,->,red] (4,2.25) -- (10.75,2.25);
\node at (7.25,1.4) {\parbox{6cm}{ \small microscopic degrees of freedom replaced by more appropriate effective ones:\\ mean field approximations, averaging, beta-functions, flow equations,...}};

\begin{scope}[yshift=0.3cm]
\node at (2.5,-1.8) {\parbox{5cm}{+ based on 'first principles'}};
\node at (2.5,-2.2) {\parbox{5cm}{- symmetries not manifest}};
\node at (2.5,-2.6) {\parbox{5cm}{input: basic properties of the }};
\node at (3.5,-3) {\parbox{5cm}{disease and the way it spreads}};
\end{scope}

\begin{scope}[xshift=-1cm,yshift=0.3cm]
\node at (13.55,-1.8) {\parbox{5cm}{+ based on manifest symmetries}};
\node at (13.55,-2.2) {\parbox{5cm}{+ computationally simpler}};
\node at (14.1,-2.6) {\parbox{6cm}{- modelling requires more intuition  }};
\node at (14.4,-3) {\parbox{6cm}{about the system and/or data}};
\node at (14.3,-3.4) {\parbox{6.5cm}{input: 'effective' properties of the }};
\node at (14.8,-3.8) {\parbox{6cm}{disease in a specific population}};

\end{scope}
\end{tikzpicture}}}
\end{center}
\caption{Schematic overview of different approaches to describe the time evolution of pandemics and their relation to field theoretical methods.}
\label{Fig:OverviewModels}
\end{figure}
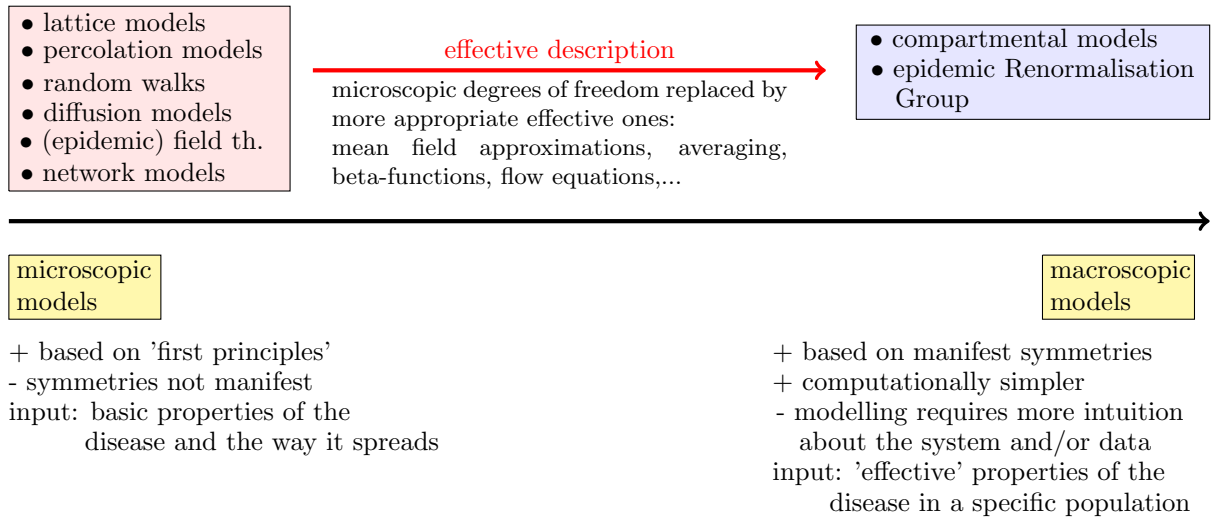

\begin{enumerate}
\item[{\emph{(i)}}] \emph{Criticality:} depending on the parameters of the model and the starting conditions, solutions of microscopic models feature either a quick eradication of the disease, where the total cumulative number of infected individuals
remains relatively low, or a fast and widespread diffusion of the disease, leading to a much larger total number of infected. Which of these two classes of solutions is realised is usually governed by a single ordering parameter (\emph{e.g.} the average number of susceptible individuals infected by a single infectious, also known as \emph{reproduction number} $R_0$), and the transition from one type to the other can be very sharp.

\item[{\emph{(ii)}}] \emph{Self-similarity and waves:} depending on the disease in question, solutions of microscopic models may exhibit distinct phases in their time evolution in the form of a wave pattern, where phases of exponential growth of the number of infected individuals are followed by intermediate periods of slow, approximately linear, growth. Each wave typically looks similar to the previous and following ones. Furthermore, certain classes of solutions may also exhibit spatial self-similarities, \emph{i.e.} the solutions describing the temporal spread of the disease among individuals follow similar patterns as the spread among larger clusters (\emph{e.g.} cities, countries \emph{etc.}).

\item[{\emph{(iii)}}]\emph{Time-scale invariance:}  several microscopic models exhibit a (nearly) time-scale invariant behaviour, which is a symmetry under rescaling of the time variable and of the rates (infection, removal, \emph{etc.}). If the solution exhibits a wave-structure, these near-symmetric regions can appear in specific regimes, \emph{e.g.} in between two periods of exponential growth.

\end{enumerate}

\noindent
These properties are familiar from field theoretical models in physics, \emph{e.g.} in solid state and high energy physics, which exhibit phase transitions. Indeed, over the years, it has been demonstrated that the various approaches mentioned above can be reformulated (or at least related to) field theoretical descriptions. The latter are typically no longer sensitive to microscopic details of the spread of the disease at the level of individuals, but instead capture \emph{universal} properties of their solutions. They are therefore an ideal arena to study properties of the dynamics of diseases and the mechanisms to counter their spread.

\subsection{Relating different scales in Field Theory}

The dynamics of physical phenomena, ranging from the fundamental laws of nature to quantum and ordinary matter phase transitions including protein behaviour, is well captured by effective descriptions in terms of fields and their interactions. These fields are meant to capture the overall features of the phenomenon in question, describe the interaction between (elementary) constituents and even predict the evolution of the system. Once the field theoretical dynamics is married to underlying approximate or exact symmetries, it becomes an extremely powerful tool that, in a given range of energy, provides a faithful representation of the microscopic physics underlying many phenomena. Zooming in or out of the relevant physical scales involved in the dynamics of a given process generically requires a modification of the degrees of freedom needed to describe that specific process. This property is  captured by the  renormalisation group (RG) framework \cite{Wilson:1971bg,Wilson:1971dh}.  Within this approach, in order to take into account  the change in degrees of freedom, one modifies (renormalises) the interaction strengths and rescales the fields. In fact, the idea of scale transformations and scale invariance is ancient, dating back to the Pythagorean school. The concept was  used in the work by Euclid and much later by Galileo. The idea received renewed popularity  towards the end of the 19th century with the idea of enhanced viscosity of O.~Reynolds to address turbulence in fluids \cite{Reynolds1,Reynolds2}. 

However, the seed-idea of the RG initially started in 1953 with the work of  Ernst Stueckelberg and Andr\'e Petermann  \cite{Stueckelberg}. They noted that the renormalisation procedure in quantum field theory exhibits a group of transformations, which acts on parameters that govern basic interactions of the system, \emph{e.g.} changing the bare couplings in the Lagrangian by including (counter) terms needed to correct the theory. For example, the application to quantum electro-dynamics (QED) was elucidated by Murray Gell-Mann and Francis E. Low in 1954 \cite{PhysRev.95.1300}: this led to the renown determination of the variation of the electromagnetic coupling in QED with the energy of the physical processes.  Hence, the basic idea at the heart of the RG approach stems from the property that, as the scale of the physical process varies, the theory displays a self-similar behaviour and any scale can be described by the same physics. In mathematical terms, this properties is reproduced by a group transformation acting on the interaction strengths of the theory. Thanks to Gell-Mann and Low a computational method based on a mathematical flow function of the interaction strength parameter was introduced. This function determines the differential change of the interaction strength with respect to a small change in the energy of the process through a differential equation known as the renormalisation group equation (RGE).  Although mainly devised for particle physics, nowadays its applications extend to solid-state physics, fluid mechanics, physical cosmology, and even nanotechnology.

In certain cases, such as in particle physics, the field theoretical description can be elevated to the ultimate description of  fundamental interactions if short distance scale invariance occurs.  Once scale invariance is married to relativity the group of invariance generically enlarges to the conformal group.

\subsection{Organisation of the Review}

In the following we shall start by presenting examples of microscopic and effective (respectively deterministic and stochastic) approaches and show how they can be related to field theoretical models. We start in Section~\ref{Sect:Percolation} with analysing the direct percolation approach, which is based on a microscopic stochastic description of the diffusion processes. We shall see that the approach, in the mean field approximation, naturally leads to compartmental models. The latter (as well as generalisations thereof) are reviewed in Section~\ref{Sect:CompartmentalModels}: we commence this investigation with a basic review of the SIR model and then investigate how to incorporate multi-wave epidemic dynamics paying particular attention to the inter-wave period. After highlighting further possible extensions of compartmental models, we finally provide a formulation of the SIR model in terms of flow equations, which resembles the $\beta$-function familiar from the RG approach to particle and high-energy physics.

We use this last result to motivate the most recent approach to epidemic dynamics, \emph{i.e.} the epidemiological renormalisation group (eRG) \cite{DellaMorte:2020wlc,DellaMorte:2020qry} in Section~\ref{Sect:RGapproach}.  The latter is inspired by the Wilsonian renormalisation group approach \cite{Wilson:1971bg,Wilson:1971dh} and uses the approximate short and long time dilation invariance of the system to organise its description. For the COVID-19, the eRG has been shown  to be very efficient  when describing the epidemic and pandemic time evolution across the world \cite{Cacciapaglia:2020mjf} and in particular when predicting the emergence of new waves and the interplay across different regions of the world \cite{cacciapaglia2020second,cacciapaglia2020better}. 

The discussion in Sections~\ref{Sect:Percolation}, \ref{Sect:CompartmentalModels} and \ref{Sect:RGapproach} is general in the sense that the methods apply to generic infectious diseases and populations. In Section~\ref{Sect:CovidSpecific} we consider particular features of the current ongoing COVID-19 pandemic, and discuss how the different approaches can be adapted to it.

Several excellent reviews already exist in the literature \cite{PERC20171,WANG20151,WANG20161,HETHCOTErev}. Our work complements and integrates them, adds to the literature on the field theoretical side and further incorporates more recent approaches.


\section{Percolation Approach}\label{Sect:Percolation}

\begin{summary}{Executive Summary}
\begin{itemize}
\item We introduce percolation and lattice models as \emph{stochastic} approaches to directly simulate microscopic interactions down to the individual level.
\item The models are characterised through a set of probabilities (related for example to the infection and recovery rates of individuals) and the geometry of the system. Time is assumed to be a quantised variable.
\item The approach naturally models the spatial as well as the temporal evolution of a disease.
\item The models feature a (sharp) \emph{phase transition} in terms of the asymptotic infected fraction of the population. The latter is the order parameter of the system.
\item Compartmental models (see next Section) emerge as a mean field description of percolation models.
\end{itemize}

\end{summary}

\subsection{Lattice and Percolation Models}\label{Sect:GeneralPercolation}
Arguably the most direct way to (theoretically) study the spread of a communicable disease is via systems that simulate the process of infection at a microscopic level, \emph{i.e.} at the level of individuals in a (finite) population. The most immediate such models are lattice simulations, in which the individuals are represented by the lattice sites on a spatial grid, some of which may be infected by the disease. These lattice sites can spread the disease with a certain probability to neighbouring sites, following an established set of rules. Lattice models, therefore, allow to track the spread of the disease in discretised time steps and, after taking the average of several simulations, allow to make statements about the time evolution (and asymptotic values) of the number of infected individuals. As we shall see in the following, even simple models of this type show particular time-scaling symmetries, as well as criticality (\emph{i.e.} the fact that the asymptotic number of infected individuals changes rapidly, when a certain parameter of the model approaches a specific critical value).

A larger class of models that work with a discrete number of individuals (as well as discretised time) consists of \emph{percolation models}, which broadly speaking consist of points (sites) scattered in space that can be connected by links. Depending on the specific details, one distinguishes \cite{Essam}:
\begin{itemize}
\item \emph{Bond percolation models}: in this case the points are fixed and the links between them are created randomly. Examples of this type are (regular) lattices in various spatial dimensions with nearest neighbour sites being linked.
\item \emph{Site percolation models}: in this case the position of the points is random, while the links between different points are created based on rules that depend on the positions of the points.
\end{itemize}
More complex models can also incorporate both aspects. An important quantity to compute in any percolation model is the so-called \emph{pair connectedness}, \emph{i.e.} the probability that two points are connected to each other (through a chain of links with other points). Assuming the system to extend infinitely (\emph{i.e.}  there are infinitely many sites), we can importantly distinguish whether it is made of only local clusters (in which finitely many sites are connected) or whether it is in a \emph{percolating state} (where infinitely many sites are connected). The probability of occurrence of these two situations usually depends on the value of a single parameter (typically related to the probability $p$ that a link exists between two `neighbouring' sites), in such a way that the transition from local connectedness to percolation can be described as a \emph{phase transition} (see \emph{e.g.} \cite{Domb}). The system close to this critical value $p_c$ lies in the same universality class of several other models in molecular physics, solid state physics and epidemiology: this implies that the behaviour of certain quantities follows a characteristic power law behaviour that is the same for all the theories in the same universality class. For example, the probability $P(p)$ for a system to be in the percolating state (as a function of $p$) takes the form
\begin{align}
\lim_{p\to p_c}P(p)\sim (p-p_c)^\nu\,,\label{DefCriticalExponent} 
\end{align}
where $\nu$ is called \emph{critical exponent}. Models within the same universality class share the same critical exponents despite the fact that the concrete details of the theory, in particular the concrete meaning of the quantity $P$ in Eq.~(\ref{DefCriticalExponent}), may be very different. This connection makes percolation models very versatile and many of them have been studied extensively (see \cite{Essam} and references therein).

In the following, we shall first present a simple lattice simulation model, which allows us to reveal important properties of the time evolution of the infection (notably criticality and time-rescaling symmetry). Furthermore, we shall discuss a percolation model that, near criticality, is in the same universality class as time-honoured epidemiological models, along with some of its extensions and generalisations. 

\subsection{Numerical Simulations and Criticality}
Lattice simulations of reaction-diffusion processes are a well established tool to study the epidemic spreading of a disease since the
original work by P.~Grassberger in~\cite{Grassberger1983}.
In specific realisations the models have been studied to very high precision and the critical values of the parameters are known
with an accuracy reaching the six digits, see for example Ref.~\cite{PhysRevE.82.051921} and references therein.
Different geometries have been considered as well as different ranges of interactions, including random long-range couplings among sites,
see~\cite{SANTOS2020126063, Mukhamadiarov_2021} for recent discussions. All of these follow a \emph{Markov decision process \cite{Markov1,Markov2}, \emph{i.e.} the population is represented by a discrete lattice and the time evolution of the disease is organised in discretised time steps (so-called Markov iterations) between each of which the state of the lattice is changed based on a set of stochastic decisions.}
Here we consider a synchronous algorithm (i.e., we update all the lattice sites in each Markov iteration), and isotropic
interactions of range $r$ (in lattice units).

\subsubsection{The principle} \label{Sect:theprinciple}

For our purposes, the simplest and most direct way to study percolation models is to simulate the time evolution of the spread of a disease via stochastic processes on a finite dimensional lattice. The individuals, represented by each lattice site, can be in one and only one of the three given states: susceptible, infectious or removed. 
They are defined as follows:
\begin{itemize}
\item \emph{Susceptible}: these are individuals that are currently not infectious, but can contract the disease. We do not distinguish between individuals who have never been infected and those who have recovered from a previous infection, but are no longer immune. 
\item \emph{Infectious}: these are individuals who are currently infected by the disease and can actively transmit it to a susceptible individual. 
\item \emph{Removed (recovered)}: these are individuals who currently can neither be infected themselves, nor can infect susceptible individuals. This comprises individuals who have (temporary) immunity (either natural, or because they have recovered from a recent infection), but also all deceased individuals. 
\end{itemize}

The time evolution of the lattice configurations follows a set of rules, which implements the following two basic mechanisms into an algorithm that models the spread of the disease within a  finite and isolated population in discretised time steps: 
\begin{itemize}
\item[\emph{i)}] the infection of susceptible individuals in the vicinity of an infectious one; 
\item[\emph{ii)}] the removal (recovery) of an infectious individual (so that it can no longer infect other individuals).
\end{itemize}
The infection process depends on the reach of an infectious site over potential nearby susceptible ones. This reach depends on the geometry of the lattice (here we always use square lattices) and on the range $r$. The removal instead depends on the site itself and on an intrinsic removal probability.

Starting from the two principles above, there are two ways to let the lattice evolve and to define the elementary time steps, starting from a given initial spatial distribution of infectious and susceptible sites. On the one hand, we can randomly choose an infectious site and begin the infection process within its surrounding sites (\emph{i.e.} determine how many susceptible neighbouring sites are turned infectious). Once the process is over, another infectious site is chosen randomly, defining the next time step. 
Such a sequence forbids multiple infections, as only one infected site is considered at each step of time. On the other hand, we could take into account all the possible infections at the same time 
and consider the susceptible sites that may become infected by them, according to the rules of the algorithm. The lattice is then updated with the new infected and thus the next time step begins. This process allows multiple infections to be considered, as susceptible sites can have multiple infected neighbours infecting them at a single time step. The first method is called ``asynchronous'' as opposed to the second ``synchronous'' algorithm.

Having discussed the temporal structure of the simulation, we can turn now to the specific mechanism of the spread, 
which, in our setup, depends on three parameters:
\begin{itemize}
\item The \emph{coordination radius} $r\in\mathbb{R}_+$, which is a measure for the distance (on the lattice) over which direct infections between individuals can take place, \emph{i.e.} only sites within a distance $r$ from the infectious one can be infected. We illustrate $r$ in \figref{Fig:CoordinationNumber2D} within a 2-dimensional squared lattice.

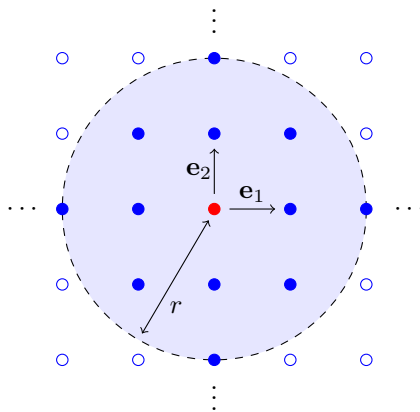
\begin{figure}[h]
\begin{center}
\vspace{-0.5cm}
\scalebox{1}{\parbox{5.8cm}{\begin{tikzpicture}
\draw[dashed,fill=blue!10!white] (0,0) circle (2cm);
\node at (0,-2.4) {$\vdots$};
\node at (0,2.6) {$\vdots$};
\node at (-2.5,0) {$\cdots$};
\node at (2.6,0) {$\cdots$};
\draw[blue] (-2,-2) circle (0.075);
\draw[blue] (-1,-2) circle (0.075);
\draw[blue,fill=blue] (0,-2) circle (0.075);
\draw[blue] (1,-2) circle (0.075);
\draw[blue] (2,-2) circle (0.075);
\draw[blue] (-2,-1) circle (0.075);
\draw[blue,fill=blue] (-1,-1) circle (0.075);
\draw[blue,fill=blue] (0,-1) circle (0.075);
\draw[blue,fill=blue] (1,-1) circle (0.075);
\draw[blue] (2,-1) circle (0.075);
\draw[blue,fill=blue] (-2,0) circle (0.075);
\draw[blue,fill=blue] (-1,0) circle (0.075);
\draw[red,fill=red] (0,0) circle (0.075);
\draw[blue,fill=blue] (1,0) circle (0.075);
\draw[blue,fill=blue] (2,0) circle (0.075);
\draw[blue] (-2,1) circle (0.075);
\draw[blue,fill=blue] (-1,1) circle (0.075);
\draw[blue,fill=blue] (0,1) circle (0.075);
\draw[blue,fill=blue] (1,1) circle (0.075);
\draw[blue] (2,1) circle (0.075);
\draw[blue] (-2,2) circle (0.075);
\draw[blue] (-1,2) circle (0.075);
\draw[blue,fill=blue] (0,2) circle (0.075);
\draw[blue] (1,2) circle (0.075);
\draw[blue] (2,2) circle (0.075);
\draw[->] (0.2,0) -- (0.8,0);
\node at (0.5,0.2) {$\mathbf{e}_1$};
\draw[->] (0,0.2) -- (0,0.8);
\node at (-0.2,0.5) {$\mathbf{e}_2$};
\draw[<->] (-0.075,-0.15) -- (-0.95,-1.65);
\node at (-0.5,-1.3) {$r$};
\end{tikzpicture}}}
\end{center}
\caption{Two-dimensional cubic lattice generated by the lattice vectors $(\mathbf{e}_1,\mathbf{e}_2)$. The blue circle of coordination radius $r$ ($r=2$ in the current example) contains all susceptible sites (blue) that may become infected by a single infectious one (red) at its centre.}
\label{Fig:CoordinationNumber2D}
${}$\\[-1.2cm]
\end{figure} 
\item The \emph{infection probability} $\mathfrak{g}\in[0,1]$ for an infectious individual to infect a neighbour site. In practice, the probability of a single individual in  the neighbourhood (defined in terms of the coordination radius) to be infected is equal to $\mathfrak{g}$
divided by the number of sites within a radius $r$ from the infectious one. This choice, as we shall see in Section~\ref{FromLatticeToSIR},
allows us to draw a more direct relation between $\mathfrak{g}$ and the infection rate parameter defined in other approaches.
\item The \emph{removal probability} $\mathfrak{e}\in[0,1]$  for an infectious individual to become removed.
\end{itemize}

\noindent

In the following we shall highlight some of the key-features of this approach and study their dependence on the three parameters above.
To do so, we consider a 2-dimensional lattice with periodic boundary conditions. We follow the ``synchronous'' algorithm with a slightly different path compared to the common approach in the literature~\cite{Arashiro_2007}. Usually, in order to determine the time-evolution of the lattice configuration, one needs to go through all infectious sites and individually apply the infection algorithm to all susceptible sites within their coordination radius: each contact is simulated by the call of a randomly generated number $x$, between 0 and 1. If $x\leq \mathfrak{g}$, an infection occurs and the considered susceptible site will become infectious at the next time step. Else, nothing happens --  the site will stay susceptible and the whole process is repeated for each of the sites surrounding a given infectious one. 

Instead of this infectious-site-centred procedure, we will consider an algorithm centred on the susceptible sites: 
for each susceptible site, we count the number $n$ of infectious sites within the coordination radius and calculate the cumulated probability of infection. One can show that, on average, the probability $\mathcal{P}\left( n\right)$ for this site to become infectious in the next Markov iteration is given by $ \displaystyle \mathcal{P}\left( n\right) = 1- (1-\mathfrak{g})^n$. We use this probability to determine the fate of each susceptible site. This improved procedure speeds-up the algorithm and reduces stochastic fluctuations, as it is equivalent to performing a local average at each time step. We turn now to the presentation of our results.

\subsubsection{Results}

\begin{wrapfigure}{r}{0.6\textwidth}
\begin{center}
\vspace{-0.5cm}
\includegraphics[scale=0.8]{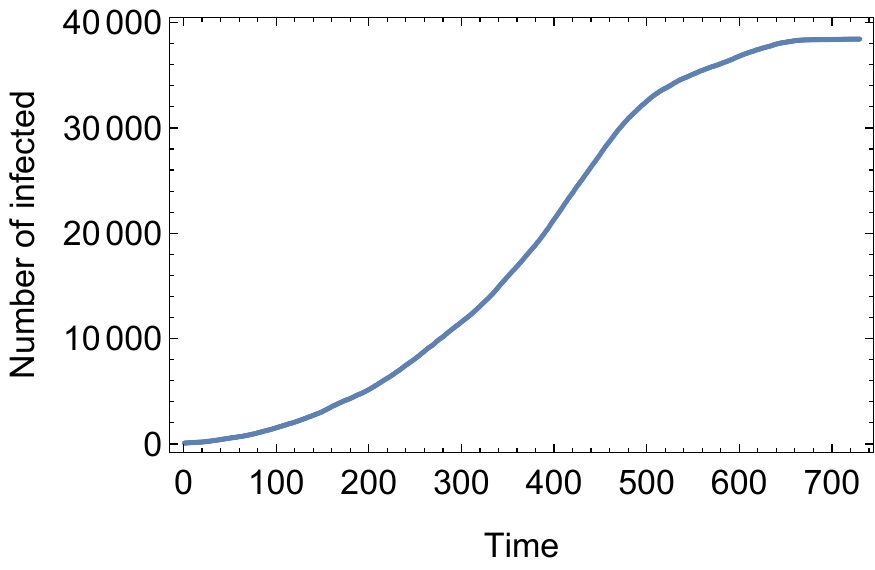}
    \caption{ Number of infected individuals as a function of the discretised time for a lattice with $40401$ sites, $\mathfrak{g}=0.7$, $\mathfrak{e}=0.1$ and coordination radius $r=1$.}
    \label{example}
    \end{center}
${}$\\[-1.5cm]
\end{wrapfigure}

\begin{figure}
\centering
\subfloat[$r=1$]{\includegraphics[scale=0.8]{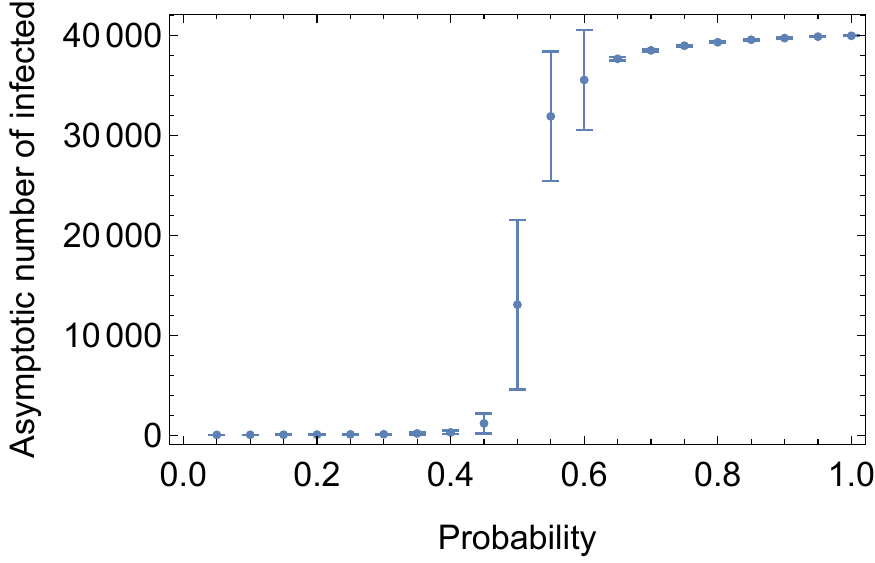}}
\subfloat[$r=2$]{\includegraphics[scale=0.8]{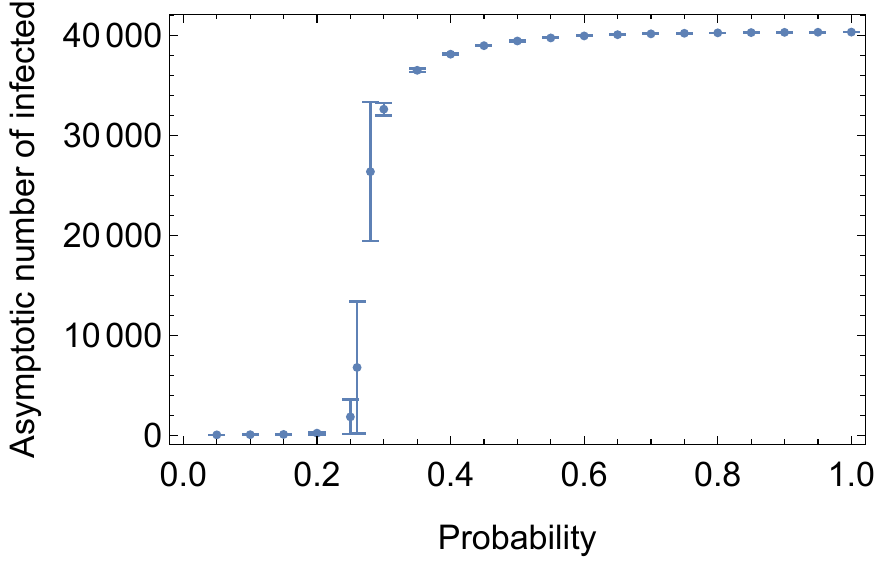}}\\
\subfloat[$r=5$]{\includegraphics[scale=0.8]{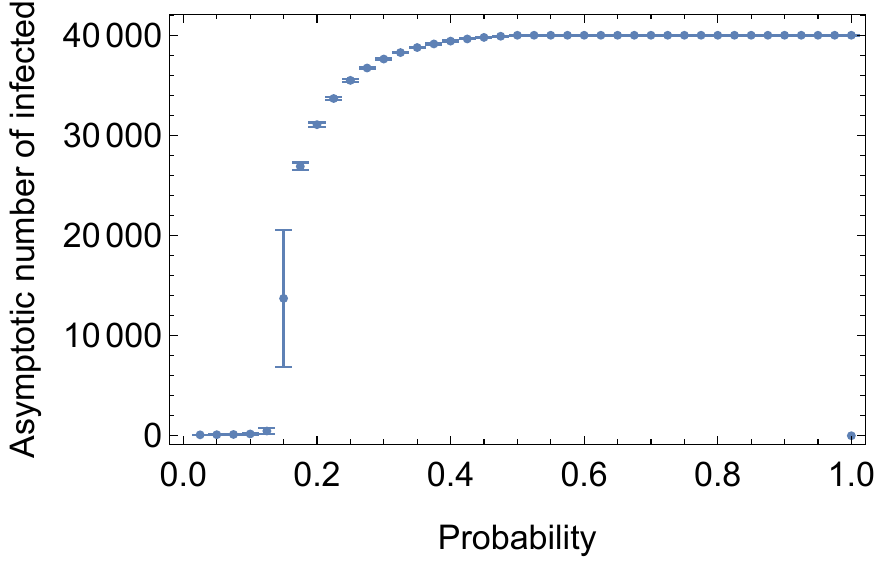}}
\subfloat[$r=50$]{\includegraphics[scale=0.8]{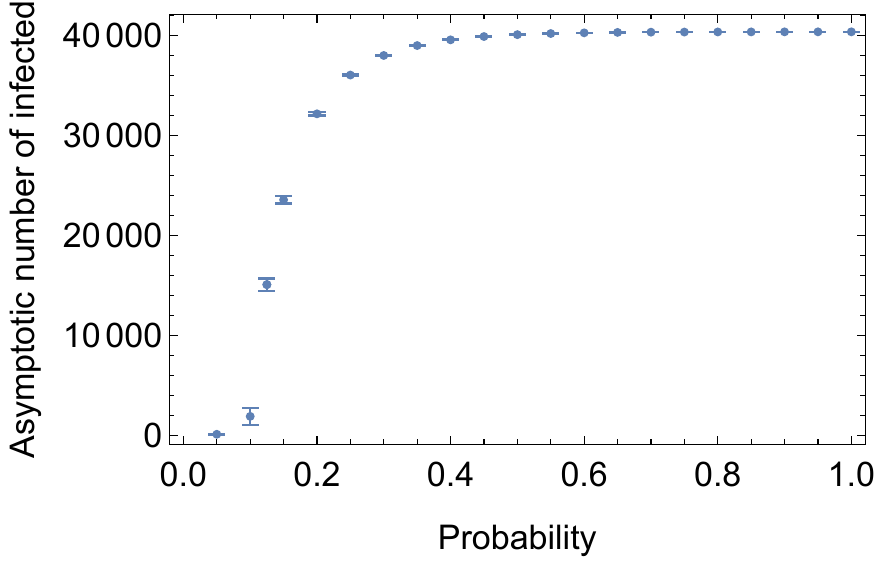}}
\caption{Evolution of the asymptotic number of infected sites as a function of the infection probability $\mathfrak{g}$ for different coordination radii $r$. The removal probability is fixed to $\mathfrak{e}=0.1$.}
\label{phasetransition}
\end{figure}

\noindent
A plot of the evolution of the cumulative number of infected sites as a function of the discretised time-steps is shown in \figref{example} for a sample choice of the parameters $\mathfrak{g}=0.7$, $\mathfrak{e}=0.1$ and $r=1$ and for a square lattice with $201$ sites on each side (\emph{i.e.} $40401$ sites in total).  At large $t$, the cumulated number of infected approaches an asymptotic value, which, averaged over a sufficient number of simulations, is a function of the probabilities $(\mathfrak{g},\mathfrak{e})$ as well as of the coordination radius $r$. Varying these parameters leads to substantially different asymptotic values, as is shown in \figref{phasetransition}: in the four panels, we plot the asymptotic values as a function of the infection probability $\mathfrak{g}$. We use the same lattice as before and fix $\mathfrak{e}=0.1$. For each point, we repeat the process $50$ times to compute the shown mean and standard  deviation. As expected, the larger $\mathfrak{g}$, the higher the number of infected sites at the end of the process. The plots also show the critical behaviour of the system, as the asymptotic value jumps from a very small value at small $\mathfrak{g}$ to a value of the same order of the total population (\emph{i.e.} the number of sites in the lattice).
For each value of $r$, one can define a critical value $\mathfrak{g}_c (r)$: increasing $r$ reduces the value of $\mathfrak{g}_c$.

\begin{figure}
\centering
\includegraphics[scale=0.8]{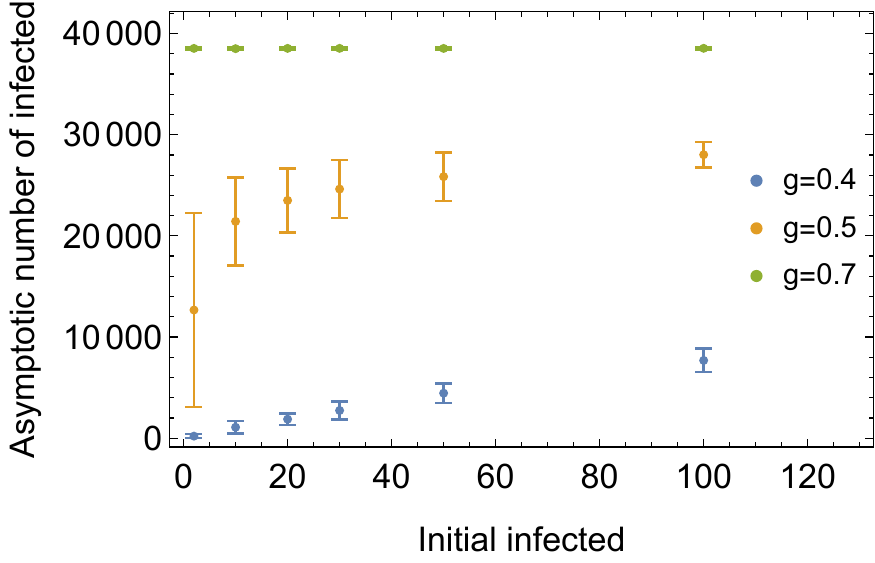}
\includegraphics[scale=0.83]{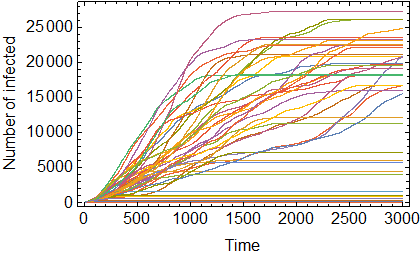}
\caption{Left panel: Evolution of the final number of infected sites as a function of the initial infectious ones. The mean and the standard deviation are computed over $50$ simulations for each point with $\mathfrak{e}=0.1$ and different  values of $\mathfrak{g}$. Right panel: Time evolution of the infected cases for $50$ simulations with $\mathfrak{e}=0.1$, $\mathfrak{g}=0.5$, $N_I=2$ and $r=1$.}
\label{initialinfected}
\end{figure}

In the simulations of \figref{phasetransition} we use the same initial condition, where all the sites within a radius $5$ (in lattice units) from the centre of the lattice are set to the infectious state, thus having initially $81$ cases. Due to the stochastic nature of the process, the final number of infected cases does depend non-trivially on the initial state, especially for small coordination radius $r$. For $r=1$ and $ \mathfrak{e}=0.1$, this dependence on the initial infected $N_I$ is shown in the left panel of \figref{initialinfected}, where we plot the asymptotic value of infected as a function of $N_I$, randomly distributed on the lattice. We plot the results for three different values of $\mathfrak{g}=0.4,\;0.5$ and $0.7$, where $\mathfrak{g}=0.5$ is close to the critical $\mathfrak{g_c}$. 
The critical behaviour described above seems to be also sensitive to $N_I$. This could be due to finite volume effects, as the evolution of the infection is expected to depend crucially on the density (rather than on the actual number) of initial infectious cases on the lattice as well as on their spatial distribution.
The dependence on the initial state is consistent with the result obtained for the SIR compartmental models discussed in Sec.~\ref{Sect:SIRQualitative}. This effect should disappear in the infinite volume limit. 
Especially near the critical value, we observe a large spread  of results for the asymptotic numbers. This is particularly evident for small densities of initial infections, where stochastic effects become relevant. As an example, we show a bundle of 50 solutions near the critical value in the right panel of \figref{initialinfected}.

\subsection{Master Action and Field Theory}\label{Sect:MasterEquationPercolation}
Here we briefly summarise the percolation approach and the  derivation via field theory of the reaction diffusion processes. We follow G.~Pruessner's lectures~\cite{Pruessner} and borrow part of his notation.  The overarching goal is to reproduce and extend the action given in the seminal work of J.L.~Cardy and P.~Grassberger~\cite{Cardy_1985}. 

We, therefore, consider a model of random walkers described by a field $W$ that diffuse
through a lattice, reproduce themselves and drop some poison $P$ as they stroll around. The poison field $P$ does not diffuse but kills walkers
if they hit a poisoned location. Interpreting the positions of the walkers as infected sites and those of the poison as simultaneously representing either the immune or removed individuals, the model effectively describes a disease diffusion process featuring infection and immunisation dynamics. The microscopic processes considered in~\cite{Cardy_1985} (see also \cite{Mollison,Bailey}) can be schematically summarised as follows:
\begin{align}
 & W \, \rightarrow \, W + W\;, && {\rm with \; rate \;} \sigma\;,  \nonumber\\
 & W \, \rightarrow \, W + P\;, && {\rm with \; rate \;} \alpha \;, \nonumber\\
 &W+P \, \rightarrow \, P  \;, && {\rm with \; rate \;} \beta \;.
\label{microCardy}
\end{align}
The first branching process corresponds to infection, while the last two processes describe immunisation. In addition we will consider a process of spontaneous creation, by which infected can appear at one site independently
from the presence of other infected at neighbouring sites, with a rate $\xi$.

\begin{wrapfigure}{r}{0.35\textwidth}
\begin{center}
\vspace{-0.5cm}
\scalebox{0.95}{\parbox{3.25cm}{\begin{tikzpicture}
\draw (0,0) circle (0.1);
\draw (1.5,0) circle (0.1);
\draw (3,0) circle (0.1);
\draw (0,1.5) circle (0.1);
\draw[fill=black] (1.5,1.5) circle (0.1);
\draw (3,1.5) circle (0.1);
\draw (0,3) circle (0.1);
\draw (1.5,3) circle (0.1);
\draw (3,3) circle (0.1);
\node at (1.5,1.1) {\footnotesize $n^W_\fx,n^P_\fx$};
\draw[thick,->] (1.6,1.5) -- (2.85,1.5);
\node at (2.25,1.65) {\footnotesize $e_1$};
\draw[thick,->] (1.5,1.6) -- (1.5,2.85);
\node at (1.3,2.25) {\footnotesize $e_2$};
\end{tikzpicture}}}
\end{center}
\caption{Schematic presentation of the state $\{n_\fx^W,n_\fx^P\}$ with $e_i$ the basis vectors of $\Gamma$.}
\label{Fig:SchemLattice}
\end{wrapfigure}
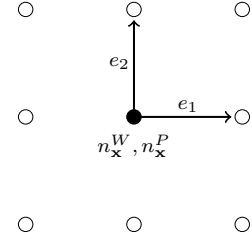

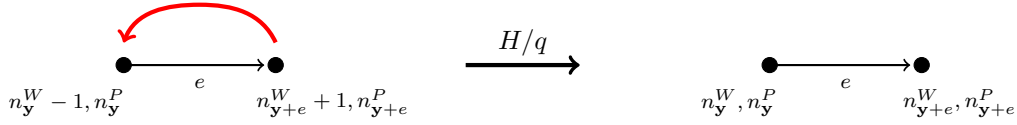
\begin{figure}[t]
\begin{center}
\scalebox{1}{\parbox{14cm}{\begin{tikzpicture}
\draw[fill=black] (0,0) circle (0.1);
\draw[fill=black] (2,0) circle (0.1);
\node at (-0.75,-0.5) {\footnotesize $n^W_\fy-1,n^P_\fy$};
\node at (2.75,-0.5) {\footnotesize $n^W_{\fy+e}+1,n^P_{\fy+e}$};
\draw[thick,->] (0,0) -- (1.85,0);
\node at (1,-0.25) {\footnotesize $e$};
\draw[ultra thick,red,->] (2,0.3) to [out=115,in=0] (1,0.8) to [out=180,in=70] (0,0.3);
\draw[ultra thick,->] (4.5,0) -- (6,0);
\node at (5.25,0.3) {$H/q$};
\begin{scope}[xshift=8.5cm]
\draw[fill=black] (0,0) circle (0.1);
\draw[fill=black] (2,0) circle (0.1);
\node at (-0.4,-0.5) {\footnotesize $n^W_\fy,n^P_\fy$};
\node at (2.5,-0.5) {\footnotesize $n^W_{\fy+e},n^P_{\fy+e}$};
\draw[thick,->] (0,0) -- (1.85,0);
\node at (1,-0.25) {\footnotesize $e$};
\end{scope}
\end{tikzpicture}}}
\end{center}
\caption{Schematic representation of the process leading to the first line of Eq.\eqref{MEquation}: a single walker moving to a neighbouring lattice site (with $n^W_\fy\geq 1$ and $n^P_\fy\,,n^W_{\fy+e}\,,n^P_{\fy+e}\geq 0$).}
\label{Fig:MasterWalking}
\end{figure}

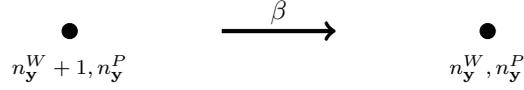
\begin{figure}[t]
\begin{center}
\scalebox{1}{\parbox{7cm}{\begin{tikzpicture}
\draw[fill=black] (0,0) circle (0.1);
\node at (0,-0.5) {\footnotesize $n^W_\fy+1,n^P_\fy$};
\draw[ultra thick,->] (2,0) -- (3.5,0);
\node at (2.75,0.25) {$\beta$};
\draw[fill=black] (5.5,0) circle (0.1);
\node at (5.5,-0.5) {\footnotesize $n^W_\fy,n^P_\fy$};
\end{tikzpicture}}}
\end{center}
\caption{Schematic representation of the branching process leading to the fourth line of \eqref{MEquation}: a single walker 'dying' from poison at the lattice site $\fy$ (with $n^P_\fy\,,n^W_{\fy}\geq 0$).}
\label{Fig:MasterWPtoW}
\end{figure}

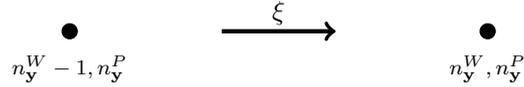
\begin{figure}[t]
\begin{center}
\scalebox{1}{\parbox{7cm}{\begin{tikzpicture}
\draw[fill=black] (0,0) circle (0.1);
\node at (0,-0.5) {\footnotesize $n^W_\fy-1,n^P_\fy$};
\draw[ultra thick,->] (2,0) -- (3.5,0);
\node at (2.75,0.25) {$\xi$};
\draw[fill=black] (5.5,0) circle (0.1);
\node at (5.5,-0.5) {\footnotesize $n^W_\fy,n^P_\fy$};
\end{tikzpicture}}}
\end{center}
\caption{Schematic representation of the branching process leading to the fifth line of \eqref{MEquation}: a single walker is spontaneously created at the lattice site $\fy$ (with $n^W_\fy\geq 1$ and $n^P_{\fy}\geq 0$).}
\label{Fig:MasterCreation}
\end{figure}

The field theory is derived from a discretised version of the model, eventually taking the continuum limit.
The starting point is a \emph{Master Equation} that directly leads to the action through a process of second-quantisation.
Let $\Gamma\subset \mathbb{Z}^d$ be a $d$-dimensional hypercubic lattice with coordination number $q$, which is generated by a set of vectors $\mathbf{e}$.  We denote by $\{n^W_\fx,n^P_\fx\}$ a state
with site $\fx$ occupied by $n^W_\fx$ and $n^P_\fx$ particles of type $W$ and $P$ $\forall \fx\in\Gamma$
(for a schematic representation see \figref{Fig:SchemLattice}). The probability that such state  is realised at time $t$ is denoted by $P(\{n^W_\fx,n^P_\fx\};t)$. Configurations can change via the different mechanisms
described above.
The probability thus satisfies the first order differential equation (Master Equation):
\begin{align}
  \frac{dP(\{{n}^W_\fx,n^P_\fx\};t)}{dt}=&\frac{H}{q}\,\sum_{\fy\in\Gamma}\sum_{e\in\mathbf{e}}\left[(n^W_{\fy+e}+1)P(\{n^W_\fy-1,n^W_{\fy-e}+1,n^P_\fx\};t)-n^W_\fy P(\{n^W_\fx,n^P_\fx\};t)\right]\nonumber\\
  &+\sigma \sum_{\fy\in\Gamma}\left[(n^W_\fy-1)P(\{n^W_\fy-1,n^P_\fx\};t)-n^W_\fy P(\{n^W_\fx,n^P_\fx\};t)\right]\nonumber\\
  &+\alpha \sum_{\fy\in\Gamma}\left[n^W_\fy P(\{n^W_\fx,n^P_\fy-1\};t)-n^W_\fy P(\{n^W_\fx,n^P_\fx\};t)\right]\nonumber\\
  &+\beta\sum_{\fy\in\Gamma}\left[(n^W_\fy+1)n_\fy^P P(\{n^W_\fy+1,n^P_\fx\};t)-n^W_\fy n^P_\fy P(\{n^W_\fx,n^P_\fx\};t)\right]\nonumber\\
&+\xi\sum_{\fy\in\Gamma}\left[P(\{n^W_\fy-1,n^P_\fx\};t)-P(\{n^W_\fx,n^P_\fx\};t)\right]\,.\label{MEquation}
\end{align}
The first line describes diffusion of walkers from one lattice site to one of its $q$ nearest neighbours with frequency $H/q$. This process is schematically shown in \figref{Fig:MasterWalking}. 
There $\{n^W_\fy-1,n^W_{\fy-e}+1,n^P_\fx\}$ denotes the state differing from   $\{n^W_\fx,n^P_\fx\}$ by having a walker less at $\fy$ and a walker more at $\fy-e$.
The second and third lines produce the first two branching processes in Eq.~\eqref{microCardy} respectively and are schematically shown in Figs~\ref{Fig:MasterWtoWW} and \ref{Fig:MasterWtoWP}.
\begin{figure}[t]
\begin{center}
\scalebox{1}{\parbox{7cm}{\begin{tikzpicture}
\draw[fill=black] (0,0) circle (0.1);
\node at (0,-0.5) {\footnotesize $n^W_\fy-1,n^P_\fy$};
\draw[ultra thick,->] (2,0) -- (3.5,0);
\node at (2.75,0.25) {$\sigma$};
\draw[fill=black] (5.5,0) circle (0.1);
\node at (5.5,-0.5) {\footnotesize $n^W_\fy,n^P_\fy$};
\end{tikzpicture}}}
\end{center}
\caption{Schematic representation of the branching process leading to the second line of \eqref{MEquation}: a single walker creating a copy of itself at the the site $\fy$ (with $n^W_\fy\geq 2$ and $n^P_\fy\geq 0$).}
\label{Fig:MasterWtoWW}
\end{figure}
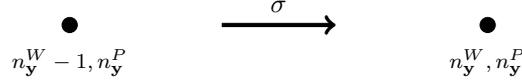
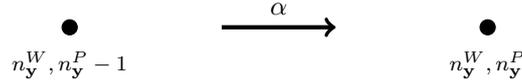
\begin{figure}[t]
\begin{center}
\scalebox{1}{\parbox{7cm}{\begin{tikzpicture}
\draw[fill=black] (0,0) circle (0.1);
\node at (0,-0.5) {\footnotesize $n^W_\fy,n^P_\fy-1$};
\draw[ultra thick,->] (2,0) -- (3.5,0);
\node at (2.75,0.25) {$\alpha$};
\draw[fill=black] (5.5,0) circle (0.1);
\node at (5.5,-0.5) {\footnotesize $n^W_\fy,n^P_\fy$};
\end{tikzpicture}}}
\end{center}
\caption{Schematic representation of the branching process leading to the third line of \eqref{MEquation}: a walker 'drops' poison at the lattice site $\fy$ (with $n^P_\fy\geq 1$ and $n^W_{\fy}\geq 0$).}
\label{Fig:MasterWtoWP}
\end{figure}
The fourth line accounts for the third process there and is graphically represented in \figref{Fig:MasterWPtoW}.
Finally, the last line gives the spontaneous creation of one walker at site $\fy$ and is schematically shown in \figref{Fig:MasterCreation}.

In view of a second quantisation, following the Doi-Peliti approach \cite{Doi1,Doi2,Peliti}, it is natural to interpret the state $\{n^W_\fx,n^P_\fx\}$ as obtained by the action of creation operators $a^\dagger (\fx)$ (for $W$) and $b^\dagger(\fx)$ (for $P$) on a vacuum state. One introduces also the corresponding annihilation operators, $a (\fx)$ and $b (\fx)$, such that
\begin{align}
&a^\dagger(\fx) | \{n^W_\fx,n^P_\fx\} \rangle =  | \{n^W_\fx +1,n^P_\fx\} \rangle \,,&&b^\dagger(\fx)| \{n^W_\fx,n^P_\fx\} \rangle=| \{n^W_\fx,n^P_\fx+1\} \rangle\,, \\
&a(\fx) | \{n^W_\fx,n^P_\fx\} \rangle =  n^W_\fx \, | \{n^W_\fx - 1,n^P_\fx\} \rangle \,,&& b(\fx) | \{n^W_\fx,n^P_\fx\} \rangle =  n^P_\fx \, | \{n^W_\fx ,n^P_\fx-1\} \rangle\,,\\
&\left[a (\fx), a^\dagger (\fy) \right]  = \delta_{\fx,\fy}\,,&&\left[b (\fx), b^\dagger (\fy) \right]  = \delta_{\fx,\fy} \,,
\end{align}
with all other possible commutators between $(a,a^\dagger)$ and $(b,b^\dagger)$ vanishing. The field theory is realised by considering the time-evolution of the state
\begin{equation}
  | \Psi(t) \rangle = \sum_{\{n_\fx^W,n_\fx^P\}} P(\{n^W_\fx,n^P_\fx\};t) \,| \{n_\fx^W,n_\fx^P\} \rangle \;, 
  \end{equation}
which can be derived from the Master Equation~\eqref{MEquation}.
Upon mapping each operator to conjugate fields
\begin{eqnarray}
  a \rightarrow W\,&,& \quad \tilde{a}=a^\dagger -1 \rightarrow W^+\;, \nonumber \\
  b \rightarrow P\,&,& \quad \tilde{b}=b^\dagger -1 \rightarrow P^+\;,
\end{eqnarray}
where the tilded operators are known as Doi-shifted operators, one finds that the evolution
is controlled by $\exp\{-\int d^dx dt \, S(W^+,W,P^+,P)\}$, with the action density $S$ given by
\begin{align}
S= & W^+\partial_t W +P^+\partial_t P + D \nabla W^+ \nabla W -\sigma(1+W^+)W^+W\nonumber \\
& -\alpha(1+W^+)P^+W + \beta(1+P^+)W^+WP  -\xi W^+ \;,
\label{actionCardy}
\end{align}
where$D=\lim_{\mathsf{a}\to0} H\mathsf{a}^2/q$ is the hopping rate in the continuum ($\mathsf{a}$ is the lattice spacing). The action in Eq.\eqref{actionCardy} corresponds to
the result in~\cite{Cardy_1985} augmented here by the last source term due to spontaneous generation.
This produces a background of infected and it is responsible in this approach for a `strolling' dynamics,
as we motivate in Section~\ref{Sect:SpontaneousCreation}. 

The renormalisation group equations stemming from the action in Eq.\eqref{actionCardy}, which follow closely those of other theories such as directed percolation models or reggeon field theory \cite{reggeon1,reggeon2}, have been analysed in~\cite{Cardy_1985}. In particular, the Fourier transform of the correlation function of a field $W$ and a field $W^+$  was computed and shown to satisfy the following scaling law near criticality
\begin{align}
\mathcal{F}\left(\langle W(\vec{x},t)\,W^+(0,0)\rangle\right)(\omega,\vec{k})=|\vec{k}|^{\eta-2}\,\Phi(\omega\,\Delta^{\nu_t}\,,\vec{k}\,\Delta^{\nu})\,,
\end{align}
for some function $\Phi$. Here $\Delta$ is a measure for the proximity to criticality (\emph{i.e.} it is proportional to $p-p_c$ of Eq.~(\ref{DefCriticalExponent}) in the context of the percolation model) and $(\eta,\nu_t,\nu)$ are critical exponents determining the universality class of the model.\footnote{In a dimensional regularisation scheme, they were found to be \cite{Cardy_1985}
\begin{align}
&\eta=-\frac{\epsilon}{21}\,,&&\nu_t=1+\frac{\epsilon}{28}\,,&&\nu=\frac{1}{2}-\frac{5}{84}\,\epsilon\,,
\end{align} 
where $\epsilon=6-d$.} The quantity above is a measure for the probability of finding a walker at some generic time and position $(\vec{x},t)\in\mathbb{R}^{6}$ if there was one at the origin, where $d=6$ corresponds to the critical dimension of the system \cite{Cardy_1985}.


\subsection{Relation to Compartmental Models}\label{Sect:RelationPercSIR}
As mentioned before, the model described by the action in Eq.\eqref{actionCardy} is in the same universality class as numerous other models that are directly relevant for the study of epidemic processes. As shown in~\cite{Cardy_1985} the particular choice $\xi=0$, in fact, includes the SIR model, which is the most prominent representative of compartmental models. To make the connection more concrete, we return to studying the time evolution of a disease on a lattice $\Gamma$ and divide the individuals that are present at a given lattice site $\fx\in\Gamma$ into three classes  or compartments~\cite{Grassberger1983}, as defined in Section~\ref{Sect:theprinciple}. We shall denote ${\msl{\fx}, \mil{\fx}, \mrl{\fx}}$ the number of susceptible, infectious and removed individuals at $\fx$, respectively.~\footnote{The occupation numbers $(\msl{\fx},\mil{\fx},\mrl{\fx})$ are denoted $(X(\fx), Y(\fx),Z(\fx))$ respectively in~\cite{Grassberger1983}. }

Concretely, for $\xi=0$, the model in~\cite{Grassberger1983} is very suitable for numerical Markovian simulations and can be connected to the SIR model. The processes of the model in~\cite{Grassberger1983} are
\begin{align}
 \msl{\fx}+\mil{\fx'} &\rightarrow  \mil{\fx}+\mil{\fx'} \;, && {\rm infection\; with \; rate \;} \hat\gamma\;,  \nonumber\\
  \mil{\fx} &\rightarrow  \mrl{\fx} \;, && {\rm recovery \; with\; rate \;} \hat\epsilon\;,  
\label{microGrass}
\end{align}
where $\fx$ and $\fx'$ are nearest neighbour sites on $\Gamma$ (\emph{i.e.} $\fx'=\fx+e$ for some basis vector $e\in\mathbf{e}$). 
As discussed in~\cite{Grassberger1983}, treating the process as deterministic (in particular, interpreting $(\msl{\fx},\mil{\fx},\mrl{\fx})$ as continuous functions of time) one obtains the following equations of motion
\begin{eqnarray}
  \frac{d \msl{\fx}}{dt}(t) &=& -\hat\gamma\, \msl{\fx}(t)\sum_{e\in\mathbf{e}}\mil{\fx+e}(t)                        \;, \nonumber \\
  \frac{d \mil{\fx}}{dt}(t) &=&  \hat\gamma\, \msl{\fx}(t)\sum_{e\in\mathbf{e}}\mil{\fx+e}(t) -\hat\epsilon\, \mil{\fx}(t)       \;, \nonumber \\
  \frac{d \mrl{\fx}}{dt}(t) &=&  \hat\epsilon\, \mil{\fx}(t)  \;, 
  \label{Grasseom}
\end{eqnarray}
where the sums on the right hand side extend over the nearest neighbours of $\fx$. Since the sum of all three equations in (\ref{Grasseom}) implies $\frac{d}{dt}(\msl{\fx}+\mil{\fx}+\mrl{\fx})(t)=0$, the total number of individuals is conserved and we denote its value by
\begin{align}
N=\sum_{\fx\in\Gamma}(\msl{\fx}(t)+\mil{\fx}(t)+\mrl{\fx}(t))\,.
\end{align}
Furthermore, we introduce the relative number of susceptible, infectious and removed individuals respectively
\begin{align}
&\ms(t)=\frac{1}{N}\,\sum_{\fx\in\Gamma}\msl{\fx}(t)\,,&&\mi(t)=\frac{1}{N}\,\sum_{\fx\in\Gamma}\mil{\fx}(t)\,,&&\mr(t)=\frac{1}{N}\,\sum_{\fx\in\Gamma}\mrl{\fx}(t)\,,
\end{align}
which satisfy
\begin{align}
\ms(t)+\mi(t)+\mr(t)=1\,.\label{ConstraintsEom}
\end{align}
Finally, by taking a \emph{mean-field approximation} for the infected field in Eq.\eqref{Grasseom} (\emph{i.e.} replacing $\mil{\fx}$ by $\mi(t)$ $\forall \fx\in\Gamma$, such that the sums $\sum_{e\in\mathbf{e}}\mil{\fx+e}$ in Eq.\eqref{Grasseom} are replaced by $\frac{q}{N}\sum_{\fx\in\Gamma} \mil{\fx}=q\mi(t)$) and summing
over all $\fx\in\Gamma$, one obtains the following coupled first order differential equations:
\begin{align}
&\frac{d \ms}{dt}(t) = -q\,\hat\gamma\, \ms(t)\,\mi(t)\,,&&  \frac{d \mi}{dt}(t) = q\,\hat\gamma\, \ms(t)\,\mi(t) -\hat\epsilon\,\mi(t) \,,&&
\frac{d \mr}{dt}(t) =  \hat\epsilon\,\mi(t) \;, \label{SIReoms}
\end{align}
where $q$ is the coordination number, \emph{i.e.}, the number of nearest neighbours for each site ($4$ in a two-dimensional rectangular lattice).  As we shall discuss in the next section, this system of differential equations, which has to be solved under the constraint in Eq.\eqref{ConstraintsEom} and with suitable initial conditions, is structurally of the same form as the SIR model~\cite{Kermack:1927}, one of the oldest deterministic models to describe the spread of a communicable disease.

Spontaneous generation can be included in Eq.\eqref{SIReoms} as an additional process
\begin{align}
&\msl{\fx} \rightarrow \mil{\fx}\,,&&\text{ with rate }\hat\xi \,.
\end{align} 
In the deterministic and mean-field equations, this amounts to a term $-\hat\xi \ms(t)$ in the first equation of~\eqref{SIReoms}, and the corresponding one with opposite sign in the second equation, as we shall discuss in the context of the SIR model in Section~\ref{Sect:SpontaneousCreation}.


 
 \section{Compartmental Models}\label{Sect:CompartmentalModels}


\begin{summary}{Executive Summary}
\begin{itemize}
\item We introduce compartmental models as \emph{deterministic} approaches that describe the diffusion of infectious diseases through coupled differential equations in time.
\item These models are characterised through a set of diffusion rates and initial conditions. Time is assumed to be a continuous variable.
\item We discuss mathematical aspects of different models capturing the asymptotic behaviour of their dynamics.
\item We show how to extract several epidemiologically relevant phenomena from compartmental models, including the endemic behaviour of diseases, the impact of superspreaders, the possibility of re-infection and multi wave patterns.
\item We show that compartmental models can be naturally related to the renormalisation group framework. 
\end{itemize}

\end{summary}

\subsection{SIR(S) Model, Basic Definitions}\label{Sect:BasicSIRmodel}
Independently of percolation models and epidemic field theory descriptions, the differential equations \eqref{SIReoms} have been proposed as early as 1927 to describe the dynamic spread of infectious diseases in an isolated population of total size $N\gg 1$. As reviewed in the Historical Overview (Section~\ref{Sect:HistoO}), a major breakthrough in the systematic study of the time evolution of infectious diseases was the application of the mass-action law \cite{Hamer,HamerLect1,HamerLect2,HamerLect3,Ross1911,Ross1916,RossHudson1916II,RossHudson1916III,McKendrick1912,McKendrick1914,McKendrick1926,Kermack:1927}, stating roughly that the rate at which individuals of two different types meet is proportional to the product of their total numbers.~\footnote{In fact H.~Heesterbeek \cite{Heesterbeek} remarks that 'In short, mass-action turned epidemiology into a science.'} This law underlies many chemical reactions, in which different agents mix. In the context of epidemiology, it leads to a class of deterministic approaches that are called \emph{compartmental models}, whose hallmark is to divide the population into several distinct classes. Each class or compartment is comprised of individuals that have peculiar behaviour in the context of the disease. The simplest of this class of models is called SIR and, as the name indicates, it includes three compartments, as already described in detail in Section~\ref{Sect:RelationPercSIR}: 
\begin{itemize}
\item \emph{{\bf S}usceptible:} the total number of susceptible individuals at time $t$ shall be denoted $N\,S(t)$.
\item \emph{{\bf I}nfectious:} the total number of infectious individuals at time $t$ shall be denoted $N\,I(t)$.
\item \emph{{\bf R}emoved (recovered):} the total number of removed individuals at time $t$ shall be denoted $N\,R(t)$.
\end{itemize}

\noindent
Depending on the type of disease under consideration, other compartments can be included to make the model more realistic, \emph{e.g.} \cite{WANG20161,HETHCOTErev}
\begin{itemize}
\item \emph{Passively immune ($M$):} infants that have been born with (temporary) passive immunity ($M$ stands for maternally-derived immunity).
\item \emph{{\bf E}xposed ($E$):} individuals in the latent period, who are infected but not yet infectious.
\item \emph{{\bf D}eceased ($D$):} individuals who have died from the disease (in some models $D$ is considered to be part of $R$).
\item \emph{{\bf C}arrier ($C$):} individuals in a state where they have not completely recovered and still carry the disease, but do not suffer from it  (examples of diseases for which this compartment is of relevance are tuberculosis and typhoid fever, see \emph{e.g.} \cite{Carrier}).
\item \emph{{\bf Q}uarantine ($Q$):} individuals who have been put under quarantine or lockdown measures (see \emph{e.g.} \cite{Liu2020.03.09.20033498}).
\item \emph{{\bf V}accinated ($V$):} individuals who are vaccinated against the disease, thus acquiring partial or total immunity.
\end{itemize}
Models are usually named/classified according to the compartments they contain, \emph{e.g.} SIR, MESIR, \emph{etc.} Repetition of labels indicates that individuals may return into a given compartment several times, \emph{e.g.} SIS denotes a model in which infectious individuals may become susceptible again after an infection. Furthermore, each compartment can be generalised to include dependences on biological (\emph{e.g.} age, gender, \emph{etc.}) and/or geometric parameters (\emph{e.g.} parameters measuring geographic mobility, \emph{etc.}). To better model social and behavioural particularities among the population (but also to simulate different variants of a given disease), models can include multiple copies of a compartment with slightly different properties (see \emph{e.g.} \cite{HethcoteYorke2}, which includes several different classes of susceptible, each of which with a different infection rate, to model the spread of gonorrhoea). Finally, depending on the duration of the epidemic, the birth and death dynamic needs to be taken into account \cite{Omran1, Omran2}.  That is, into each class, new individuals may be born, or individuals of each compartment can die from causes other than the disease.

In the following, for simplicity, we shall only consider models including the compartments S, I and R. We assume that the total size of the population remains constant, \emph{i.e.} we impose the algebraic relation
\begin{align}
&1=S(t)+I(t)+R(t)\,,&&\forall t\in\mathbb{R}_+\,,\label{ConstraintSIR}
\end{align}
where, without restriction of generality, we assume that the outbreak of the epidemic starts at $t=0$. We shall also refer to $\ms$, $\mi$ and $\mr$ as the \emph{relative} number of susceptible, infectious and removed individuals, respectively. Furthermore, we assume that $N$ is sufficiently large such that we can treat $S$, $I$ and $R$ as continuous functions of time:
\begin{align}
&\ms\,,\mi\,,\mr\,:\hspace{1cm} \mathbb{R}_+\longrightarrow [0,1]\,.
\end{align}
While in Section \ref{Sect:RelationPercSIR} the differential equations \eqref{SIReoms} were a consequence of the basic microscopic processes in Eq.\eqref{microGrass} on the lattice $\Gamma$, within compartmental models they are independently argued on the basis of dynamical mechanisms that change $(S,I,R)$ as functions of time:
\begin{itemize}
\item Infectious individuals can infect susceptible individuals, turning the latter into infectious individuals themselves. We call an `infectious contact' any type of contact that results in the transmission of the disease between an infectious and a susceptible and we denote the average number of such contacts per infectious individual per unit of time by $\rin$. In the original SIR model \cite{Kermack:1927}, $\rin$ is considered to be constant (\emph{i.e.} it does not change over time), however, in the following sections we shall not always limit ourselves to this restriction. The total number of susceptible individuals that are infected per unit of time (and thus become infectious themselves) is thus $\rin\,N\,\ms\,\mi$.
\item Infectious individuals can be removed by recovering (and thus gaining temporary immunity) or by being given immunity (\emph{e.g} via vaccinations), by death or via any other form of removal. We shall denote $\rhe$ the rate at which infected individuals become removed. As before, we consider $\rhe$ as a function that may change with time.
\item Removed individuals may become susceptible again after some time or, conversely, susceptible individuals may become directly removed. In both cases we shall denote the respective rate by $\rsu$, which may be positive or negative. If removed individuals are only temporarily immune against the disease, they can become susceptible again. In this case $\rsu>0$, which corresponds to the rate at which removed individuals become susceptible again. Susceptible individuals may become immunised against the disease (\emph{e.g.} through vaccinations). In this case $\rsu<0$. We remark that this is not the only way to implement vaccinations to compartmental models,
as the most direct way is to add a specific compartment.
\end{itemize}

\begin{wrapfigure}{r}{0.45\textwidth}
\begin{center}
\vspace{-0.8cm}
\scalebox{1}{\parbox{7cm}{\begin{tikzpicture}
\draw[thick] (0,0) rectangle (1,1);
\draw[ultra thick,->] (1.2,0.5) -- (2.8,0.5);
\node at (2,0.1) {$\rin\, N\,\mi \,\ms $};
\node at (0.5,0.5) {$N\,S$};
\begin{scope}[xshift=3cm]
\draw[thick] (0,0) rectangle (1,1);
\draw[ultra thick,->] (1.2,0.5) -- (2.8,0.5);
\node at (0.5,0.5) {$N\,I$};
\node at (2,0.1) {$\rhe N\,\mi $};
\end{scope}
\begin{scope}[xshift=6cm]
\draw[thick] (0,0) rectangle (1,1);
\node at (0.5,0.5) {$N\,R$};
\end{scope}
\draw[ultra thick,->] (6.5,1.2) -- (6.5,1.6) -- (0.5,1.6) -- (0.5,1.2);
\node at (3.5,1.9) {$\rsu\,N\,\mr $};
\end{tikzpicture}}}
\end{center}
\caption{Flow between susceptible, infectious and removed individuals.}
\label{Fig:SIRschem}
${}$\\[-2.9cm]
\end{wrapfigure}
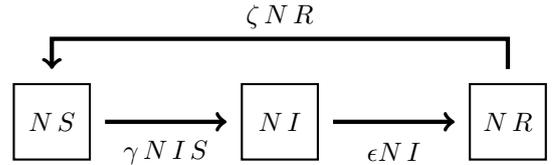

\noindent
The flow among susceptible, infectious and removed is schematically shown in \figref{Fig:SIRschem}. 
The dynamics of the system is also crucially determined by the initial conditions in each compartment. 
As already mentioned, we consider $t=0$ as the start of the epidemic diffusion, where a non-zero 
number of infectious individuals is needed for the diffusion to start. Without loss of generality, we start with zero removed at the initial time.
Hence, the initial conditions are given by
\begin{align}
&S(t=0)=\ms_0\,, && I(t=0)=\mi_0\,, & R(t=0)=0\,,\label{SIRInitial}
\end{align}
where $\ms_0,\mi_0\in[0,1]$ are constants that satisfy $\ms_0+\mi_0=1$. With this notation, the time dependence of $\ms$, $\mi$ and $\mr$ is described by the following set of coupled first order differential equations~\footnote{These equations coincide with Eq.\eqref{SIReoms} upon identifying $q\hat\gamma \equiv \gamma$, $\hat\epsilon \equiv \epsilon$, and for $\zeta = 0$. Spontaneous generation of infectious individuals can be added straightforwardly.}

\noindent
\begin{align}
&\frac{d\ms}{dt}=-\rin\,\mi \,\ms +\rsu\,\mr\,,\nonumber\\
& \frac{d\mi}{dt}=\rin\,\mi\,\ms -\rhe\,\mi \,,\nonumber\\
& \frac{d\mr}{dt}=\rhe\,\mi -\rsu\,\mr \,,\label{SIReqs}
\end{align}
together with the initial conditions (\ref{SIRInitial}). Notice that $\frac{d}{dt}(\ms(t)+\mi(t)+\mr(t))=0$ such that the initial conditions (\ref{SIRInitial}) with $\ms_0=\mi_0=1$ guarantee the algebraic relation (\ref{ConstraintSIR}).

For $\zeta=0$, the system of equations (\ref{SIReqs}) is indeed the same model as described in Section~\ref{Sect:RelationPercSIR}, which is called the SIR-model~\cite{Kermack:1927}. For $\zeta> 0$, this model is sometimes referred to as the SIRS model, since it holds the possibility that recovered individuals may become susceptible again.

\subsection{Numerical Solutions and their Qualitative Properties}\label{Sect:SIRQualitative}

\begin{figure}[h]
\begin{align}
&\includegraphics[width=7cm]{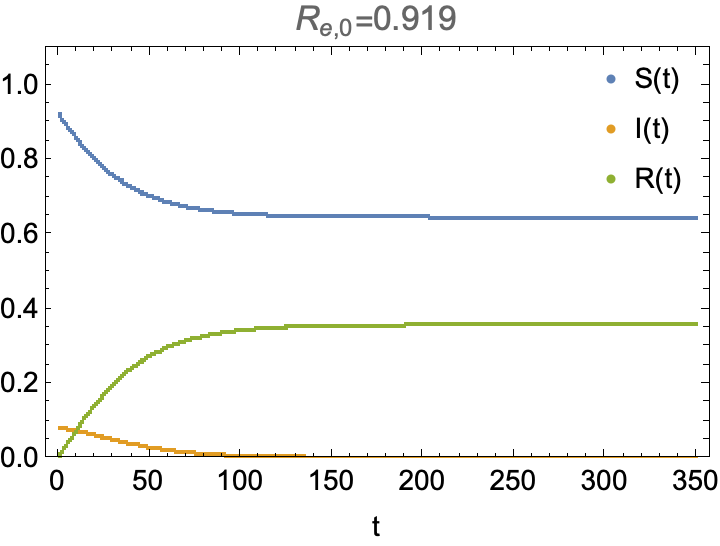} && \includegraphics[width=7cm]{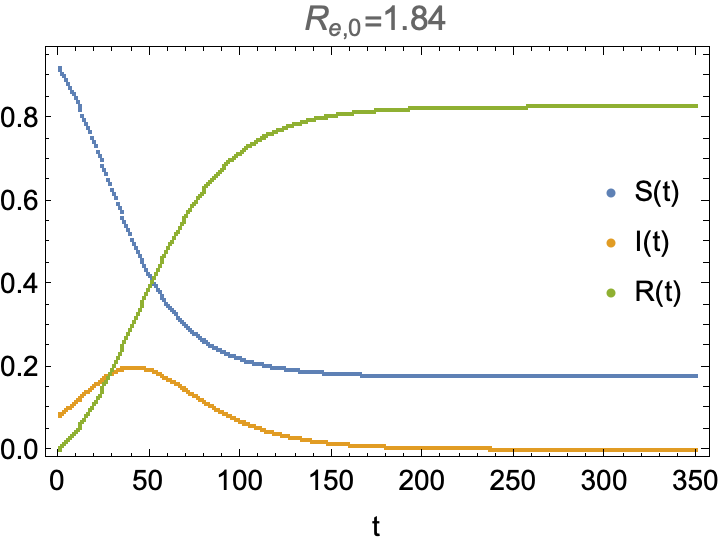}\nonumber
\end{align}
\caption{Numerical solution of the differential equations (\ref{SIReqs}) for $\ms_0=0.92$, $\rin=0.1$ and $\rsu=0$ for two different choices of $\epsilon$: $\epsilon=0.1001$ such that $R_{e,0}=0.919$ (left) and $\epsilon=0.05$ such that $R_{e,0}=1.84$ (right). }
\label{Fig:SIRnumerical}
\end{figure}

The Eqs~(\ref{SIReqs}) can be solved analytically for $\zeta=0$, as we will discuss in the next subsection. First, we shall present some qualitative remarks that can be deduced by considering numerical solutions, which we obtained by using a simple forward Euler method (see \emph{e.g.} \cite{Press,Butcher}). We first consider $\zeta=0$, for which the temporal evolution of $(\ms,\mi,\mr)$ is illustrated in \figref{Fig:SIRnumerical} in two qualitatively different scenarios, depending on the value of the \emph{initial effective reproduction number} $R_{e,0}$, that we define as \cite{Delamater_2019} (see also \cite{HeesterbeekR0,KEELING200051,HeesterbeekR02,HeffernanR,PMID:17490942,PELLIS201285} for further discussion of the effective reproduction number)
\begin{align}
R_{e,0} = \ms_0\, \sigma\,, \qquad \sigma = \frac{\rin}{\rhe}\,.\label{DefSIRsigma}
\end{align}
The quantity $\sigma$, often called \emph{basic reproduction number} ($R_0$), can be interpreted as the average number of infectious contacts of a single infectious individual during the entire period they remain infectious. In other words, $\sigma$ is the average number of susceptible individuals infected by a single infectious one. In the left panel of \figref{Fig:SIRnumerical}, $(\rin,\rhe,S_0)$ have been chosen such that $R_{e,0}<1$: in this case, even though at initial time a significant fraction of the population ($8\%$) is infectious, the function $\mi(t)$ decreases continuously, leading to a relatively quick eradication of the disease. This is also visible directly from Eqs~(\ref{SIReqs}): since (for $\zeta=0$) $\ms(t)$ is a monotonically decreasing function (\emph{i.e.} $\ms(t)\leq \ms_0$ $\forall t>0$), then $\frac{d\mi}{dt}(t)<0$ $\forall t>0$ such that the number of infectious individuals is continuously decreasing.
In the right panel of \figref{Fig:SIRnumerical}, we chose $R_{e,0}>1$: the number of infectious cases grows to a maximum and starts decreasing once only a small number of susceptible individuals remain available. This maximum is reached once $\ms(t)=\frac{1}{\sigma}$ such that $\frac{d\mi}{dt}=0$.

This behaviour is more clearly visible in the asymptotic number of susceptible (\emph{i.e.} $\ms(\infty)=\lim_{t\to \infty}\ms(t)$) or (equivalently) the cumulative number of individuals that have become infected throughout the entire epidemic. Both quantities are a measure of how far the disease has spread among the population. For later use, we define the function $\Ic(t):\,[0,\infty) \mapsto [0,N]$ as
\begin{align}
\Ic(t)=N\,\mi _0+\int_0^t dt'\,\rin\, N\,\mi (t')\,\ms (t')\,.\label{CumulativeInfected}
\end{align}
It quantifies the cumulative total number of individuals who have been infected by the disease up to time $t$. The definition \eqref{CumulativeInfected} can be used for generic $\rsu$ as a function of time. For $\rsu=0$, using Eqs~\eqref{SIReqs}, we obtain the identity $\rin\,\mi \,\ms = \frac{d}{dt}(\mi +\mr )$ that allows to simplify Eq.\eqref{CumulativeInfected} to:
\begin{align}
&\Ic(t)=N(\mi (t)+\mr (t))=N(1-\ms(t))\,,&&\text{for} &&\zeta=0\,.
\end{align}
For $\rsu=0$, we also have that $\lim_{t\to \infty}\mi(t)\to 0$, thus we find the following relations at infinite time: 
\begin{align}
\Ic(\infty)=\lim_{t\to \infty}\Ic(t)=1-\ms(\infty)=\mr(\infty)=\lim_{t\to\infty} \mr(t)\,.
\end{align}

\begin{wrapfigure}{r}{0.45\textwidth}
\vspace{-0.3cm}
\parbox{7cm}{\includegraphics[width=7cm]{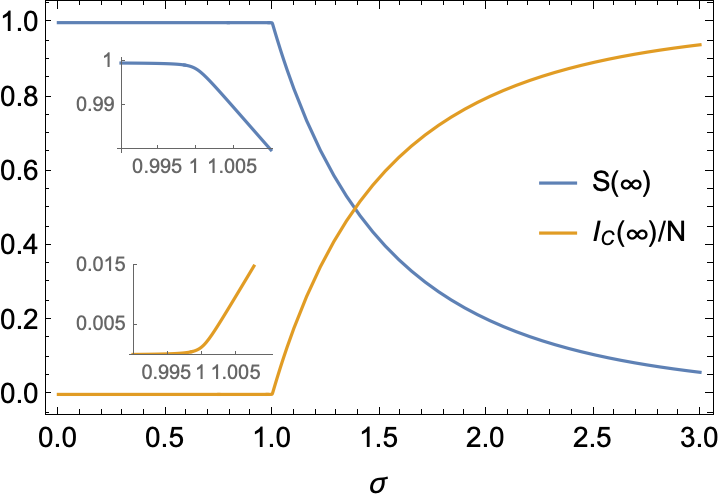}}
\caption{Asymptotic number of susceptible and cumulative number of infectious as a function of $R_{e,0}$ for $\ms_0=1-10^{-6}$.}
\label{Fig:SIRasymptotic}
\end{wrapfigure} 

The limit $\ms(\infty)$ can be computed analytically, by realising that 
\begin{align}
G(t)=\ms(t)\,e^{\sigma\,\mr(t)}\,,
\end{align}
is conserved, \emph{i.e.} $\frac{dG}{dt}(t)=0$ $\forall t\in\mathbb{R}$. This implies
\begin{align}
\ms(t)=\ms_0\,e^{-\sigma(1-\mi(t)-\ms(t))}\,.\label{SformIS}
\end{align}
With $\lim_{t\to\infty} \mi(t)=0$, this equation can be solved for the asymptotic number of susceptible in the limit $t\to\infty$, giving
\begin{align}
\ms(\infty)=-\frac{\ms_0}{R_{e,0}}\,W(-R_{e,0}\,e^{-\frac{R_{e,0}}{\ms_0}})\,,\label{SIRanalyticAsymptotic}
\end{align}
where $W$ is the Lambert function. The limiting values $\ms(\infty)$ and $\Ic(\infty)/N$ are shown in \figref{Fig:SIRasymptotic} as functions of $R_{e,0}$ for the initial conditions of $\ms_0=1-10^{-6}$, \emph{i.e.} a starting configuration with one infectious individual per million. A kink seems to appear for $R_{e,0} = 1$, however both functions are smooth (continuous and differentiable) for $\ms_0<1$, as highlighted in the subplots. In the limit $\ms_0 \to 1$, the solutions discontinuously jump to constants, as the absence of initial infectious individuals prevents the spread of the disease. Qualitatively, this plot shows that for $R_{e,0}<1$, the disease becomes eradicated before a significant fraction of the population can be infected. However for $R_{e,0}>1$ the cumulative number of infected grows rapidly.

For $\rsu\neq 0$, we can distinguish two different cases, depending on the sign:
\begin{itemize}
\item {\bf Re-infection} $\rsu>0$: a positive $\rsu$ implies that removed individuals become susceptible again after some time. This can be interpreted to mean that recovery from the disease only grants temporary immunity, such that a re-infection at some later time is possible. At large times $t\to\infty$, the system enters into an equilibrium state, such that $(\ms(t)\,,\mi(t)\,,\mr(t))$ approach constant values $(\ms(\infty)\,,\mi(\infty)\,,\mr(\infty))$. To find the latter, we impose the equilibrium conditions
\begin{align}
&\lim_{t\to\infty}\frac{d^n\ms}{dt^n}(t)=\lim_{t\to\infty}\frac{d^n\mi}{dt^n}(t)=\lim_{t\to\infty}\frac{d^n\mr}{dt^n}(t)=0\,,&&\forall n\in\mathbb{N}\,,
\end{align}
which have as solution
\begin{align}
(\ms(\infty),\mi(\infty),\mr(\infty))=\left\{\begin{array}{lcl}(1,0,0) & \text{if} & \sigma\leq 1\text{ or }\ms_0=1\,,\\[10pt] \left(\frac{\rhe}{\rin}\,,\frac{(\rin-\rhe)\rsu}{\rin(\rhe+\rsu)}\,,\frac{(\rin-\rhe)\rhe}{\rin(\rhe+\rsu)}\right) & \text{if} & \sigma>1\,,\end{array}\right.&&\text{for} &&\zeta>0\,.\label{ZetaAsympt}
\end{align}
Here we have used that $0\leq (\ms(t)\,,\mi(t)\,,\mr(t))\leq 1$ (in particular that $(\ms(t)\,,\mi(t)\,,\mr(t))$ cannot become negative) as well as the fact that the equilibrium point $(1,0,0)$ cannot be reached for $\ms_0<1$ and $\rin>\rhe$: indeed, this would require
\begin{align}
&\ms(t)> \frac{\rhe}{\rin}\,,&&\text{and} &&\frac{d\mi}{dt}(t)<0\,,
\end{align}
which are not compatible with Eqs~\eqref{SIReqs}.~\footnote{Furthermore, the only solutions of the conditions $\frac{d^2\ms}{dt^2}(t)=\frac{d\mi}{dt}(t)=\frac{d^2\mr}{dt^2}(t)=0$ are in fact the two equilibrium points (\ref{ZetaAsympt}) (where in fact all derivatives of $(\ms\,,\mi\,,\mr)$ vanish). This therefore suggests that there are no solutions that are continuous oscillations with non-decreasing amplitudes and the system indeed reaches an equilibrium at $t\to \infty$. The numerical solutions in~\figref{Fig:SIRzetaNumerical} comply with this expectation.} The two qualitatively different solutions of Eqs~\eqref{SIReqs} that lead to the asymptotic equilibria (\ref{ZetaAsympt}) are plotted in \figref{Fig:SIRzetaNumerical}: for $\sigma <1$ (left panel), the disease is eradicated and the individuals that have been infected eventually move back to be susceptible; for $\sigma > 1$ (right panel), after some oscillations, an equilibrium is reached between the infections and the end of immunity and the number of infectious individuals tends to the non-zero constant given in Eq.\eqref{ZetaAsympt} (this corresponds to an endemic state of the disease). The distinction between eradication of the disease and the endemic phase does not depend on $S_0$ (except for the trivial initial condition $\ms_0=1$) but only on the basic reproduction number $\sigma$. This fact can be intuitively understood as the rate $\zeta$ dynamically increases the number of susceptible individuals, thus the regime becomes independent of the initial condition. 

\begin{figure}[ht]
\begin{align}
&\includegraphics[width=7cm]{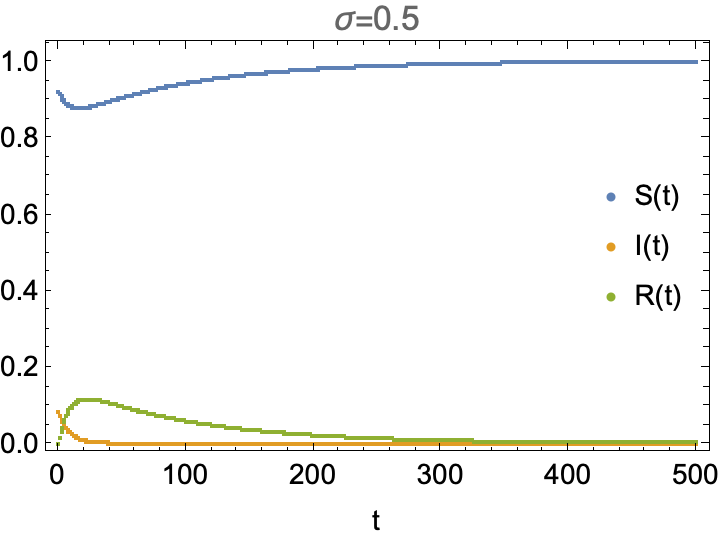} && \includegraphics[width=7cm]{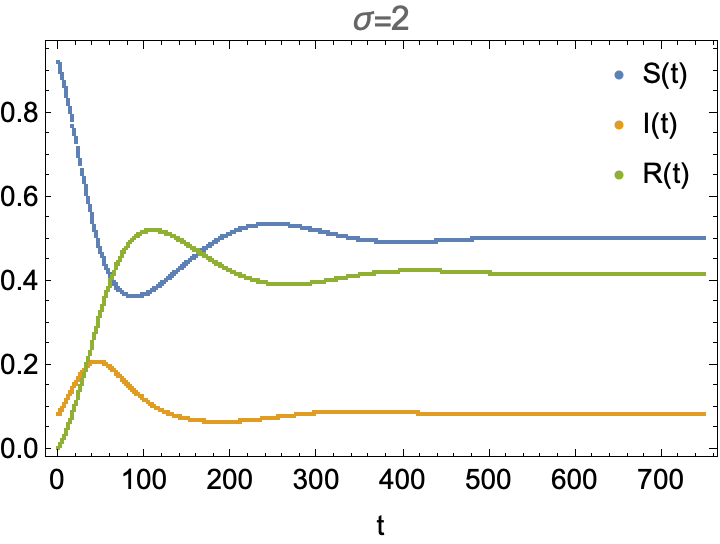}\nonumber
\end{align}
\caption{Numerical solution of the differential equations (\ref{SIReqs}) for $\ms_0=0.92$, $\rin=0.1$ and $\rsu=0.01$ for two different choices of $\rhe$: $\rhe=0.2$ implying $\sigma=0.5$ (left) and $\rhe=0.05$ implying $\sigma=2$ (right). }
\label{Fig:SIRzetaNumerical}
\end{figure}

\item {\bf Direct immunisation} $\rsu<0$: a negative $\rsu$ implies the possibility that over time susceptible individuals can become removed and thus immune to the disease, proportionally to the number of removed individuals. 
Schematically, different solutions are shown in \figref{Fig:SIRNegzetaNumerical}. For $\zeta<0$ the dynamics always leads to the asymptotic values $(\ms(\infty)\,,\mi(\infty)\,,\mr(\infty))=(0,0,1)$ at large $t\to\infty$.

\begin{figure}[ht]
\begin{align}
&\includegraphics[width=7cm]{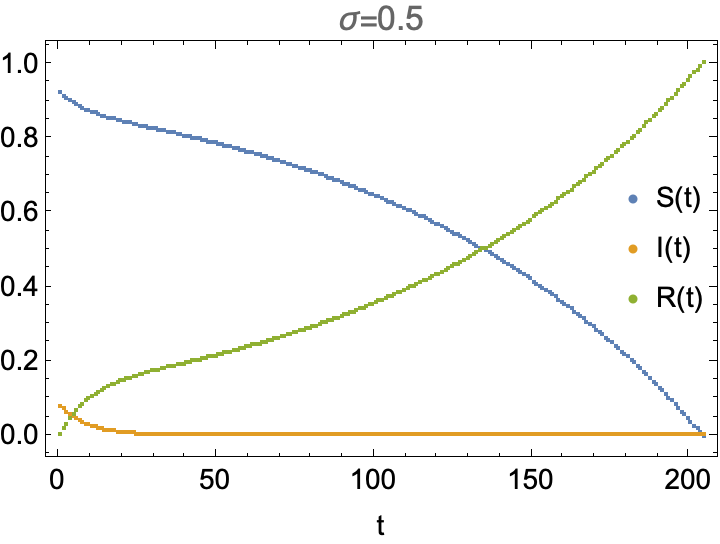} && \includegraphics[width=7cm]{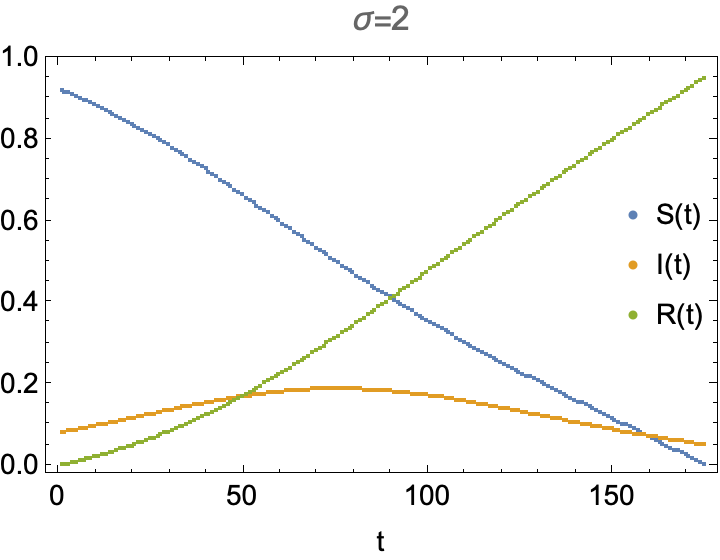}\nonumber
\end{align}
\caption{Numerical solution of the differential equations (\ref{SIReqs}) for $\ms_0=0.92$, $\rin=0.1$ and $\rsu=-0.01$ for two different choices of $\rhe$: $\rhe=0.2$ implying $\sigma=0.5$ (left) and $\rhe=0.05$ implying $\sigma=2$ (right). }
\label{Fig:SIRNegzetaNumerical}
\end{figure}

\end{itemize}

\subsection{From Lattice to SIR}\label{FromLatticeToSIR}
The relation between Compartmental Models and Percolation Field Theory has already been established in Section \ref{Sect:RelationPercSIR}. However it is also possible to link the numerical simulations to the SIR model directly, as the microscopic processes in the lattice simulations are in one-to-one correspondence with the transfer mechanisms among compartments in the SIR model.

\begin{figure}
\centering
\subfloat[$r=1$]{\includegraphics[width=0.5\textwidth]{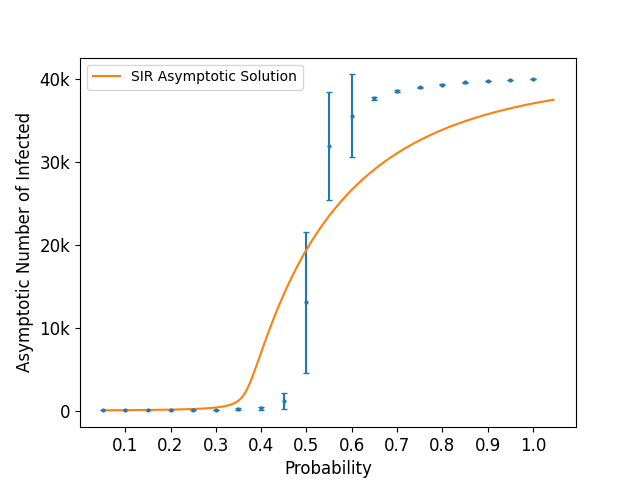}}
\subfloat[$r=2$]{\includegraphics[width=0.5\textwidth]{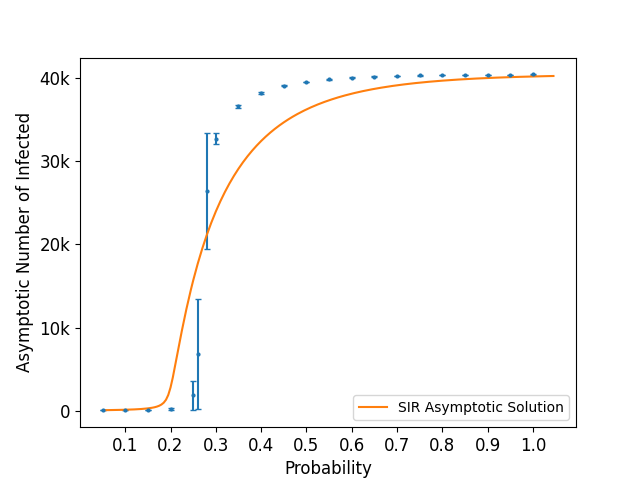}}\\
\subfloat[$r=5$]{\includegraphics[width=0.5\textwidth]{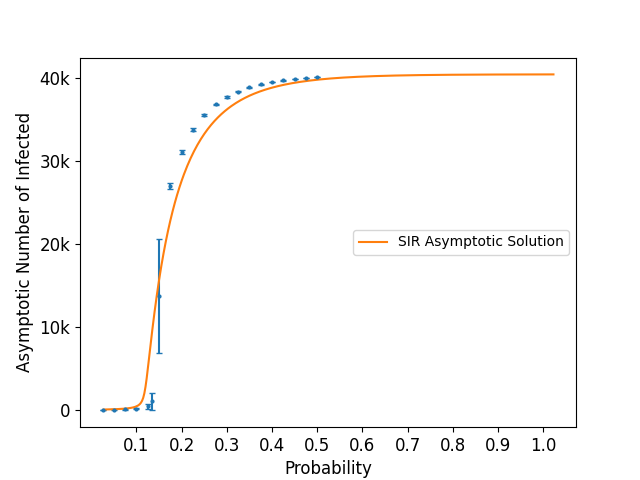}}
\subfloat[$r=50$]{\includegraphics[width=0.5\textwidth]{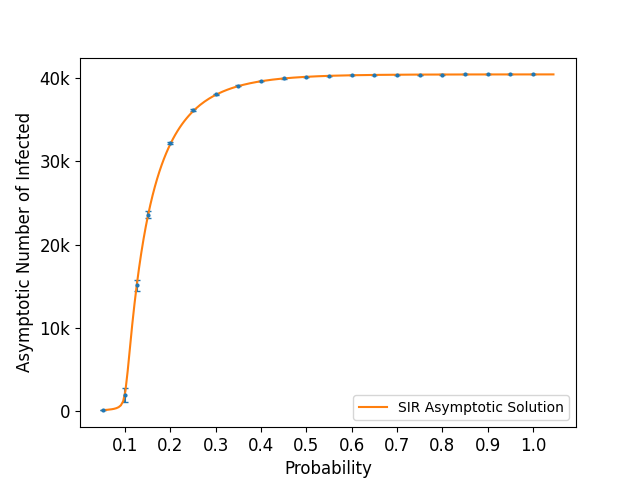}}
\caption{Evolution of the final number of infected cases as a function of the infection probability for different coordination radii $r$, compared to the asymptotic solution of the SIR model. The optimal factor found for the cases (a),(b),(c) and (d) are respectively: $\rho= 0.27, \; 0.42, \; 0.50, \; 0.99 $. }
\label{compare}
\end{figure}

To visualise this we used the results in \figref{phasetransition}, where the lattice is of size $201 \times 201$ (\emph{i.e.} a population of $40401$) and the recovery probability is fixed to $0.1$. Once the recovery rate and the initial number of susceptible individuals $S_0$ is fixed, in the SIR model the value of the infection rate completely determines the asymptotic number of total infected via Eq.\eqref{SIRanalyticAsymptotic}. For each coordination radius, we look for the best rescaling of the infection probability that could reproduce the behaviour in \figref{phasetransition}, \emph{i.e.} we compute the optimal $\rho$ such that changing $\mathfrak{g} \longrightarrow \rho \mathfrak{g}$ gives the best fit of the numerical results. We show the solution in \figref{compare}.

The results clearly show that increasing the coordination radius improves the match between the lattice and the SIR model results. The reason for this is simple: for maximal coordination radius, the mean-field approximation applied to Eq.~\eqref{Grasseom} leads directly to the SIR equations. The reason is that any infectious site can infect any susceptible site on the lattice with equal probability. Numerical lattice simulations of compartmental models, and in particular of the SIR type, have been widely used in the literature (see \emph{e.g.} \cite{de_Souza_2010,Tom__2010,Santos_2020,alves2019epidemic}).

\subsection{Parametric Solution of the Classical SIR Model}\label{Sect:SIRparamet}
\begin{figure}[h]
\begin{align}
&\includegraphics[width=7cm]{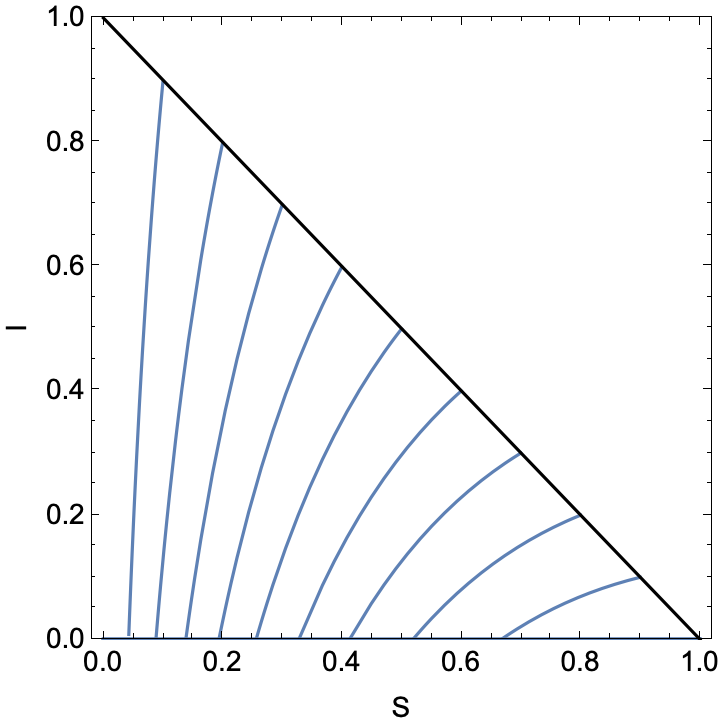} && \includegraphics[width=7cm]{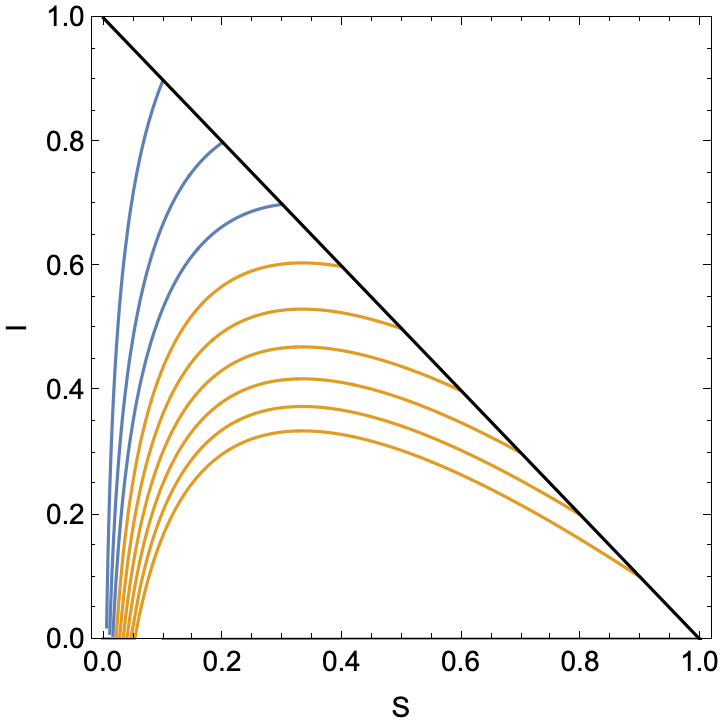}\nonumber
\end{align}
\caption{Relative number of infectious $\mi$ as a function of the relative number of susceptible $\ms$ for $\ms_0\in\{0.1,0.2,0.3,0.4,0.5,0.6,0.7,0.8,0.9\}$ and $\sigma=0.9$ (left) as well as $\sigma=3$ (right). Curves with a local maximum are drawn in orange while curves that are monotonically growing within $\mathbb{P}$ are drawn in blue.}
\label{SIRanalytIS}
\end{figure}

Apart from the numerical solutions, we can also gain insight into analytical aspects by discussing a parametric solution of the classical SIR model \cite{HETHCOTE1976335}. For simplicity, we assume $\rsu=0$, such that the system in Eqs \eqref{SIReqs}, \eqref{ConstraintSIR} and \eqref{SIRInitial} reduces to  
\begin{align}
&\begin{array}{l}\frac{d\ms}{dt}(t)=-\rin\,\mi(t) \,\ms(t)\,, \\[2pt] \frac{d\mi}{dt}(t)=\rin\,\mi(t) \,\ms(t) -\rhe\,\mi(t)\,, \\[2pt] \frac{d\mr}{dt}(t)=\rhe\,\mi (t)\,,\end{array}&&\text{with} && (\ms +\mi +\mr)(t) =1 &&\text{and} &&\begin{array}{l}\ms (t=0)=\ms _0>0\,,\\[2pt]\mi (t=0)=\mi _0>0\,,\\\mr (t=0)=0\,.\end{array}\label{ReducedSIRclassical}
\end{align}
Since the constraint in Eq.\eqref{ConstraintSIR} allows to remove one function, \emph{e.g.} $\mr (t)=1-\ms (t)-\mi (t)$, it is sufficient to consider the differential equations for $\ms $ and $\mi $. Dividing the latter by the former, we obtain a differential equation for $\mi $ as a function of $\ms $
\begin{align}
&\frac{d\mi }{d\ms }=-1+\frac{1}{\sigma\,\ms }\,,\label{SIRparametric}
\end{align}
which can be integrated to
\begin{align}
&\mi (\ms )=-\ms +\frac{1}{\sigma}\,\ln\ms +c\,,&&\text{for} &&c\in\mathbb{R}\,.\label{SolIfctS}
\end{align}
The parameter $\sigma$ is defined in Eq.\eqref{DefSIRsigma} and the constant $c$ appearing in Eq.\eqref{SolIfctS} can be fixed by the initial conditions at $t=0$ and gives $c=\mi _0+\ms _0-\frac{1}{\sigma}\,\ln\ms _0$, such that
\begin{align}
\mi (\ms )=1-\ms +\frac{1}{\sigma}\,\ln\frac{\ms }{\ms _0}\,.\label{SIRsolAnalyt}
\end{align}
A plot of this function in the allowed region
\begin{align}
\mathbb{P}=\{(\ms,\mi)\in[0,1]\times [0,1]|\ms+\mi\leq 1\}\,,
\end{align}
for different initial conditions and $\sigma=0.9$ (left) and $\sigma=3$ (right) is shown in \figref{SIRanalytIS}.
These plots once more highlight the qualitatively different solutions: the solution $\mi(S)$ in Eq.\eqref{SIRsolAnalyt} has a maximum at $\mi_\text{max}=1-\frac{1}{\sigma}\left(1+\ln(\sigma \ms_0)\right)$, which lies inside of $\mathbb{P}$ only if the initial effective reproduction number defined in Eq.~\eqref{DefSIRsigma} is $R_{e,0} \equiv \sigma\ms_0\geq 1$. Since $\ms(t)$ is a monotonically decreasing function of time, as demonstrated in \cite{HETHCOTE1976335}, this implies that:
\begin{itemize}
\item If $R_{e,0} \leq 1$, then $\mi(t)$ tends to $0$ monotonically for $t\to\infty$, as already established before.
\item If  $R_{e,0} >1$, $\mi(t)$ first increases to a maximum equal to $1-\frac{1}{\sigma}\left(1+\ln(\sigma \ms_0)\right)$ and then decreases to zero for $t\to\infty$. The limit $\ms(\infty)=\lim_{t\to\infty} S(t)$ is the unique root of
\begin{align}
1-S(\infty)+\frac{1}{\sigma}\,\ln\left(\frac{S(\infty)}{\ms_0}\right)=0\,,
\end{align}
in the interval $[0,\tfrac{1}{\sigma}]$, which is explicitly given in terms of the Lambert function in Eq.\eqref{SIRanalyticAsymptotic}. 
\end{itemize}
Furthermore, inserting the solution \eqref{SIRsolAnalyt} into Eq.\eqref{ReducedSIRclassical}, we obtain the following non-linear, first order differential equation for $\ms$ (as a function of time)
\begin{align}
\frac{d\ms}{dt}=\rin\,\ms(\ms-1)-\rin\,\frac{\ms}{\sigma}\,\ln\frac{\ms}{\ms_0}\,.
\end{align}
The latter can be solved numerically using various methods.

\subsection{Generalisations of the SIR Model}\label{Sect:ExtensionsSIR}
The SIR model, with 3 compartments $(\ms,\mi,\mr)$ and constant rates $\rin$, $\rhe$ and $\rsu$, provides a simple, but rather crude, description of the time evolution of an epidemic in an isolated population. This description can be refined and extended in various fashions. The most common way consists in adding more compartments, with more refined properties, giving birth to models like SIRD (including Deceased separately), SEIR (including Exposed individuals, in presence of a substantial incubation period), SIRV \cite{Ghostine2021AnES,Meng_2021} (see also \cite{SIRV}) (including vaccinated individuals), an so on \cite{HETHCOTErev}. Here, as an illustration, we shall discuss some generalisations of the SIR model that do not introduce fundamentally new compartments: in Section~\ref{Sect:SIRtimeDependence} we shall allow for time-dependent infection and recovery rates, in Section \ref{Sect:SpontaneousCreation} we shall include new terms in the differential equations~\eqref{SIReqs} that simulate the spontaneous appearance of new infectious (\emph{e.g.} from outside of the population), while in Section~\ref{Sect:Superspreaders} we allow for multiple different types of infectious individuals in an attempt to model inhomogeneous spreading of the disease among the population. While these variations add new compartments to the system, these are not of a completely new nature but simply copy an already existing compartment. In all cases we shall  motivate how these modifications can be used to describe specific features of certain diseases. For more general compartmental models (notably with the addition of completely new compartments) we refer the reader to the above mentioned literature (see \emph{e.g.} \cite{HETHCOTErev} for an overview). Another generalisation is the inclusion of the spatial evolution of the disease. This generally leads to coupled differential equations which are of first order in the time variable and of second order in the spatial variable. We shall not discuss these approaches in any detail in this review. 

\subsubsection{Time Dependent Infection and Recovery Rates}\label{Sect:SIRtimeDependence}
In the SIR model of Eqs~\eqref{SIReqs}, the rates $(\rin,\rhe,\rsu)$ are considered to be constant in time. This assumption is difficult to justify, in particular for epidemics that last over an extended period of time: many diseases show (natural) seasonal effects \cite{GrassleyFraser,Dowell2001SeasonalVI} related to the weather dependence of the effectiveness of transmission vectors or the behaviour of hosts (\emph{e.g.} it can be argued that the rate of child infections is linked to the cycle of school holidays \cite{Childhood}). Furthermore,  even in the absence of an effective vaccine, populations may take measures to prevent the spread of the disease by imposing social distancing rules or quarantine procedures, thus changing the (effective) infection rate $\rin$. Pathogen mutations and various forms of immunisations (including vaccines) can also increase or reduce the value of $\rin$ over time. With a prolonged duration of an epidemic, more data about the disease can be collected, leading to better ways to fight it on a biological and medical level, thus changing the recovery rate $\rhe$. Similarly, the disease may mutate and bypass previous immunisation strategies, thus changing the rate $\rsu$ at which removed individuals may become susceptible again. Modelling such effects and gauging their impact on the time evolution of an epidemics requires $(\rin,\rhe,\rsu)$ to change over time. In practice, this can be achieved by either interpreting them as explicit functions of $t\in\mathbb{R}$, \emph{i.e.} $(\rin(t),\rhe(t),\rsu(t))$, or by considering them to be functions of the relative number of susceptible and/or infectious individuals, \emph{i.e.} $(\rin(\ms,\mi),\rhe(\ms,\mi),\rsu(\ms,\mi))$. Since $(\ms,\mi)$ themselves are functions of time, the latter possibility induces an implicit dependence on $t$. For example, periodic and seasonal models in which these rates are assumed to be smoothly varying functions in $t$ have been developed for HIV \cite{HIVperiodic}, tuberculosis \cite{PMID:20063125} or cutaneous leishmaniasis \cite{Bacaer}, while models for pulse-vaccinations have been proposed in \cite{Agur,Stone2000TheoreticalEO,SHULGIN19981123,PMID:16806597,GaoPulse,LIU20091923,LIU20121974,Meng,Alberto,Chunjin,Zhou2003StabilityOP} (a model which in addition takes into account seasonal effects was presented in \cite{HE20138131}). The functional dependence can furthermore be used, for example, to model population-wide lockdowns, \emph{i.e.} quarantine measures that are imposed if the relative number of infectious individuals exceeds a certain value.

In the following we shall provide a simple (numerical) example of how the time dependence of different

\begin{wrapfigure}{r}{0.45\textwidth}
\vspace{-0.3cm}
\parbox{7cm}{\includegraphics[width=7cm]{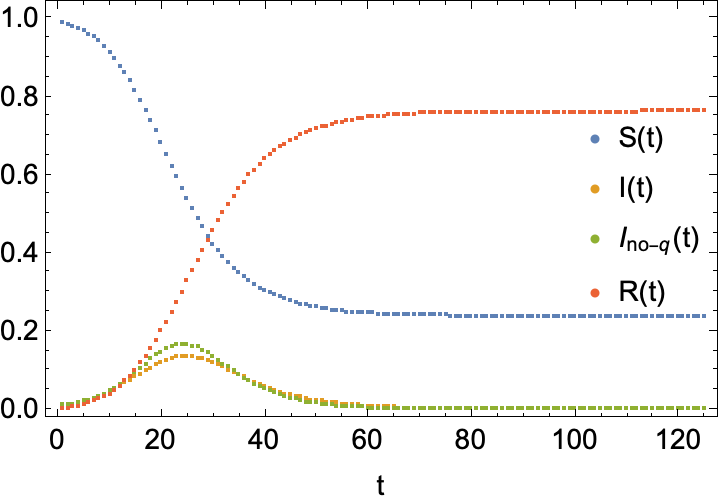}}
\caption{Numerical solution of the SIR equations (\ref{SIReqs}) for the time-dependent infection rate (\ref{TimeDepInfectionLockdown}) with $\ms_0=0.99$, $\rhe=0.05$, $\rsu=0$, $\rin_0=0.1$, $w=0.1$ and $\Delta \mi=0.05$.}
\label{Fig:SIRLockdownSchematic}
${}$\\[-0.5cm]
\end{wrapfigure} 

\noindent
parameters affects the time-evolution of the pandemic. We start by a simple model that can be used to qualitatively assess the efficiency of lockdown measures. To this end, we assume a `base' infection rate $\rin_0=$const., but assume that the population takes measures (social distancing, lockdowns, \emph{etc.}) to ensure that the actual infection rate $\rin(t)$ is reduced by a percentage $w$ if the number of (active) infectious individuals exceeds a certain value $\Delta\mi$. To model such social distancing measures in a very simplistic fashion, we introduce the following implicit time-dependence:
\begin{align}
\rin(\mi)=\rin_0\,\left[1-w\,\theta(\mi(t)-\Delta \mi)\right]\,,\label{TimeDepInfectionLockdown}
\end{align}
where $\theta$ is the Heaviside theta-function.~\footnote{To be mathematically rigorous, since $\theta$ is not a continuous function, using this infection rate in Eqs~\eqref{SIReqs} would require to interpret $(\ms(t),\mi(t),\mr(t))$ as distributions. This can be circumvented by replacing $\theta(\mi(t)-\Delta \mi)$ by $1+\tanh(\kappa_0(\mi(t)-\Delta \mi))$ with $\kappa_0$ a parameter that `smoothens' the step function. For the following discussion, however, this point shall not be relevant.} 
We hasten to add that Eq.\eqref{TimeDepInfectionLockdown} offers a very crude depiction of lockdown and quarantine measures taken by societies in the real-world: indeed, decisions on whether or not to impose a lockdown (or other social distancing measures) are usually based on numerous indicators which would (at least) require a more complicated dependence of $\gamma$ on $\mi$ (\emph{e.g.} its derivatives or averages of $\mi$ over a certain period of time prior to $t$). Furthermore, the conditions when a lockdown is lifted are typically independent of those when it is imposed.

\begin{figure}[h]
\begin{align}
&\includegraphics[width=7cm]{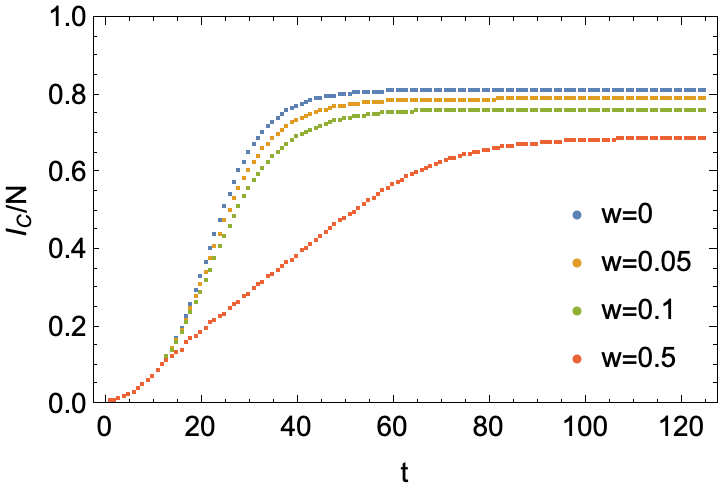} && \includegraphics[width=7cm]{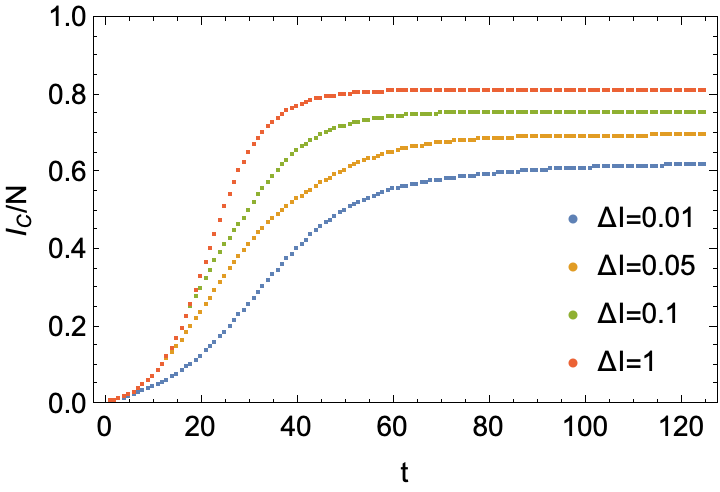}\nonumber
\end{align}
\caption{Numerical solution of the SIR equations (\ref{SIReqs}) for the time-dependent infection rate (\ref{TimeDepInfectionLockdown}), with $\ms_0=0.99$, $\rhe=0.05$, $\rsu=0$, $\rin_0=0.1$ and different choices of $(w,\Delta\mi)$: $w\in\{0.05\,, 0.1\,,0.5\}$ and $\Delta\mi=0.05$ (left) and $w=0.25$ and $\Delta\mi\in\{0.01\,,0.05\,,0.1\,,1\}$ (right).}
\label{Fig:SIRLockDownEffectiveness}
\end{figure}

An exemplary numerical solution of Eqs \eqref{SIReqs} for the particular $\rin$ in Eq.\eqref{TimeDepInfectionLockdown} is shown in \figref{Fig:SIRLockdownSchematic}. For better comparison we have also plotted $\mi_{\text{no-q}}(t)$, which is the solution for $\mi(t)$ in the case of constant $\rin=\gamma_0=\mbox{const.}$ (\emph{i.e.} with no reduction of the infection rate) and all remaining parameters chosen the same. Despite its simplicity and shortcomings, the model allows to make a few basic observations: the plot shows that the time-dependent infection rate leads to a reduction of the maximum of infectious individuals (`flattening of the curve'). Moreover, this simple model allows to compare the effectiveness of the quarantine measures as a function of $w$ and $\Delta\mi$. To gauge this effectiveness, we consider the cumulative number of infected individuals, which is plotted for different values of $w$ and $\Delta\mi$ in \figref{Fig:SIRLockDownEffectiveness}. These plots confirm the intuitive expectation that lockdown measures are the more effective the stronger the reduction of the infection rate is and the earlier they are introduced. However, due to its simplicity, the model also misses certain aspects compared to the time evolution of real-world communicable diseases in the presence of measures to prevent its spread: for example, possibly due to non-zero incubation time of most infectious diseases, the effect of quarantine measures on the number of infectious individuals can be detected only a certain time after the measures have been imposed (see \cite{Lai2020,liautaud2020fever,Huang2020.07.31.20143016,cacciapaglia2020mining} where this has been established for the COVID-19 pandemic). To include the latter would require a refinement of the model.

\begin{figure}[ht]
\begin{align}
&\includegraphics[width=7cm]{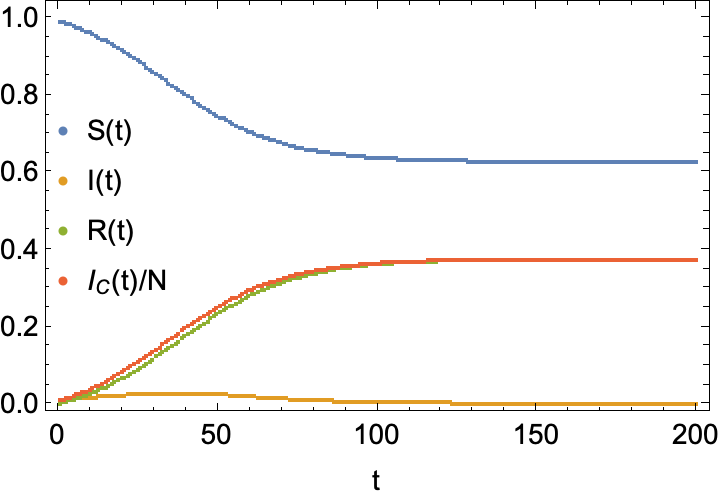} && \includegraphics[width=7cm]{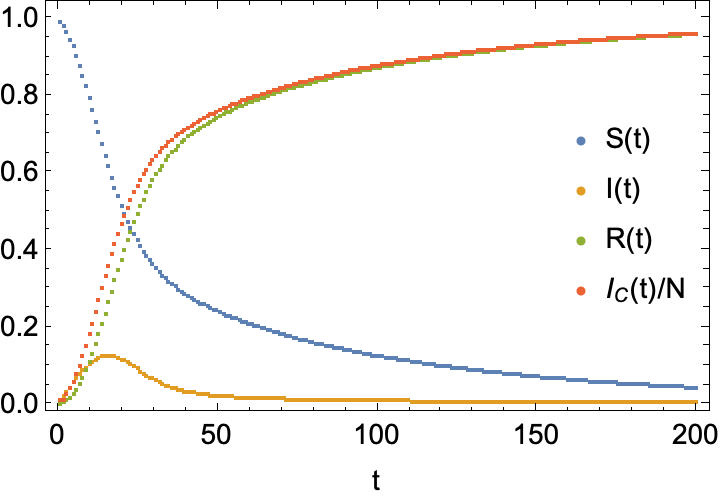}\nonumber
\end{align}
\caption{Numerical solution of the differential equations (\ref{SIReqsSpont}) for $\ms_0=0.99$, $\rin=0.055$ and $\rsu=0.045$ for two different choices of $\rgb$: $\rgb=0$ (left) and $\rgb=0.002$ (right).}
\label{Fig:SIRSpontCompare}
\end{figure}


\subsubsection{Spontaneous Creation and Multiple Waves}\label{Sect:SpontaneousCreation}
In Section~\ref{Sect:MasterEquationPercolation}, in the context of percolation models, we have discussed microscopic processes that correspond to the spontaneous creation of infected individuals. Such processes can simulate, for example, the infection of individuals through external sources (\emph{e.g.} pathogen sources, contaminated food sources, wildlife, \emph{etc}.), but may also be used to model the infection of susceptible individuals through asymptomatic infectious individuals or the appearance of infectious individuals from outside of the population through travel. How to introduce this process in SIR-type models has been discussed at the end of Section~\ref{Sect:RelationPercSIR}. Mathematically, the SIR equations \eqref{SIReqs} can be extended to
\begin{align}
&\frac{d\ms}{dt}=-\rin\,\mi \,\ms +\rsu\,\mr-\rgb\,\ms\,,&& \frac{d\mi}{dt}=\rin\,\mi\,\ms -\rhe\,\mi +\rgb\,\ms \,,&& \frac{d\mr}{dt}=\rhe\,\mi -\rsu\,\mr \,,\label{SIReqsSpont}
\end{align}

This is schematically shown in \figref{Fig:SIRSpontCompare}, where we show the solutions for $\rgb=0$ (left panel) compared to 

\begin{wrapfigure}{r}{0.45\textwidth}
\vspace{-0.3cm}
\parbox{7cm}{ \includegraphics[width=7cm]{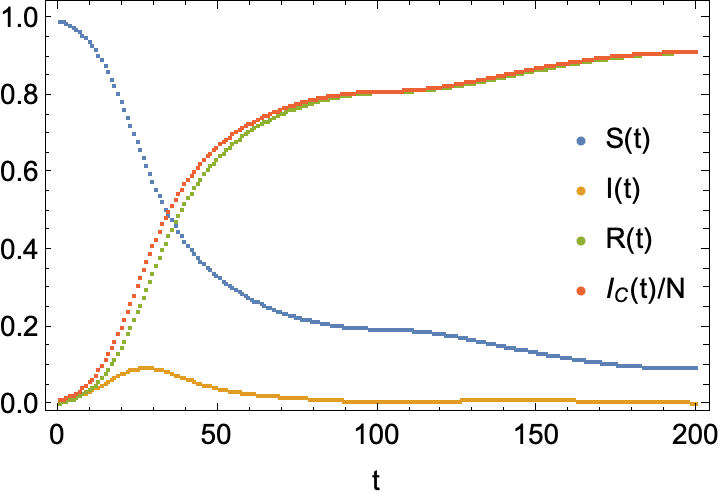}}
\caption{Numerical solution of the differential equations (\ref{SIReqsSpont}) for $\ms_0=0.99$, $\rin=0.055$, $\rsu=0.045$ and $\rgb=0.002\,\left|\sin\left(\tfrac{2\pi t}{200}\right)\right|$.}
\label{Fig:SIRCreateTime}
${}$\\[-0.5cm]
\end{wrapfigure} 

\noindent
where the rate $\xi = \hat\xi$ of Section~\ref{Sect:RelationPercSIR}.
The system still needs to be solved with the initial conditions \eqref{SIRInitial}. Here $\rgb\in\mathbb{R}_+$ is a constant that governs the rate at which new infectious individuals appear in the population, corresponding to a qualitative change in the basic infection mechanism: since susceptible individuals can contract the disease even if there are no infectious individuals present in the population, the epidemic can not be stopped before the entire population becomes infected. As a consequence, the cumulative number of infected tends to $N$ for $t\to\infty$. 

the solution for  $\rgb\neq 0$ (right panel). In the former case, the number of cumulative infected tends to a finite value, while in the latter case, $\lim_{t\to\infty} \ms(t)\to 0$.

Following the discussion in Section~\ref{Sect:SIRtimeDependence}, we can also analyse the effect of a time-dependent rate $\rgb(t)$. This can be used to model a time-dependent rate of the spontaneous creation of new infectious individuals, \emph{e.g.} induced by quarantine measures or geographical restrictions of the population. As a simple example, we have plotted the numerical solution for a periodic function $\rgb$ in \figref{Fig:SIRCreateTime}. Since $\rgb$ does not remain zero after finite time, the relative number of susceptible tends to $0$ (indicating that the entire population is infected for $t\to\infty$). Moreover, the solution features oscillations in time, which could be interpreted as different waves of the epidemic spreading in the population. 

\subsubsection{Heterogeneous Transmission Rates and Superspreaders}\label{Sect:Superspreaders}
As another generalisation of compartmental models, we consider adding multiple versions of the compartments $S$, $I$, $R$ \cite{KEMPER1980111,Szapudi2020.07.02.20145490} to model the heterogeneity of social interactions and their impact on the spread of a disease: indeed, indications for \emph{superspreaders} (\emph{i.e.} individuals who transmit the disease with a significantly higher rate than average) have been found in many diseases (\emph{e.g.} influenza \cite{FoxKilbourne,ElvebackFox}, rubella \cite{Hattis}) and for certain diseases it has in fact been suggested that only a small fraction of the population is responsible for most infections (see \emph{e.g.} \cite{Dillon,Schuchat} for a study of COVID-19). Similarly, the gender of individuals plays an important role in the modelling of sexually transmitted diseases (see \emph{e.g.}  \cite{Wilcox1,Wilcox2,Wilcox3,Wilcox4} for the study of gonorrhoea, which also suggests the necessity of an extended range of contact rates \cite{KEMPER1980111}). To account for these modified contact rates, modifications of the SIR model (as described above) have been suggested, which consist in adding multiple compartments of infectious individuals, \emph{i.e.} new subgroups that allow to refine the study of the disease spread in a not-so-uniform population. These additional compartments can,

\begin{wrapfigure}{r}{0.57\textwidth}
\begin{center}
\vspace{-0.3cm}
\scalebox{1}{\parbox{8cm}{\begin{tikzpicture}
\begin{scope}[xshift=-1cm]
\draw[thick] (0,0) rectangle (1,1);
\node at (0.5,0.5) {$N\,S$};
\end{scope}
\begin{scope}[xshift=3cm,yshift=-1cm]
\draw[thick] (0,0) rectangle (1,1);
\node at (0.5,0.5) {$N\,\mi_2$};
\end{scope}
\begin{scope}[xshift=3cm,yshift=1cm]
\draw[thick] (0,0) rectangle (1,1);
\node at (0.5,0.5) {$N\,\mi_1$};
\end{scope}
\begin{scope}[xshift=6cm]
\draw[thick] (0,0) rectangle (1,1);
\node at (0.5,0.5) {$N\,R$};
\end{scope}

\draw[ultra thick,->] (-0.25,1.2) -- (-0.25,1.5) -- (2.8,1.5);
\node at (1.25,1.75) {\footnotesize $\beta(\rin_1\mi_1+\rin_2\mi_2)\,N\,\ms$};
\draw[ultra thick,->] (-0.25,-0.2) -- (-0.25,-0.5) -- (2.8,-0.5);
\node at (1.1,-0.8) {\footnotesize $(1-\beta)(\rin_1\mi_1+\rin_2\mi_2)\,N\,\ms$};
\draw[ultra thick,->] (4.2,1.5) -- (6.25,1.5) -- (6.25,1.2);
\node at (5.25,1.75) {\footnotesize $\rhe\,N\,\mi_1$};
\draw[ultra thick,->] (4.2,-0.5) -- (6.25,-0.5) -- (6.25,-0.2);
\node at (5.25,-0.8) {\footnotesize $\rhe\,N\,\mi_2$};
\draw[ultra thick,->] (6.75,1.2) -- (6.75,2.6) -- (-0.75,2.6) -- (-0.75,1.2);
\node at (3,2.9) {\footnotesize $\rsu\,N\,\mr$};
\end{tikzpicture}}}
\end{center}
\caption{Flow between susceptible, 2 compartments of infectious and removed individuals.}
\label{Fig:SIRschemSuper}
${}$\\[-1cm]
\end{wrapfigure}
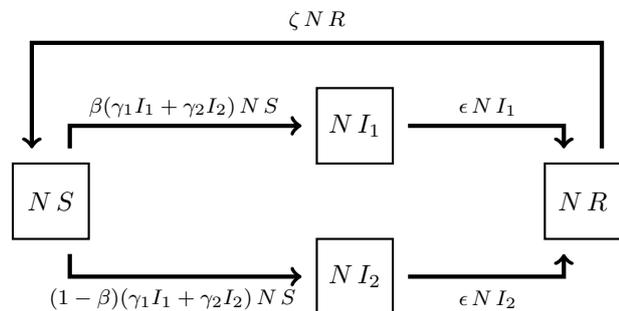 

\noindent
therefore, distinguish individuals based on biological/medical indicators (\emph{e.g.} gender, age, preexistent medical conditions, \text{etc.}), geographic distribution, social behaviour and/or may be used to introduce additional stages in the progression of the disease, such as latency periods or different stages of symptoms. Inclusion of more compartments naturally renders the relevant set of differential equations more complicated and is more demanding in terms of computational costs (see \cite{STD} as an example). Furthermore, the increase in the number of parameters (rates) leads to a loss of predictive power compared to simpler models.

\begin{figure}[h]
\begin{align}
&\includegraphics[width=7cm]{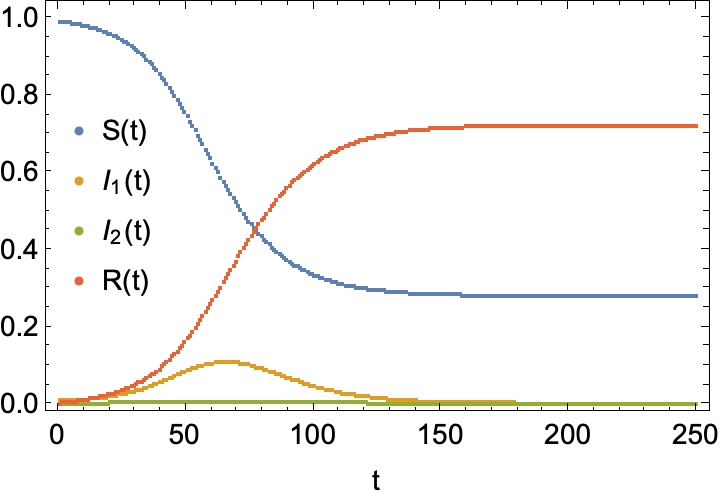} && \includegraphics[width=7cm]{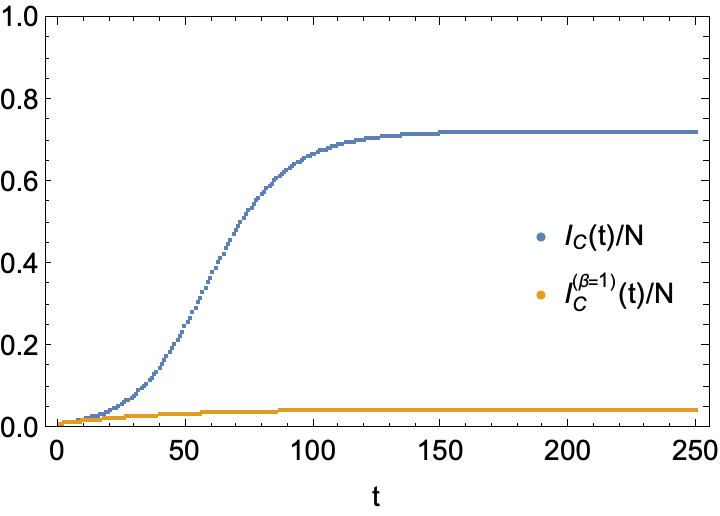}\nonumber
\end{align}
\caption{Numerical solution of the SIR equations in the presence of superspreaders, Eqs~\eqref{SIRsuperspreaders}: time evolution for $\ms_0=0.99$, $\mi_{0,1}=0.01$, $\mi_{0,2}=0$, $\rin_1=0.04$, $\rin_2=1$, $\rhe=0.05$ and $\beta=0.95$ (left) and comparison of the cumulative number of infected with the `usual' SIR model without superspreaders (\emph{i.e.} $\beta=1$) (right).}
\label{Fig:SIRsuperspreaders}
\end{figure}

In the following we shall present one simple example that includes one additional class of infectious individuals. This model is useful in characterising different (social) behaviours among individuals. Indeed, in general, the infection rate $\rin$ is not homogeneous throughout the entire population, since it depends on various factors such as geographical mobility, social behaviour \emph{etc.}, which may vary considerably. A particular effect in this regard is the existence of so-called \emph{superspreaders}. These are individuals who are capable of transmitting the disease to susceptible individuals at a rate that significantly exceeds the average.  
The presence of superspreaders can be described by introducing two groups of infectious individuals $\mi_{1,2}$, with different infection rates $\rin_{1,2}$ and  appearing with a relative ratio $\beta\in[0,1]$. Extending \figref{Fig:SIRschem}, the new flow among compartments 
is shown in \figref{Fig:SIRschemSuper} (for $\zeta=0$), and can be described by the following differential equations \cite{KEMPER1980111}:
\begin{align}
&\frac{d\ms}{dt}=-(\rin_1\,\mi_1+\rin_2\,\mi_2)\,\ms\,,
&&\frac{d\mi_1}{dt}=\beta(\rin_1\,\mi_1+\rin_2\,\mi_2)\,\ms-\rhe\,\mi_1\,,\nonumber\\
&\frac{d\mi_2}{dt}=(1-\beta)(\rin_1\,\mi_1+\rin_2\,\mi_2)\,\ms-\rhe\,\mi_2\,,&&\frac{d\mr}{dt}=\rhe(\mi_1+\mi_2)\,,\label{SIRsuperspreaders}
\end{align}
together with the initial conditions
\begin{align}
&\ms(t=0)=\ms_0\,,&&\mi_1(t=0)=\mi_{0,1}\,,&&\mi_2(t=0)=\mi_{0,2}\,,&&\mr(t=0)=0\,,
\end{align}
with
\begin{align}
&0\leq \ms_0,\mi_{0,1},\mi_{0,2}\leq 1\,,&& 1=\ms_0+\mi_{0,1}+\mi_{0,2}\,.
\end{align}
In \cite{KEMPER1980111} the parameters $\beta$, $\rin_{1,2}$, and $\rhe$ were assumed to be constant in time. 
By defining an effective infectious population $J=(\rin_1\,\mi_1+\rin_2\,\mi_2)/\lambda$, we can extract the following differential equations for $(\ms,J)$~\footnote{Note that our definition of $J$ differs from the definition of the infective potential $J=\rin_1\,\mi_1+\rin_2\,\mi_2$  in \cite{KEMPER1980111} by a constant normalisation.}
\begin{align}
&\frac{dS}{dt}=-\lambda\, J\,\ms\,,&&\frac{d J}{dt}=\lambda\,J\,\ms-\rhe\,J\,,&&\text{with} &&\lambda=\rin_1\,\beta+(1-\beta)\,\rin_2\,.\label{SIRsuperConst}
\end{align}
Thus, for $S$ and $J$ we obtain the same equations as in the classical SIR model, which can be solved along the lines of Section~\ref{Sect:SIRparamet}: we extract the following non-linear first-order equation for $\ms$:
\begin{align}
&\frac{d\ms}{dt}=\lambda\,\ms^2-\rhe\,\ms\,\ln\ms+\mathfrak{c}_0\,\ms\,,&&\text{with} &&\mathfrak{c}_0=\rhe\,\ln\ms_0-\lambda\,\ms_0-(\rin_1\mi_{0,1}+\rin_2\mi_{0,2})\,.
\end{align}
which leads to the asymptotic number of susceptible $\ms(\infty)$ implicitly given by
\begin{align}
0=\lambda\,\ms(\infty)-\rhe\,\ln\ms(\infty)+\mathfrak{c}_0\,.
\end{align}
As was pointed out in \cite{KEMPER1980111}, the SIR model with superspreaders leads to the same dynamics as the classical SIR models, albeit with a larger-than-average infection rate $\lambda$, due to the contribution of superspreaders.
With constant infection and recovery rates and monotonically diminishing number of susceptible (\emph{i.e.} for $\zeta=0$), the impact of superspreaders is conceptually not detectable. Nevertheless, from the perspective of the total number of infected, superspreaders may have a significant impact in driving the epidemics. In \figref{Fig:SIRsuperspreaders} (left) we have plotted the time evolution of a typical solution, which indeed follows the same pattern as the usual SIR model. However, as visible from \figref{Fig:SIRsuperspreaders} (right), even the presence of a relatively small number of superspreaders can have a strong impact on the cumulative number of infected.

Finally, it was argued in \cite{KEMPER1980111} that in situations in which the number of susceptible individuals is no longer a monotonical function (which can for example be achieved by allowing for a non-trivial $\rsu$), the time evolution of the SIR model looks qualitatively different in the presence of superspreaders. 


\subsection{The SIR model as a set of Renormalisation Group Equations}

As we have seen from simple numerical studies in Section~\ref{Sect:SIRQualitative}, solutions $(\ms(t),\mi(t),\mr(t))$ of the classical SIR equations \eqref{SIReqs} exhibit interesting properties as functions of time, which structurally remain valid for many of the generalisations discussed in Section~\ref{Sect:ExtensionsSIR}. In particular, the solutions show a qualitatively different behaviour when a key parameter (in the classical SIR model, the initial effective reproduction number $R_{e,0} = \ms_0 \sigma$) exceeds a critical value. This seems to play a similar role to an ordering parameter in physical systems undergoing a phase transition. A further related observation is the fact that Eqs \eqref{SIReqs} are invariant under a re-scaling of the time-variable, if simultaneously all the rates are also re-scaled:
\begin{align}
&t\rightarrow \frac{1}{\mu}\,t\,,&&\rin\to\mu\,\rin\,,&&\rhe\to\mu\,\rhe\,,&&\rsu\to \mu\,\rsu\,,&&\forall \mu\in\mathbb{R}\setminus\{0\}\,.
\end{align}
This rescaling of the time-variable is structurally not unlike the change of the energy scale in quantum field theories that is used to describe the \emph{Wilsonian renormalisation} of the couplings among elementary particles \cite{Wilson:1971bg,Wilson:1971dh}. The renormalisation flow can also feature similar symmetries to the ones of the solutions of the SIR equations. Compartmental models can be formulated in a way that is structurally similar to Renormalisation Group Equations (RGEs) \cite{DellaMorte:2020wlc,mcguigan2020pandemic}, and this analogy lead to the formulation of an effective description called \emph{epidemiological Renormalisation Group} \cite{DellaMorte:2020wlc,DellaMorte:2020qry}, which we will introduce in the next section.

To understand the analogy, we recall that most (perturbative) quantum field theories are effective models: they are typically based on an action that encodes fundamental interactions of certain `bare' fundamental fields, whose strength is described by a set of coupling constants $\{\lambda_i\}$ (where $i$ takes values in a suitable set $\{\mathcal{S}\}$). Each effective description, however, is generally well adapted only at a certain energy scale, beyond which new degrees of freedom are more appropriate and new interactions may become important. In practice, one introduces a cut-off parameter (or some other regularisation form), beyond which the effective description is no longer valid. The effective theory  can thus be interpreted as encoding all effective interactions, after having integrated out all interactions at energy scales higher than the cut-off. From this perspective it is clear that changing the energy scale (and thus the cut-off) will lead to different interactions being integrated out and thus has a strong impact on the theory, along with the fundamental degrees of freedom and the couplings used to describe it. The process of arriving at the new effective theory is called \emph{renormalisation}. To describe it, we study universal quantities that are invariant under the renormalisation, first and foremost the partition function $\mathcal{Z}(\{\lambda_i\})$, which encodes the statistical properties of the quantum system and depends on the set of coupling constants mentioned before. For the purpose of this review, we can think of $\mathcal{Z}$ as a mathematical function that encodes all the physical properties of the system and its symmetries, independently on its explicit definition. One of the symmetries is, as already mentioned, the invariance under renormalisation, \emph{i.e.} the change in the energy scale of the physical interactions. 
If $\{\lambda'_a\}$ (with $a$ taking values in a new set $\{\mathcal{S}'\}$) is the new set of renormalised couplings and $\mathcal{Z}'$ the partition function of the renormalised theory, invariance of the partition function implies
\begin{align}
\mathcal{Z}(\{\lambda_i\})=\mathcal{Z}'(\{\lambda'_a\})\,.\label{InvariancePartitionFunction}
\end{align}
Hence, by continuously changing the energy scale, the theory sweeps out a trajectory in the space of all possible effective theories, called the \emph{renormalisation group flow}, which is governed by the invariance Eq.\eqref{InvariancePartitionFunction}. From the perspective of the interactions, the theory  sweeps out a trajectory in the space of all couplings $\lambda_i$. This is governed by the \emph{beta-functions} $\beta_i(\lambda_k)$, defined as the derivatives of the couplings $\lambda_i$ with respect to the logarithm of the cut-off parameter, and are functions of the couplings $\lambda_i$ themselves. The flow is thus described in terms of a system of differential equations, like the SIR model does, whose fixed points (\emph{i.e.} zeros of the beta functions) denote critical (\emph{i.e.} scale invariant) points of the theory.

Before making the connection to epidemiology, we remark that physical theories in general allow for field redefinitions, which means that they can be equivalently formulated using different bare fields. This implies that the coupling set $\{\lambda_i\}$ is not uniquely determined, but should rather be thought of as a (local) choice of basis in the space of couplings. A specific choice of a set of $\{\lambda_i\}$ is called a renormalisation \emph{scheme}. While a priori the specific form of the beta-functions depend on the scheme (in particular their perturbative expansions as functions of the $\{\lambda_i\}$), a change of scheme can be understood as an analytic transformation in the space of couplings.

In \cite{DellaMorte:2020wlc}, and subsequent works \cite{DellaMorte:2020qry,Cacciapaglia:2020mjf,cacciapaglia2020evidence}, it was suggested to interpret the time evolution of the spread of a disease (specifically COVID-19) within the framework of the Wilsonian renormalisation group equation. We shall explain this description in more detail in Section~\ref{Sect:RGapproach}.
In the following, however, we shall show how such a description can at least qualitatively be obtained from the SIR equations by allowing time-dependent infection and removal rates, as first pointed out in \cite{DellaMorte:2020qry}.

\subsubsection{Beta Function}\label{Sect:BetaFunctionSIR}
In preparation to Section~\ref{Sect:RGapproach}, we notice that the SIR model (with $\rsu=0$, but time-dependent infection and recovery rates $\rin(t)$ and $\rhe(t)$) can be written in a form which is strongly reminiscent of a RGE. To this end, we return to Eqs~\eqref{ReducedSIRclassical} and repeat the same steps as in Section~\ref{Sect:SIRparamet}, except for allowing $\sigma:\,[0,1]\to \mathbb{R}_+$ to be a priori a function of $\ms$. Thus, we can integrate Eq.\eqref{SIRparametric} in the following form
\begin{align}
\mi(S)=1-\ms+\int_{\ms_0}^{\ms}\frac{du}{u\,\sigma(u)}\,,
\end{align}
which is compatible with the initial conditions in Eq.\eqref{ReducedSIRclassical} at $t=0$. Inserting this relation into the first equation of \eqref{SIReqs}, for $\zeta=0$ it  yields
\begin{align}
\frac{d\ms}{dt}=-\rin(t)\,\ms(t)\,\left[1-\ms+\int_{\ms_0}^{\ms}\frac{du}{u\,\sigma(u)}\right]\,.\label{ParametricSeq}
\end{align}
Instead of the relative number of susceptible, this equation can be re-written in terms of the cumulative number of infected individuals $\Ic$, as defined in Eq.~\eqref{CumulativeInfected}. Thus, Eq.\eqref{ParametricSeq} can be rewritten as
\begin{align}
\frac{d\Ic}{dt}=N\,\rin\,\left(1-\frac{\Ic}{N}\right)\left[\frac{\Ic}{N}+\int_{\ms_0}^{1-\frac{\Ic}{N}}\frac{du}{u\,\sigma(u)}\right]\,.\label{IcEq}
\end{align}
Next, generalising what was proposed in \cite{DellaMorte:2020wlc,cacciapaglia2020evidence}, we define an \emph{epidemic coupling} $\alpha(t)$ as a function of the cumulative number of infected individuals:
\begin{align}
&\alpha(t)=\phi(\Ic(t))\,,
\end{align}
where $\phi:\,[0,N]\rightarrow \mathbb{R}$ is a strictly monotonically growing, continuously differentiable function with non-vanishing first derivative. A priori, $\phi$ could also explicitly depend on $t$ (not only through $\Ic(t)$), but in the following we shall not explore this possibility. In \cite{DellaMorte:2020wlc}, in the context of the COVID-19 pandemic, $\phi$ was chosen to be the natural logarithm, while in \cite{cacciapaglia2020evidence,cacciapaglia2020multiwave} $\phi(x)=x$ was chosen. For the moment, we shall leave $\phi$ arbitrary, which mimics the liberty to choose different renormalisation schemes in the framework of the Wilsonian approach.  Upon defining formally the $\beta$-function as
\begin{align}
\beta(\Ic(t))=-\frac{d\alpha}{dt}\,,\label{FormalDefBetaFunction}
\end{align}
Eq.~\eqref{IcEq} can  be re-formulated as
\begin{align}
-\beta=\left(\frac{d\phi}{d\Ic}\right)\frac{d\Ic}{dt}=\left(\frac{d\phi}{d\Ic}\right)\,N\,\rin\,\left(1-\frac{\Ic}{N}\right)\left[\frac{\Ic}{N}+\int_{\ms_0}^{1-\frac{\Ic}{N}}\frac{du}{u\,\sigma(u)}\right]\,.\label{FormBetaFromSIR}
\end{align}
An explicit example that is designed to make contact with the work in  \cite{cacciapaglia2020evidence} is discussed in Section~\ref{App:CalcBetaFct}. Eq.\eqref{FormBetaFromSIR}, at least structurally, resembles a RGE and has several intriguing properties to support this interpretation. Note that with Eq.\eqref{CumulativeInfected}, we can also write
\begin{align}
\beta(t)=-\left(\frac{d\phi}{d\Ic}\right)\,\frac{d\Ic}{dt}=-\left(\frac{d\phi}{d\Ic}\right)\,N\,\rin(t)\,\mi(t)\,\ms(t)\,,\label{DefBetaInteg}
\end{align}
which vanishes when:
\begin{itemize}
\item the infection rate vanishes $\rin(t)=0$,
\item or there are no susceptible individuals left $\ms(t)=0$, 
\item or the number of active infected vanishes $\mi(t)=0$ and the disease is eradicated.
\end{itemize}
Further (structural) evidence can be given by considering concrete solutions: an example for the interplay between the beta-function and $\sigma$ is provided in the following Section~\ref{App:CalcBetaFct}. Furthermore, independently of its connection to compartmental models, a renormalisation group approach can be used to model and describe the dynamics of an epidemic, as we discuss in Section~\ref{Sect:RGapproach}.

\subsubsection{Connection between SIR models and the eRG approach}\label{App:CalcBetaFct}

We now discuss via a concrete example how to formulate a SIR model (with time-dependent $\sigma(t)$) in a way that reproduces the eRG framework, which will be discussed in more detail in the next section. This relation has been first discussed in \cite{DellaMorte:2020qry,Cacciapaglia:2020mjf}.
Following the logic outlined above, we will highlight the similarities between the SIR equations and RGEs. In particular, we show how a particular beta-function can be obtained from a time-dependent $\sigma$, starting from Eq.~\eqref{FormBetaFromSIR}. 
Concretely, we shall make contact with the following:
\begin{align}
- \beta_0 (\Ic) = {\rgg }\, \Ic  \left[  \left(1 - \frac{\Ic}{{A}} \right)^2  - \delta\right]^p\,,\label{LinearTargetBeta}
\end{align}
where $\phi(\Ic)=\Ic$, and $p,\delta,A$ are constant. The form of the beta-function \eqref{LinearTargetBeta} will be motivated and discussed in more detail in Section~\ref{Sect:eRGcomplex} and is used to study a single wave followed by an endemic period characterised by a quasi linear growth, which can be precursor to a next wave. We shall return on this linear period in Section \ref{Sect:AnSolLinear}.

\begin{figure}[p]
\begin{align}
&\fbox{\parbox{7.25cm}{\includegraphics[width=7.25cm]{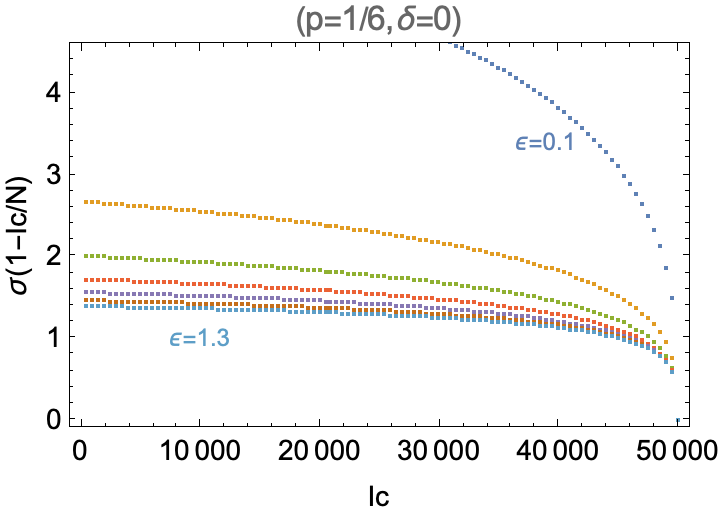}}} && \fbox{\parbox{7.25cm}{\includegraphics[width=7.25cm]{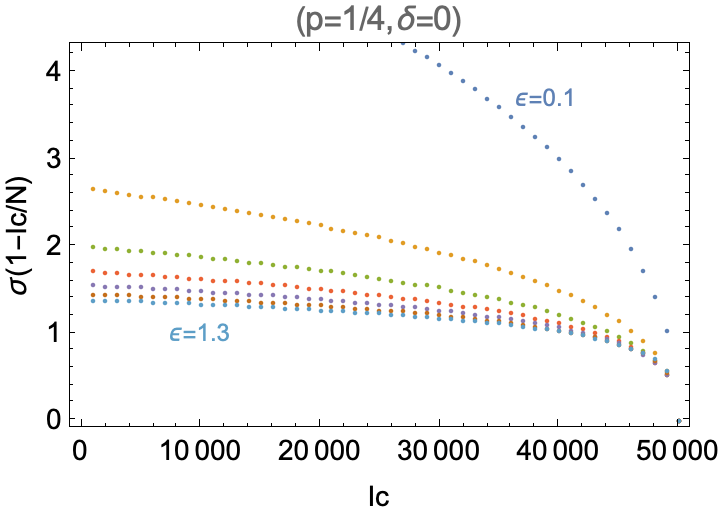}}}\nonumber\\
&\fbox{\parbox{7.25cm}{\includegraphics[width=7.25cm]{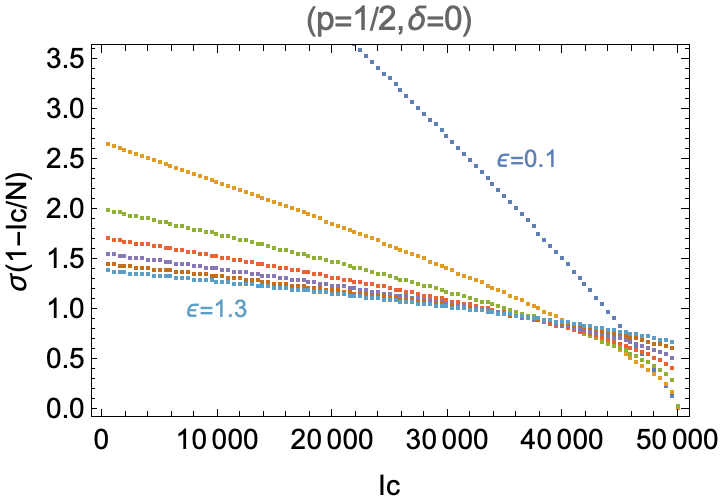}}} && \fbox{\parbox{7.25cm}{\includegraphics[width=7.25cm]{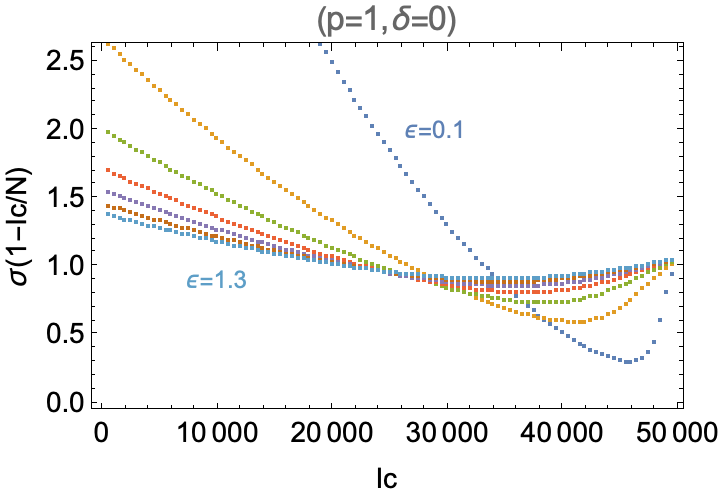}}}\nonumber\\
&\fbox{\parbox{7.25cm}{\includegraphics[width=7.25cm]{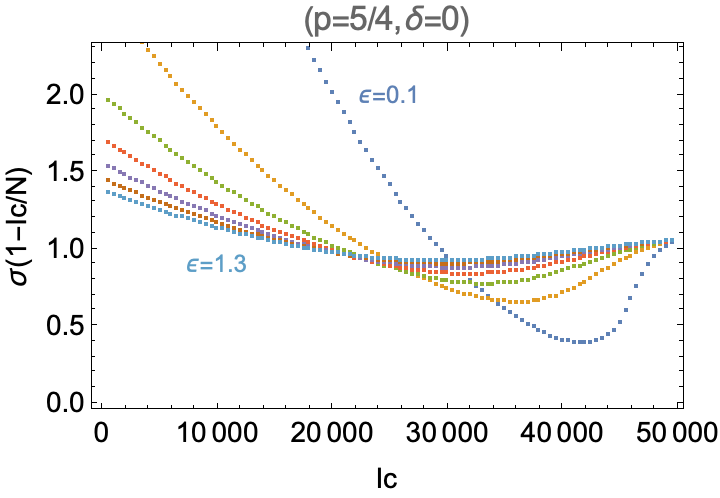}}} && \fbox{\parbox{7.25cm}{\includegraphics[width=7.25cm]{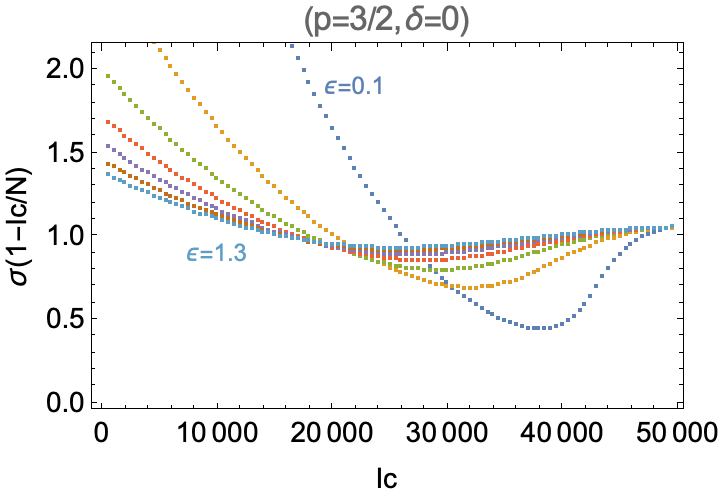}}}\nonumber
\end{align}
\caption{$\sigma$ as a function of $\Ic$ for different values of $p$ and $\delta=0$ in the limit $\ms_0\to 1$ with $N=1.000.000$, $A=50.000$, $\rgg =0.5$ and $\rhe\in\{0.1\,,0.3\,,0.5\,,0.7\,,0.9\,,1.1\,,1.3\}$.}
\label{SIRbetaSigma}
\end{figure}

As a starting point, we shall consider a SIR model where, for simplicity, $\rhe$ is constant, \emph{i.e.} the rate of recovery remains constant throughout the pandemic\footnote{$\epsilon$ depends on biological properties of the virus as well medical and pharmaceutical means of the population to cure it. Since these are difficult to change without significant effort, the value of $\epsilon$ is difficult to change.}, while $\rin$ and $\sigma=\frac{\rin}{\rhe}$ are continuous functions of $\ms$. Finally, to make contact with Eq.\eqref{LinearTargetBeta}, we shall consider the asymptotic limit $\ms_0\rightarrow 1$.
Identifying the function $\beta (t)$ in Eq.\eqref{DefBetaInteg} with $\beta_0$ leads to an integral equation that, for constant $\rhe$, can be turned into a differential equation for $\sigma (t)$ (recall that $\ms=1-\frac{\Ic}{N}$):
\begin{align}
\frac{d}{d\Ic}\left[\frac{\beta_0(\Ic)}{\rhe \,\sigma\left(1-\frac{\Ic}{N}\right)}\right]=1-\frac{1}{\left(1-\frac{\Ic}{N}\right)\,\sigma\left(1-\frac{\Ic}{N}\right)}\,.
\end{align}
The equation above can be brought into the form 
\begin{align}
&0=\sigma'(\ms)+g_1(\ms)\,\sigma(\ms)+g_2(\ms)\,\sigma^2(\ms)\,,&&\text{with} &&\begin{array}{l}g_1(\ms)=\frac{1}{S}-\frac{N}{\beta_0(N(1-\ms))}\,\left(\rhe -\beta'_0(N(1-\ms))\right)\,,\\[4pt] g_2(\ms)=\frac{N\rhe  \ms}{\beta_0(N(1-\ms))}\,.\end{array}
\end{align}
In the above and following equations, the prime indicates a derivative with respect 
to the argument of the function.
The general solution of this first order, non-linear differential equation is
\begin{align}
&\sigma(\ms)=\frac{D(\ms)}{\frac{1}{\sigma_0}+\int_{\ms_0}^{\ms}dx\,D(x)\,g_2(x)}\,,&&\text{with} &&D(\ms)=\text{exp}\left[-\int_{\ms_0}^{\ms}g_1(x)\,dx\right]\,.\label{SigSolInt}
\end{align}
Here $\sigma_0$ is an integration constant, which can be determined by comparing the first derivative of $\beta_0$ and $\beta$ at $\ms=\ms_0\rightarrow1$ (\emph{i.e.} at $\Ic=N(1-\ms_0)=0$). In fact, $\beta'_0(0)=\beta'(0)$ implies
\begin{align}
\sigma(1)=\sigma_0=1-\frac{1}{\rhe }\,\beta_0'(0)=1+\frac{\rgg }{\rhe }(1-\delta)^p\,.
\end{align}
With $\beta_0$ given in Eq.\eqref{LinearTargetBeta}, the integral over $g_1$ can be performed analytically (involving an Appell hypergeometric function). However, using this result to insert $D(S)$ into the first expression in Eq.\eqref{SigSolInt}, the integral in the denominator is more involved and we could only find analytic solutions for generic\footnote{We remark in passing that we were able compute analytic solutions for other combinations of $(p,\delta)$ for specific combinations of $(\rgg ,\rhe)$, \emph{i.e.} for certain fixed ratios $\frac{\rgg }{\rhe}$.} $\rgg ,\rhe$ for $(p=\tfrac{1}{4},\delta=0)$ and $(p=\tfrac{1}{2},\delta=0)$, whose limit $S_0\to 1$ is
\begin{align}
\lim_{S_0\to 1}\sigma(1-\tfrac{\Ic}{N})\bigg|_{{p=\frac{1}{4}}\atop{\delta=0}}&=\frac{\frac{\rgg   N }{\rhe   (N-\Ic)}\sqrt{1-\frac{\Ic }{A}}  }{1 +\frac{2^{1-\frac{\rhe  }{\rgg  }}A \rhe}{\Ic(\rgg  +\rhe)}   \left(\sqrt{1-\frac{\Ic }{A}}-1\right)
   \left(\sqrt{1-\frac{\Ic }{A}}+1\right)^{\frac{\rhe  }{\rgg  }} \, _2F_1\left(\frac{\rhe  }{\rgg },\frac{\rgg  +\rhe  }{\rgg  };\frac{\rhe  }{\rgg  }+2;\frac{
   1-\sqrt{1-\frac{\Ic }{A}}}{2}\right) }\,,\nonumber\\
\lim_{S_0\to 1}\sigma(1-\tfrac{\Ic}{N})\bigg|_{{p=\frac{1}{2}}\atop{\delta=0}}&=\frac{N  (A-\Ic ) (\rgg  +\rhe  ) \left(1-\frac{\Ic }{A}\right)^{-\frac{\rhe  }{\rgg  }}}{A \rhe  
   (N -\Ic) \, _2F_1\left(\frac{\rhe  }{\rgg  },\frac{\rgg  +\rhe  }{\rgg  };2+\frac{\rhe 
   }{\rgg  };\frac{\Ic }{A}\right)}\,.
\end{align}
However, the integration can be performed numerically, and for different values of $(p,\delta)$, $\sigma$ as a function of $\Ic$ is shown in Fig.~\ref{SIRbetaSigma}.
 
\begin{figure}[ht]
\begin{align}
&\includegraphics[width=9cm]{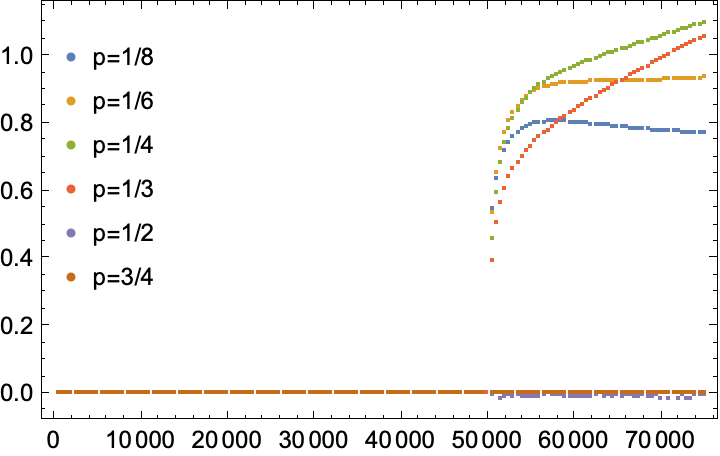} \nonumber
\end{align}
\caption{Numerical computation of the imaginary part of $\sigma(1-\tfrac{\Ic}{N})$ for $\rgg =0.5$, $\rhe=0.7$, $N=1.000.000$ and $A=50.000$ in the limit $S_0\to 1$ for various values of $p$ and $\delta=0$.}
\label{Fig:NumericImaginary}
\end{figure}

\begin{figure}[p]
\begin{align}
&\fbox{\parbox{7.25cm}{\includegraphics[width=7.25cm]{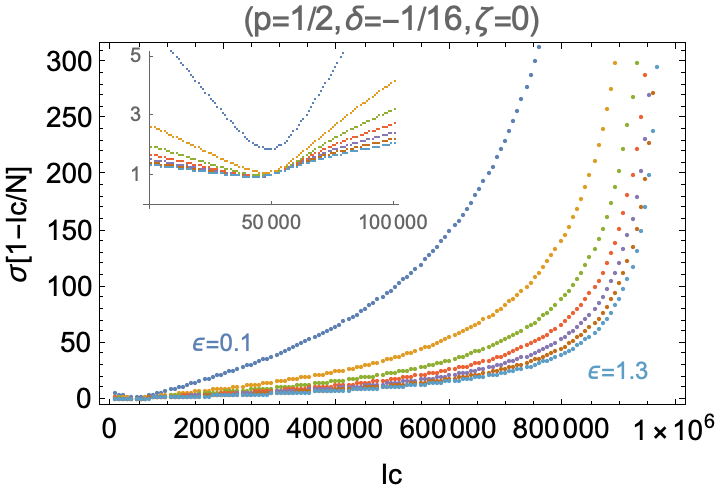}}} && \fbox{\parbox{7.25cm}{\includegraphics[width=7.25cm]{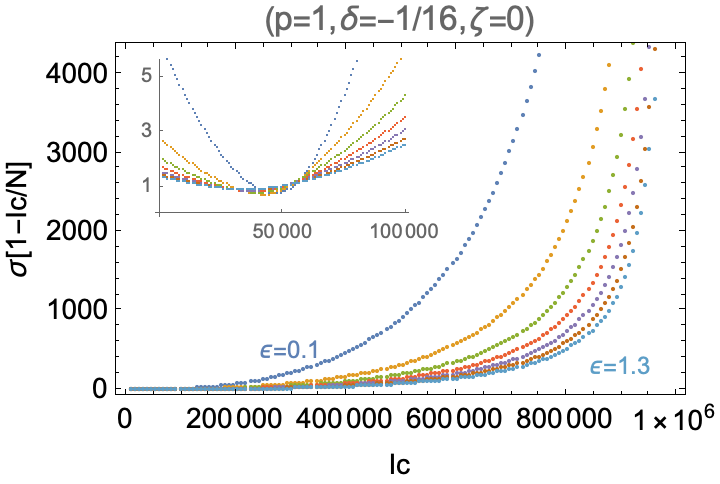}}}\nonumber\\
&\fbox{\parbox{7.25cm}{\includegraphics[width=7.25cm]{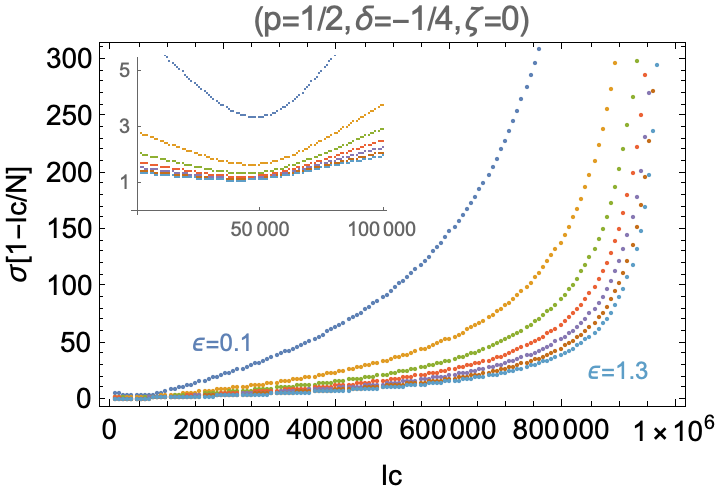}}} && \fbox{\parbox{7.25cm}{\includegraphics[width=7.25cm]{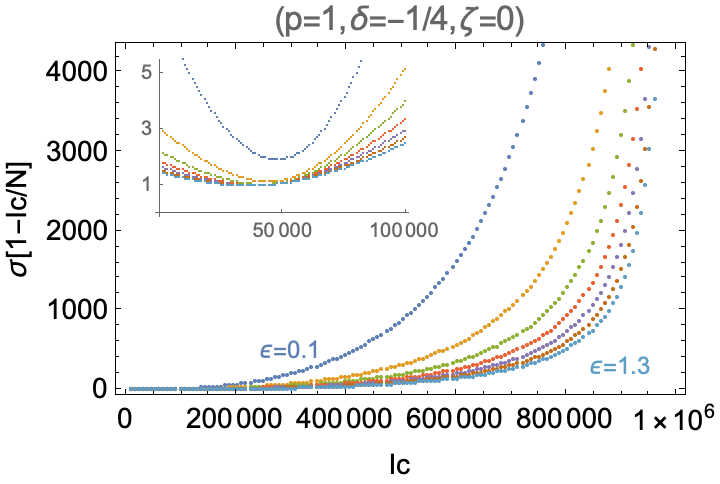}}}\nonumber\\
&\fbox{\parbox{7.25cm}{\includegraphics[width=7.25cm]{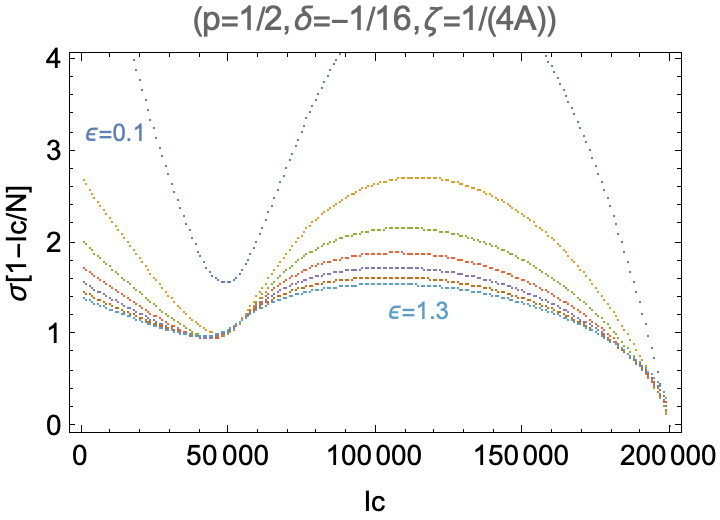}}} && \fbox{\parbox{7.25cm}{\includegraphics[width=7.25cm]{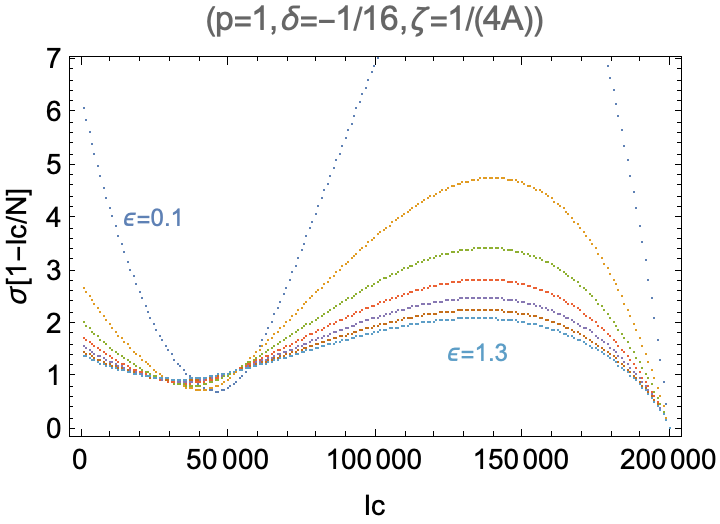}}}\nonumber
\end{align}
\caption{$\sigma$ as a function of $\Ic$ for different values of $(p,\delta,\zeta)$ in the limit $\ms_0\to 1$ with $N=1.000.000$, $A=50.000$, $\rgg =0.5$ and $\rhe\in\{0.1\,,0.3\,,0.5\,,0.7\,,0.9\,,1.1\,,1.3\}$.}
\label{SIRbetaSigmaZeta}
\end{figure}

We note that for $p\leq 1/2$, $\text{Im}(\sigma)\neq 0$ for $\Ic>A$, thus indicating that the solution does not extend beyond the maximal number of cumulative infected $\Ic=A$ (see Fig.~\ref{Fig:NumericImaginary}). Similar plots for $\delta\neq 0$ are shown in Fig.~\ref{SIRbetaSigmaZeta}.

Finally, we also remark that the numerical integration allows us to include $\delta<0$ and can even be generalised to more general classes of $\beta$-functions proposed in \cite{cacciapaglia2020evidence}
\begin{align}
- \beta_0 (\Ic) = {\rgg }\, \Ic  \left[  \left(1 - \frac{\Ic}{{A}} \right)^2  - \delta\right]^p(1-\zeta \Ic)\,,\label{TargetBeta}
\end{align}
as shown in Fig.~\ref{SIRbetaSigmaZeta}. In the case $\zeta>0$ we remark that $\text{Im}(\sigma)\neq 0$ for $\Ic>\zeta^{-1}$, indicating as above the breakdown of the assumptions. 

\subsection{Analytic Solution during a Linear Growth Phase} \label{Sect:AnSolLinear}

Many epidemics generated by an infectious disease feature a multi-wave pattern, with periods in between waves where an approximately linear growth of the number of infected is observed. As an example, COVID-19 data show this period very clearly in most of the countries, thanks to the large amount of data collected (see Section~\ref{Sect:CovidSpecific}). This phase of the epidemic, which links two consecutive waves, has found a natural explanation in the eRG framework  \cite{cacciapaglia2020evidence,cacciapaglia2020multiwave}, which we will review in Section~\ref{Sect:eRGcomplex}.

Here we attempt to describe this linear phase from the perspective of compartmental models. In fact, we have seen from the explicit solutions in Section~\ref{Sect:Percolation} that such a behaviour is not found in simple percolation models in which, notably, the probability or rate of infection remains constant throughout the entire pandemic. Similarly, this type of solutions is absent in compartmental models. However, more general approaches and extensions of these simple models might exhibit such linear growth phases. Since the phenomenon is seen in the cumulative number of infected (which is a `global' key figure pertaining to the entire population), we shall in the following analyse it from the perspective of a SIR model, with time-dependent infection and recovery rates.

\subsubsection{Simplified SIR Model with Constant New Infections}
We consider a SIR model described by the equations (\ref{SIReqs}) and the initial conditions (\ref{SIRInitial}) with time-dependent $\rin$, $\rhe$ and $\rsu$ (see Section~\ref{Sect:SIRtimeDependence}). We define a linear growth regime as a period in time $[t_1,t_2]$ during which the cumulative number of infected $\Ic$, defined in Eq.\eqref{CumulativeInfected} as
\begin{align}
\Ic(t)=N\,\mi _0+\int_0^t dt'\,\rin(t')\, N\,\mi (t')\,\ms (t')\,,
\end{align}
is a linear function of time. In other words,
\begin{align}
&\frac{d}{dt}\,\Ic(t)=N\,f=\text{const.}&&\forall t\in[t_1,t_2]\,,\label{Stroll}
\end{align}
while $0\leq \ms (t),\mi (t),\mr (t)\leq 1$, with $f\in\mathbb{R}_+$. This implies
\begin{align}
&\rin(t)\,\mi (t)\,\ms (t)=f&&\forall t\in[t_1,t_2]\,.\label{StrollingConstraint}
\end{align}
The condition above allows to analytically solve the SIR equations (\ref{SIReqs}) $\forall t\in[t_1,t_2]$ with the initial conditions at the beginning of the linear growth
\begin{align}
&\ms (t=t_1)=\ms_s\,,&&\mi(t=t_1)=\mi_s\,,&&\mr(t=t_1)=\mr_s\,,&&\text{with}&&\begin{array}{l}0\leq \ms_s,\mi_s,\mr_s\leq 1\,, \\ \ms_s+\mi_s+\mr_s=1\,.\end{array}\label{SIRboundary}
\end{align}
To see this, we define
\begin{align}
&\Dg(t)=e^{\int_{t_1}^t\rhe(t')dt'}\,,&&\text{and} &&\Dz(t)=e^{\int_{t_1}^t\rsu(t')dt'}\,,
\end{align}
which have the properties
\begin{align}
&\frac{d \Dg}{dt}(t)=\rhe(t)\,\Dg(t)\,,&&\frac{d \Dz}{dt}(t)=\rsu(t)\,\Dz(t)\,,&&\Dg(t=t_1)=1=\Dz(t=t_1)\,.\label{Dprops}
\end{align}
Next, we insert the constraint in Eq.\eqref{StrollingConstraint} into Eqs~\eqref{SIReqs} to obtain
\begin{align}
&\frac{d\mi}{dt}=-\rhe\,\mi +f\,,&&\forall t\in[t_1,t_2]\,.\label{IstrollFirst}
\end{align}
This differential equation only contains $\mi$ (hence, it is decoupled from $\ms$ and $\mr$). Multiplying by $\Dg(t)$, we find
\begin{equation}
\left[\frac{d\mi}{dt}+\rhe\,\mi \right]\,\Dg(t)=f\,\Dg(t) \qquad \qquad \Rightarrow \qquad \qquad
\frac{d}{dt}\left[\mi (t)\,\Dg(t)\right]=f\,\Dg(t)\,,
\end{equation}
which can be directly integrated, with the initial conditions \eqref{SIRboundary}, as:
\begin{align}
&\mi (t)=\frac{1}{\Dg(t)}\left[f\,\int_{t_1}^t\,\Dg(t')\,dt'+\mi _s\right]\,,&&\forall t\in[t_1,t_2]\,.\label{StrollingI}
\end{align}
For the relative number of recovered,  $\mr$, we can integrate the last equation of \eqref{SIReqs}
\begin{align}
\frac{d\mr}{dt}(t)+\rsu (t)\,\mr=\rhe(t)\,\mi (t)\,,
\end{align}
where,  inserting the solution for $\mi (t)$ in Eq.\eqref{StrollingI}, the right hand side can be understood as an inhomogeneity. Multiplying by $\Dz$ we obtain, as before,
\begin{align}
\frac{d}{dt}\left[\mr (t)\,\Dz(t)\right]=\rhe(t)\,\mi (t)\,\Dz(t)\,,
\end{align}
which can be directly integrated, with the initial conditions \eqref{SIRboundary}, to give
\begin{align}
&\mr(t)=\frac{\mr_s}{\Dz(t)}+\mi_s\,\int_{t_1}^tdt'\,\frac{\rhe(t')}{\Dg(t')}\,\frac{\Dz(t')}{\Dz(t)}+f\,\int_{t_1}^tdt'\int_{t_1}^{t'}dt''\,\rhe(t')\,\frac{\Dg(t'')}{\Dg(t')}\,\frac{\Dz(t')}{\Dz(t)}\,,&&\forall t\in[t_1,t_2]\,.\label{StrollingR}
\end{align}
Finally, $\ms(t)$ is obtained through the constraint (\ref{ConstraintSIR}): $\ms(t)=1-\mi(t)-\mr(t)$. Notice that the solutions (\ref{StrollingI}) and (\ref{StrollingR})  remain valid as long as $0\leq \ms (t),\mi (t),\mr (t)\leq 1$.

\subsubsection{Vanishing $\rsu$ and Constant $\rhe$}
To simplify the solutions found above, we can adapt the functions $\rsu$ and $\rhe$ to reflect more closely the COVID-19 pandemic: since currently only very few cases of patients contracting COVID-19 twice within a short time, \emph{i.e.} a single epidemic wave, are known in the medical literature \cite{Stokel-Walkern99} we can set $\rsu(t)=0$ to simplify the solutions \eqref{StrollingI} and \eqref{StrollingR}. Since $\rsu=0$ also implies $\Dz(t)=1$, we find for these solutions
\begin{align}
\ms(t)&=\ms_s-f (t-t_1)\,,\nonumber\\
\mi (t)&=\frac{\mi_s}{\Dg(t)}+f\,\int_{t_1}^t\,\frac{\Dg(t')}{\Dg(t)}\,dt'\,,\nonumber\\
\mr(t)&=\mr_s+\mi_s\,\int_{t_1}^tdt'\,\frac{\rhe(t')}{\Dg(t')}+f\,\int_{t_1}^tdt'\int_{t_1}^{t'}dt''\,\rhe(t')\,\frac{\Dg(t'')}{\Dg(t')}\,,\hspace{1.8cm}\forall t\in[t_1,t_2]\,.\label{SIRstrollingSolZeta0}
\end{align}
We have explicitly verified that this is indeed a solution of Eqs~\eqref{SIReqs} that satisfies the correct initial conditions.

\begin{figure}[ht]
\centering
\includegraphics[width=7.5cm]{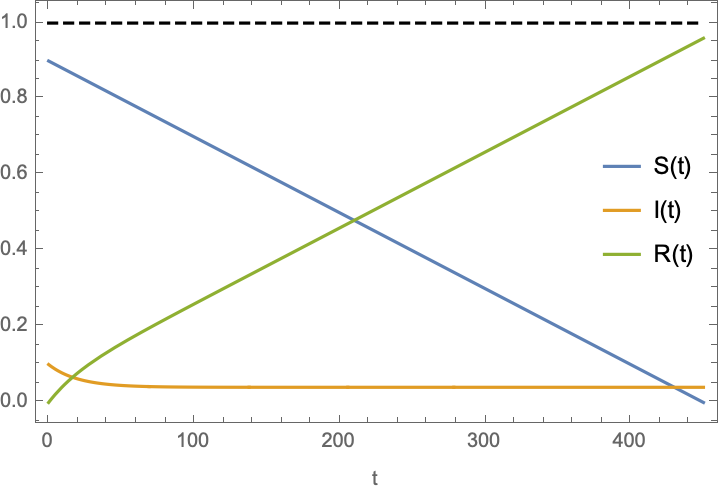}
\includegraphics[width=7.5cm]{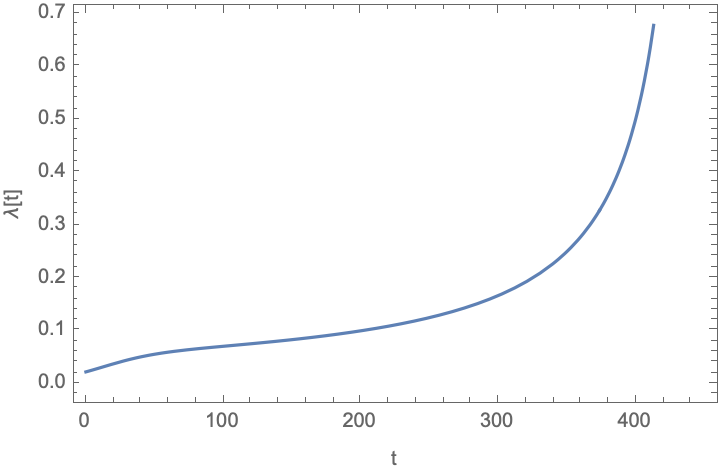}
\caption{Solutions (\ref{SIRstrollingSolZeta0}) and $\rin(t)$ for $\rhe=0.05$, $f=0.002$, $\ms_s=0.9$, $\mi_s=0.1$, $\mr_s=0$ and $t_1=0$ as a function of time $t$.}
\label{Fig:SIRzeta0gammacon}
\end{figure}

Furthermore, since the recovery rate in most cases depends on the disease in question and/or the state of medical advancement to cure it, $\epsilon$ is difficult to change throughout a pandemic without significant effort. For simplicity, we therefore consider it in the following to be constant, \emph{i.e.} $\rhe=$ const. (in addition to $\rsu=0$), such that $\Dg(t)=e^{\rhe(t-t_1)}$. In this case, we can perform the integrations in Eq.\eqref{SIRstrollingSolZeta0} to obtain 
\begin{align}
\mi (t)&=e^{-\rhe(t-t_1)}\,\left[f\int_{t_1}^tdt'\,e^{\rhe(t'-t_1)}+\mi _s\right]=e^{-\rhe(t-t_1)}\,\mi _s+\frac{f}{\rhe}\left(1-e^{-\rhe(t-t_1)}\right)\,,&&\forall t\in[t_1,t_2]\,,\label{StrollConstantI}
\end{align}
as well as the relative number of removed
\begin{align}
\mr(t)&=\mr_s+\mi_s\,\rhe\,\int_{t_1}^t\,dt'\,e^{-\rhe(t'-t_1)}+\rhe f\int_{t_1}^t dt'\,e^{-\rhe t'}\int_{t_1}^{t'}dt''\,e^{\rhe t''}\nonumber\\
&=\mr_s+f (t-t_1)+\left(\mi_s-\frac{f}{\rhe}\right)\left(1-e^{-\rhe(t-t_1)}\right)\,,&&\forall t\in[t_1,t_2]\,.
\end{align}
One can directly verify that these expressions satisfy Eqs~\eqref{SIReqs} along with
\begin{align}
&\ms(t)+\mi(t)+\mr(t)=\ms_s+\mi_s+\mr_s\,,&&\forall t\in[t_1,t_2]\,.
\end{align}
For some (random) values of $\rhe$, $f$, $\ms_s$, $\mi_s$ and $\mr_s$, the functions $\ms(t)$, $\mi(t)$ and $\mr(t)$ (for the region where $0\leq \ms (t),\mi (t),\mr (t)\leq 1$) are plotted in the left panel of~\figref{Fig:SIRzeta0gammacon}, while the associated $\rin(t)=\frac{f}{\ms(t)\,\mi(t)}$ is plotted in the right panel.

\subsubsection{Constant Active Number of Infectious Individuals}

During the linear growth periods, the COVID-19 data also shows that the number of active infectious individuals remains constant. Intriguingly, this feature is also observed in the solutions in the left panel of~\figref{Fig:SIRzeta0gammacon}. 
In this section, we will seek a solution of the SIR model with this property,~\emph{i.e.} 
\begin{align}
&\mi(t) =\mathsf{f}=\text{const.}&&\forall t\in [t_1,t_2]\,,
\end{align}
for some $\mathsf{f}\in [0,1]$, which in particular implies
\begin{align}
&\frac{d}{dt}\,I(t)=0\,,&&\forall t\in [t_1,t_2]\,.
\label{strolling-SIR}
\end{align}
Injecting this into Eqs~\eqref{SIReqs}, we obtain under the assumption $\mi(t)\neq 0$ $\forall t\in[t_1,t_2]$
\begin{align}
&\ms =\frac{\rhe}{\rin}\,,&&\forall t\in [t_1,t_2]\,, 
\label{seqepsovergamma}
\end{align}
and thus for $\rsu\neq 0$
\begin{align}
&\frac{d}{dt}\left(\frac{\rhe}{\rin}\right)=-\rhe\,\mathsf{f}\,+\rsu\,\mr \,,&&\Longrightarrow &&\mr =\frac{1}{\rsu}\left[\frac{d}{dt}\left(\frac{\rhe}{\rin}\right)+\rhe \mathsf{f}\right]\,,&&\forall t\in [t_1,t_2]\,.
\end{align}
For $\rsu=0$ we obtain the following constraint for the infection and recovery rate 
\begin{align}
&\frac{d}{dt}\left(\frac{\rhe}{\rin}\right)=-\rhe\,\mathsf{f}\,,&&\forall t\in [t_1,t_2]\,.\label{SIRConstConstraint}
\end{align}
For the classical SIR model (for which $\rhe$ and $\rin$ are time-independent $\forall t$ and $\rsu=0$), assuming that $\rin\neq 0$, the constraint (\ref{SIRConstConstraint}) implies that either 
\begin{itemize}
\item $\mathsf{f}=0$, which however is excluded since $\mi \neq 0$;
\item or $\rhe=0$, in which case $\frac{d\mr}{dt}=0$ $\forall t$ (\emph{i.e.} not just $t\in[t_1,t_2]$). However, with the initial conditions (\ref{SIRInitial}) this implies $\mr (t)=0$ and thus 
\begin{align}
&\frac{d}{dt}\ms(t)=-\rin\,\mathsf{f}\,\ms  &&\Longrightarrow&&\ms =c\,e^{-\rin \mathsf{f}\,t}\,,&&\forall t\in [t_1,t_2]\,,
\end{align}
for $c\in [0,1]$. On the other hand the relation (\ref{ConstraintSIR}) implies that $\frac{d\ms}{dt}=0$ and thus (with $\rin\neq 0$ and $\mathsf{f}\neq 0$) $\ms =0$ (consistent with Eq.\eqref{seqepsovergamma}), in which case $\mi =\mathsf{f}=1$ and the entire population is infected (and stays infected for all times).
\end{itemize}

\noindent
Thus, within the classical SIR model, the only solution with $\mi(t)=\mathsf{f}\neq 0$ constant is $\rhe=0$ (\emph{i.e.} instead of the SIR model we only consider the SI model) and $\mi=1$. This corresponds to the late phase of the SI model, where the entire population is infected.  We, therefore, see that the traditional SIR model cannot account for the linear growth of the cumulative number of infected related to  Eq.\eqref{strolling-SIR} and observed in the COVID-19 data.


\section{Epidemic Renormalisation Group}\label{Sect:RGapproach}

\begin{summary}{Executive Summary}
\begin{itemize}
\item We introduce the epidemic renormalisation group approach that efficiently captures asymptotic time symmetries of the diffusion process.
\item The framework is based on flow equations characterised by fixed point dynamics. 
\item We show the power of the approach by studying single wave epidemics, which can be naturally generalised to describe multi wave patterns via complex fixed points
\item We demonstrate how the spreading of diseases across different regions of the world can be efficiently described and predicted
\end{itemize}

\end{summary}

As anticipated in the previous section, it has been proposed in \cite{DellaMorte:2020wlc,Cacciapaglia:2020mjf} to study the spread of a communicable infectious disease within the framework of the Wilsonian renormalisation group \cite{Wilson:1971bg,Wilson:1971dh}. We have already pointed out in Section \ref{Sect:CompartmentalModels} that the SIR model, defined by the differential equations \eqref{SIReqs}, can be formulated in a fashion that is structurally similar to a set of RGEs (see \cite{mcguigan2020pandemic}).  In this section we review the new framework, first proposed in \cite{DellaMorte:2020wlc,Cacciapaglia:2020mjf}, dubbed \emph{epidemic Renormalisation Group (eRG)}. 

The eRG framework consists, effectively, in a single differential equation that captures the time evolution of the disease diffusion, after the microscopic degrees of freedom and interactions have been  `integrated out' and all the detailed effects (virulence of the disease, social measures, \emph{etc.}) are taken into account by the few parameters in the equation. This leads to a much more economical description in terms of calculation complexity as compared to microscopic or compartmental models. At this stage, the main relation with the renormalisation group is the fact that symmetries can be explicitly included in the formalism. In the case of the eRG, the symmetries are related to time scale invariance, \emph{i.e.} the presence of phases where the disease diffusion is (nearly) stable in time. In the Wilsonian renormalisation group, which describes the energy dependence of physical charges (for instance, the interaction strength among fundamental particles), the symmetries involved are related to scale invariance of distances and energies. A RGE, therefore, describes the energy flow of a charge from the Ultra-Violet (UV) regime at high energies to the Infra-Red (IR) regime at low energies.
The eRG also describes a flow, however in time instead of in energy, as we will see shortly.  The physical charge is replaced by an \emph{epidemiological charge}, which is defined as a monotonic, differentiable function of the cumulative number of individuals infected by the disease as a function of time. This discussion has already been anticipated in Section~\ref{Sect:BetaFunctionSIR}.

The economy of this approach in terms of free parameters and computing time needed to solve the flow of the disease makes it an ideal tool to study the diffusion of an infectious disease at different scales, from small regions to a global level. The eRG framework has been first used to characterise a single epidemic wave, \emph{i.e.} a single episode of exponential increase in the number of infections followed by an attenuation \cite{DellaMorte:2020wlc}, and extended to study the inter-region propagation of the disease \cite{Cacciapaglia:2020mjf}, with validation on the COVID-19 data  in Europe \cite{cacciapaglia2020second} and in the US states \cite{cacciapaglia2020better}. Mobility data have also been used to study the effect of non-pharmaceutical interventions \cite{cacciapaglia2020mining} as well as the role played by passenger flights in the US \cite{cacciapaglia2020better}. As we also review in this section, the framework can be extended to include the multi-wave pattern \cite{cacciapaglia2020evidence,cacciapaglia2020multiwave} that emerges in many communicable diseases, like the 1918 influenza pandemic, the seasonal flu and the COVID-19 pandemic of 2019. Finally, preliminary work on the inclusion of vaccinations \cite{cacciapaglia2020better} and virus mutations \cite{cacciapaglia2021epidemiological,dehoffer2021variantdriven} are present in the literature, however we will not cover them in this review.

\subsection{Beta Function and Asymptotic Fixed Points}

The main motivation behind the eRG approach to epidemiology stems from the observation that a single epidemic wave starts with a very small number of infected individuals and ends when the cumulative number of infections reaches a constant, hence no new infections are detected. This dynamics is characteristic of a system that flows from a fixed point at $t = - \infty$, when no infections are present, to a new fixed point at $t = \infty$, when the number of cumulative infections reaches a constant value again. The dynamical flow between the two fixed points can be described by the following differential equation:
\begin{align}
- \beta (\alpha) = \frac{d \alpha}{dt}(t) = \rgg \, \alpha  \left( 1 - \frac{\alpha}{A} \right)\,,\label{eq:beta0}
\end{align}
where $\alpha$ is a function of the number of infected, hence a function of time. The precise form of this equation is an ansatz, for now, and we will establish the precise relation between $\alpha$ and the number of infections later. The main feature to stress is the presence of two zeros, corresponding to the fixed points of the system: if $\alpha (t_0) = 0$ or $\alpha (t_0) = A$ at any time $t_0$, the system will remain in this state at all times. 
The zeros can be characterised through the so-called \emph{scaling exponents}:
\begin{align}
\vartheta=\frac{\partial\beta}{\partial\alpha}\bigg|_{\alpha^\ast}=\left\{\begin{array}{lcl}-\lambda & \text{for} & \alpha_1^*=0\,, \\[4pt] \lambda & \text{for} & \alpha_2^*=A\,,\end{array}\right.
\end{align} 
where $\alpha^\ast$ is the epidemic coupling constant at the fixed point.
A negative (positive) scaling exponent corresponds to a repulsive (attractive) fixed point.
Thus, a system in the repulsive fixed point at $\alpha^\ast = 0$, once perturbed (by a small initial number of infected individuals) will flow towards the attractive fixed point at $\alpha^\ast = A$. As such, $A$ is a function of the cumulative number of individuals infected during the epidemic wave.

\begin{wrapfigure}{l}{0.45\textwidth}
\vspace{-0.3cm}
\parbox{7cm}{\includegraphics[width=7cm]{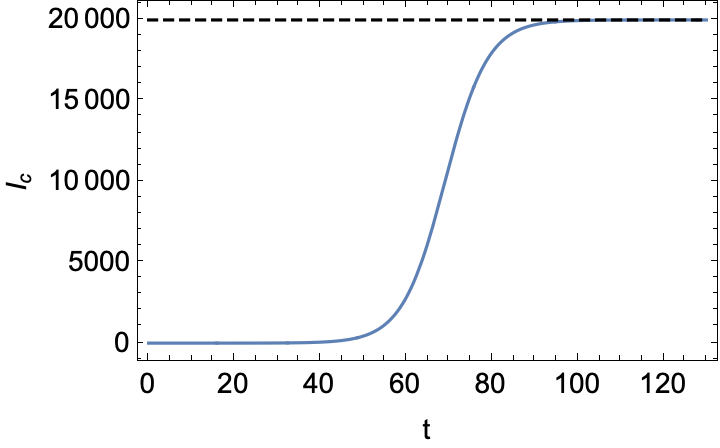}}
\caption{The logistic function schematically representing the cumulative number of infected as a function of time. With regards to \eqref{LogisticFunctionGen} we have $A=20.000$, $B=1.000.000$ and $\kappa=0.2$.}
\label{Fig:LogisticSchematic}
${}$\\[-0.5cm]
\end{wrapfigure} 

The solution of Eq.\eqref{eq:beta0} is a \emph{logistic function} (sigmoid) of the form:
\begin{align}
\alpha\,:\mathbb{R}&\longrightarrow [0,A]\nonumber\\
t&\longmapsto \alpha(t)=\frac{A}{1+B\, e^{-\lambda t}}\,,\label{LogisticFunctionGen}
\end{align}
where $A,B,\lambda\in\mathbb{R}_+\setminus\{0\}$.
This function shows a characteristic `S'-shape (see Fig.~\ref{Fig:LogisticSchematic} for a schematic representation) and has the following asymptotic values
\begin{align}
&\lim_{t\to-\infty} \alpha(t)=0\,,&&\lim_{t\to\infty} \alpha(t)=A\,,
\end{align}
corresponding to the zeros of the derivative $\frac{d\alpha}{dt}=0$.

As already mentioned, the parameter $A$ corresponds to (a function of) the asymptotic number of infected cases during the epidemic wave.
The second parameter in Eq.\eqref{eq:beta0}, $\lambda$, which has dimension of a rate, measures how fast the number of infections increases, while $B$ is an integration constant that corresponds to a shift of the entire curve in time and determines the beginning of the infection increase. More details about the properties of this function and its epidemiological interpretation can be found in \cite{DellaMorte:2020wlc} and will not be repeated here. 
It is, however, important to notice that the parameters $\lambda$ and $A$ can be removed from the differential equation by a simple rescaling of the function and of the time variable:
\begin{align}
\frac{d\tilde \alpha}{d\tau}=\tilde \alpha(\tau)\,(1-\tilde \alpha(\tau))\,, \qquad \tau = \lambda t\,, \quad \tilde \alpha (\tau) = \frac{\alpha (\tau/\lambda)}{A}\,.
\end{align}
Thus, while $A$ is a mere normalisation, $\lambda$ can be thought of as a `time dilation' parameter. Once the normalised solutions are shown in the `local time' $\tau$, therefore, all epidemic waves should reveal the same universal temporal shape.

This universality property has been first pointed out in \cite{DellaMorte:2020wlc} from data of the Hong Kong (HK) Sars-2003 outbreak as well as the COVID-19 pandemic during the spring of 2020. It has been shown that the time dependence of the cumulative total number of infected cases in various regions of the world shows the same characteristic behaviour. In \cite{DellaMorte:2020wlc}, the epidemic coupling has been defined as the logarithm of the cumulative infected, $\alpha (t) = \ln I_c (t)$, however other choices, like $\alpha (t) = I_c (t)$, can also reproduce the data. The same framework can also be applied to the number of hospitalisations or the number of deceased individuals. This feature of the epidemiological data shows that the dynamics encoded in Eq.\eqref{eq:beta0} provides an accurate description of the diffusion of an infectious diseases in terms of a flow equation.

In \cite{DellaMorte:2020wlc,Cacciapaglia:2020mjf,cacciapaglia2020evidence} the following dictionary between the spread of an epidemic and the Wilsonian renormalisation group was proposed:
\begin{itemize}

\item The time variable is identified with the (negative) logarithm of the energy scale $\mu$
\begin{align}
\frac{t}{t_0}=-\ln\left(\frac{\mu}{\mu_0}\right)\,,
\end{align}
where $t_0$/$\mu_0$ set the scale for the time and energy (for simplicity, and without loss of generality, we will fix $t_0=1$). With this identification, Eq.\eqref{eq:beta0} is similar to the RGE for the gauge coupling in a theory that features a Banks-Zaks type fixed point \cite{Banks:1981nn}, \emph{i.e.} an interactive fixed point at low energies (in the Infra-Red).
 
\item The solution can be associated to a coupling constant in the high energy physics RGEs, $\alpha$. The epidemic coupling strength is defined as a monotonic, differentiable and bijective, function $\phi$ of the cumulative number of infected cases
\begin{align}
\alpha(t)=\phi(\Ic(t))\,.
\end{align}
In \cite{DellaMorte:2020wlc,cacciapaglia2020evidence} $\phi$ was chosen as the natural logarithm $\phi(x) = \ln(x)$, while in \cite{cacciapaglia2020evidence,cacciapaglia2020multiwave} it was chosen $\phi(x)=x$. The choice was justified by a better fit to the actual data of the COVID-19 pandemic, while from the perspective of the Wilsonian renormalisation group, the difference corresponds to a different choice of scheme.

\item The \emph{beta function} is defined as the (negative) time-derivative of the epidemic coupling strength
\begin{align}
\beta\equiv\frac{d\alpha}{d\ln\left(\frac{\mu}{\mu_0}\right)}=-\frac{d\alpha}{dt}=-\frac{d\phi}{d\Ic}\,\frac{d\Ic}{dt}(t)\,.
\end{align}
\end{itemize}

In order to better model the respective data of various countries during the COVID-19 pandemic, it was furthermore proposed in \cite{cacciapaglia2020evidence,cacciapaglia2020multiwave} to consider the more general beta-function
\begin{align}
- \beta (\alpha) = \frac{d \alpha}{dt}(t) = \rgg \, \alpha  \left( 1 - \frac{\alpha}{A} \right)^{2p}\,,\label{eq:betap}
\end{align}
for $p\in[1/2,\infty]$ and $\rgg,A\in\mathbb{R}_+$. The role of the exponent $p$ is to smoothen the `S'-shape of the solution when it approaches the attractive fixed point at $\alpha^\ast = A$.

\subsubsection{Generalisation to multiple regions} \label{sec:multiregion}

The approach discussed so far assumes an isolated population of sufficient size. However, the simplicity of the eRG approach allows for a simple generalisation to study the interaction between various regions of the world \cite{Cacciapaglia:2020mjf} via the mobility of individuals. For $M$ separated populations (labelled by $i=1,\ldots,M$) of size $N_i$ whose cumulative number of infected is denoted by $I_{\text{c},i}$, it was proposed in \cite{Cacciapaglia:2020mjf} that infections can be transmitted between these populations by travellers. Hence, the epidemic diffusion can be described by $M$ coupled differential equations, in the form of Eq.\eqref{eq:betap} for each population, with the addition of an interaction term:
 \begin{align}
- \beta (\alpha_i)  = \rgg \, \alpha_i  \left( 1 - \frac{\alpha_i}{A} \right)^{2p} + \frac{d\phi}{dI_{\text{c},i}}\,  \sum_{j=1}^M\frac{k_{ij}}{N_i}  \left( I_{\text{c},j} (t) - I_{\text{c},i} (t) \right)\,,
\label{eq:deltaI}
\end{align}
where $k_{ij}\in\mathbb{R}$ is a measure for the number of travellers between populations $i$ and $j$. The contribution to the beta function can be obtained by replacing $I_{\text{c},i} \to \phi^{-1} (\alpha_i)$, where $\alpha_i$ is the epidemic coupling in each population. For more details, see Ref.~\cite{Cacciapaglia:2020mjf}.

\begin{figure}
\begin{center}
\includegraphics[width=7.5cm]{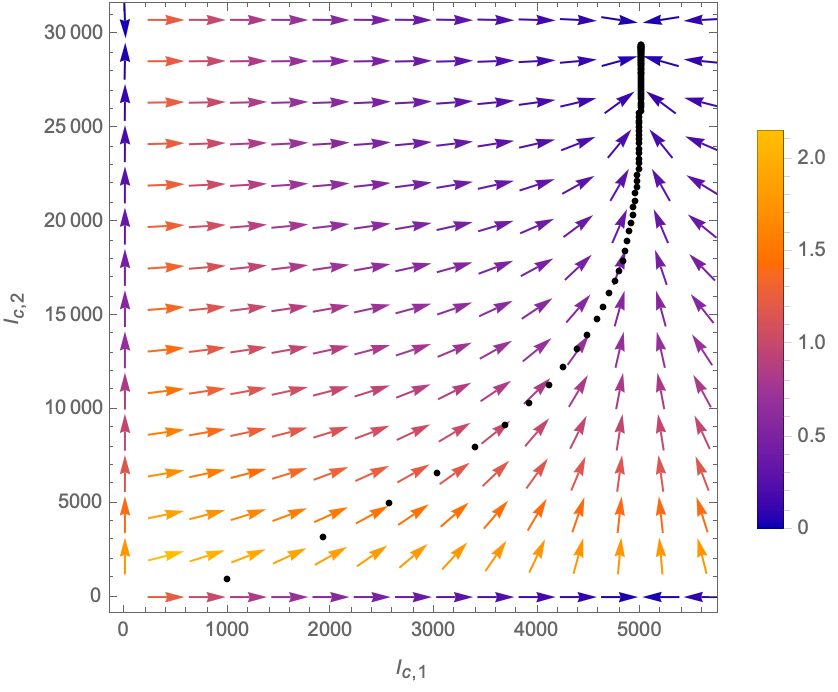} 
\end{center}
\caption{Schematic representation of the flow in a two-region coupled eRG framework of Eq.~\eqref{eq:deltaI}. In this fictitious example we fix  $\lambda_1=0.7$, $\lambda_2=0.9$, $N_1=200000$, $N_2=300000$, $A_1=\log\left(\frac{1}{40}N_1\right)$, $A_2=\log\left(\frac{N_2}{10}\right)$ and $p_1=p_2=\frac{1}{2}$. For the matrix of couplings $k_{ij}$, we use $k_{12} = k_{21}=10^{-3}$ and $k_{11} = k_{22} = 0$. The two-component vectors are given by $(-\beta(\alpha_1),-\beta(\alpha_2))$, with $I_{c}\in\mathbb{C}$ with the overall length represented by the colour-coding. The function $\phi$ was chosen $\phi(x)=x$ $\forall x\in\mathbb{R}$.}
\label{Fig:CoupledFlow}
\end{figure}

These coupled differential equations can be thought of flow equations, in the spirit of the Wilsonian renormalisation, with the second term representing a coupling between the different regions. A graphical representation of the coupled $\beta$-functions in Eq.\eqref{eq:deltaI} can be given in the form of flow in an $M$-dimensional space. In Fig.~\ref{Fig:CoupledFlow}, we provide a numerical (fictitious) example for $M=2$: choosing the scheme $\alpha_i(I_{c,i})=\ln(I_{c,i})$ for $i=1,2$, the arrows indicate the vector field $\left(\begin{array}{c}-\beta(\alpha_1) \\ -\beta(\alpha_2)\end{array}\right)$ with the colour representing the length $\sqrt{\beta(\alpha_1)^2+\beta(\alpha_2)^2}$ at any point in the $(I_{c,1},I_{c,2})$-plane. The black dots are the actual trajectory of the system calculated as the numerical solution of the coupled differential equations \eqref{eq:deltaI}. As it can be seen, the former flows along the arrows from a repulsive fixed point at $(I_{c,1},I_{c,2})=(0,0)$ (all arrows point away from it), which represents the absence of the disease in both countries, to an attractive fixed point (all arrows point towards it) which corresponds to the eradication of the disease.

The coupled eRG framework in Eq.\eqref{eq:deltaI} has been used to explain the diffusion of the COVID-19 pandemic across different regions of the world. This is one of the main mechanisms that can generate multiple waves across a geographic region, while a second one will be discussed in the next section. The method has been used to predict the arrival of a second COVID-19 wave, which has hit Europe in the fall of 2020 \cite{cacciapaglia2020second}: the new infections originate from a seed region, which can be interpreted as inflow from outside Europe or the effect of hotspots and clusters, while the number of travellers, i.e. the entries of $k_{ij}$, were generated randomly. In Ref.~\cite{cacciapaglia2020better}, the same framework was used to explain the geographical wave patterns observed in the United States, with the aid of open-source flight data to estimate the couplings.

\subsection{Complex (fixed point) epidemic Renormalisation Group}\label{Sect:eRGcomplex}
Although the beta-function in Eq.\eqref{eq:beta0} is relatively simple and contains only two parameters, it describes the time evolution of short-time epidemics (such as HK SARS-2003 and each wave of COVID-19) quite efficiently, as the flow from a repulsive to an attractive fixed point (or from an UV to an IR fixed point in the language of high-energy physics). However, this beta-function is too simple to describe correctly longer lasting pandemics with a more intricate time-evolution, such as subsequent waves of COVID-19: the attractive fixed point at $t\to \infty$ corresponds to a complete eradication of the disease and Eq.\eqref{eq:beta0} describes outbreaks that follow a single wave. 
We have already discussed the role of passenger mobility in generating further epidemiological waves. However, data from COVID-19 has unveiled a second potential mechanism that may be at the origin of multiple-waves: in fact, after the end of each wave, a period of linear growth has been observed in all regions of the world (except those where the virus has been locally eradicated thanks to aggressive isolation policies). This is characterised by a nearly-constant number of new infected cases, and it can be seen as an endemic phase of the pandemic, where the virus circulates within the local population, without an exponential increase.

\begin{figure}
\includegraphics[width=8cm]{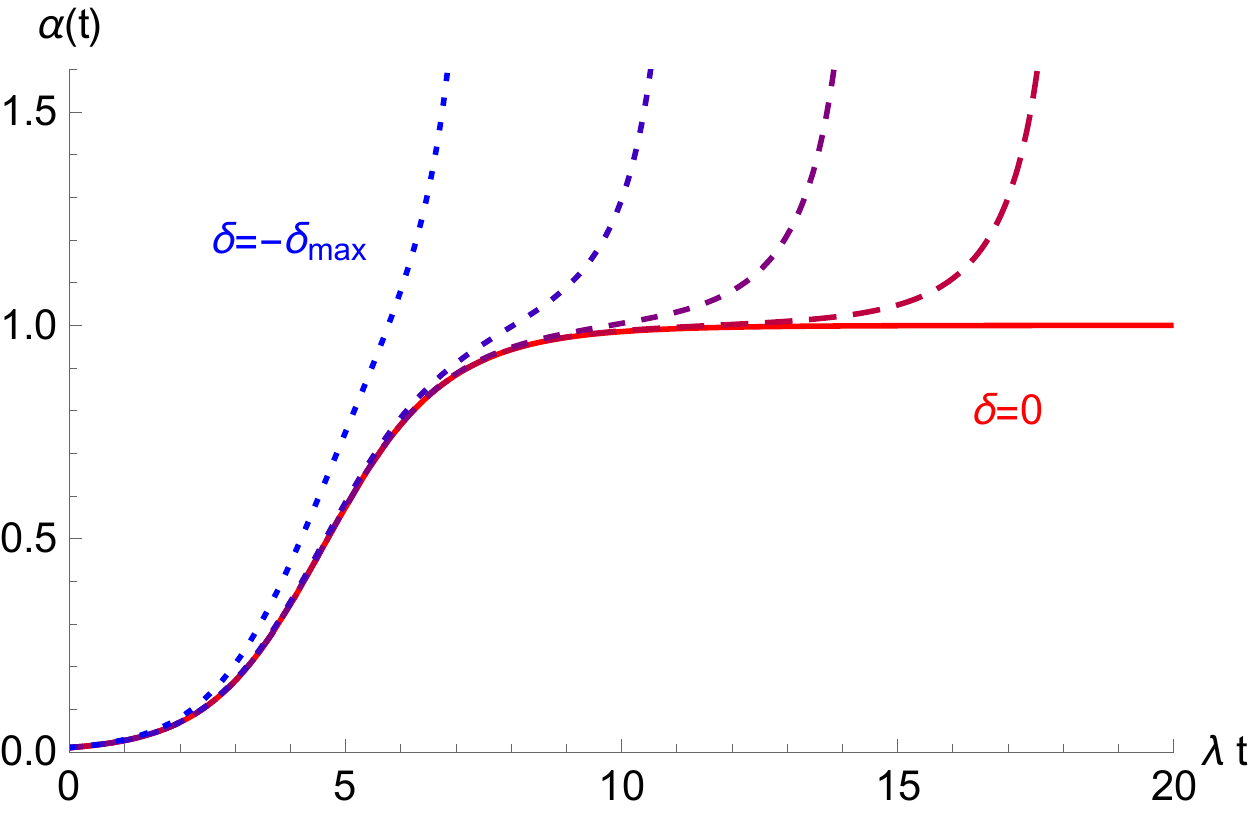} 
\includegraphics[width=8cm]{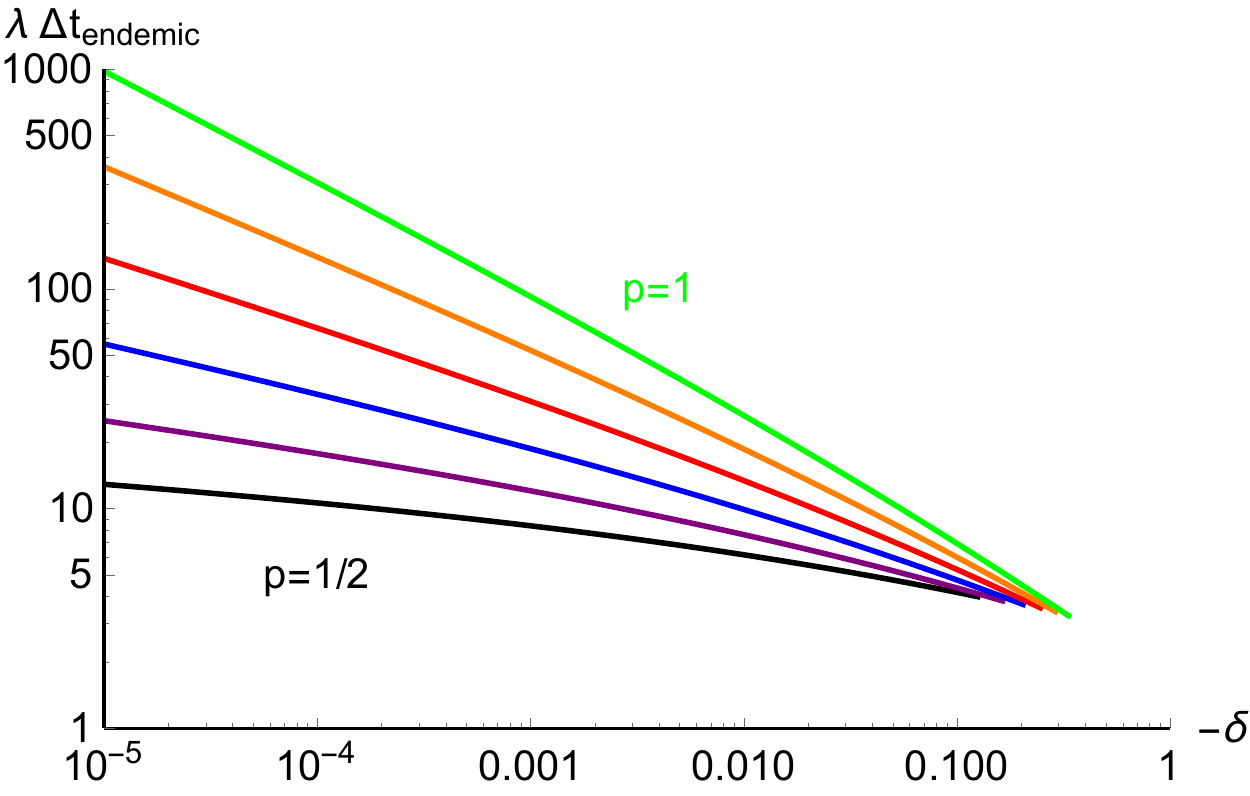} 
\caption{Right: solutions of the CeRG equation, normalised to $A=1$ and with time in units of $\rgg$, for $-\delta = 0, 10^{-4}, 10^{-3}, 10^{-2}$ and $\delta_{\rm max}$, for $p=0.55$. Left: Estimated duration of the linear growth phase, in units of $\rgg$, as a function of $-\delta$ for $p=0.5$, $0.6$, $0.7$, $0.8$, $0.9$ and $1$. The lines end for $\delta = - \delta_{\rm max}$.}
\label{Fig:Strollingtime}
\end{figure}

In \cite{cacciapaglia2020evidence} it was proposed that this linear phase is evidence for a near time-scale invariance symmetry in the dynamics governing the diffusion of the virus. 
In practice, the system does not reach the second fixed point of Eq.\eqref{eq:beta0}, instead it hits an instability that drives the system to a new exponential phase after a given amount of time.
The time-evolution of pandemics can still be described within the framework of a RGE, however with a more complicated beta-function that features a richer structure of (complex) fixed points. The new framework was called the \emph{Complex epidemic Renormalisation Group (CeRG)}. In the CeRG approach, the beta function of Eq.\eqref{eq:betap} is modified as follows:
\begin{align}
- \beta (\Ic) = \frac{d \Ic}{dt} = \rgg\, \Ic  \left[  \left(1 - \frac{\Ic}{{A}} \right)^2  - \delta\right]^p
 = \rgg \, \Ic \left(\frac{\Ic}{{A}} - 1 +  \sqrt{\delta} \right)^p   \left(\frac{\Ic}{{A}} - 1 -  \sqrt{\delta} \right)^p \, , 
\label{eq:beta1}
\end{align}
where the additional parameter $\delta\in \mathbb{R}_-$,  \emph{i.e.} $\delta=-|\delta|$. 
While this equation can be written for any epidemic coupling $\alpha$, here we commit to the case $\alpha(t) = \Ic(t)$ for reasons that will be clear in the next Section.
The eRG equation \eqref{eq:betap} can be recovered for $\delta \to 0$. For non-vanishing $\delta$, instead of only two asymptotic fixed points, this functions has three fixed points
\begin{align}
& I_{\text{c},0} =0 \,,&& I_{\text{c},\pm} = A\left(1 \pm i \sqrt{|\delta|}\right)\,,
 \end{align} 
with complex $I_{\text{c},\pm}\in\mathbb{C}$. Besides the repulsive fixed point at $\Ic^\ast = 0$, which remains, the attractive fixed point splits into two complex fixed points.  Since the (cumulative) number of infected individuals is a strictly real number, the system cannot actually reach the complex fixed points and thus cannot exactly enter into a time-scale invariant regime at infinite time. Instead, for small $|\delta|$, when the solution approaches the would-be fixed point at $\Ic \approx A$, the time evolution will be strongly slowed down due to the effect of the nearby complex fixed points. This results in a near-linear behaviour of the solution, as shown in the left panel of \figref{Fig:Strollingtime}.
Thus, the new beta function \eqref{eq:beta1} realises an approximate time-scale symmetry in the solution. Concretely, the precise form of the flow in the vicinity of these complex fixed points depends on $|\delta|$:

\begin{figure}
\includegraphics[width=9cm]{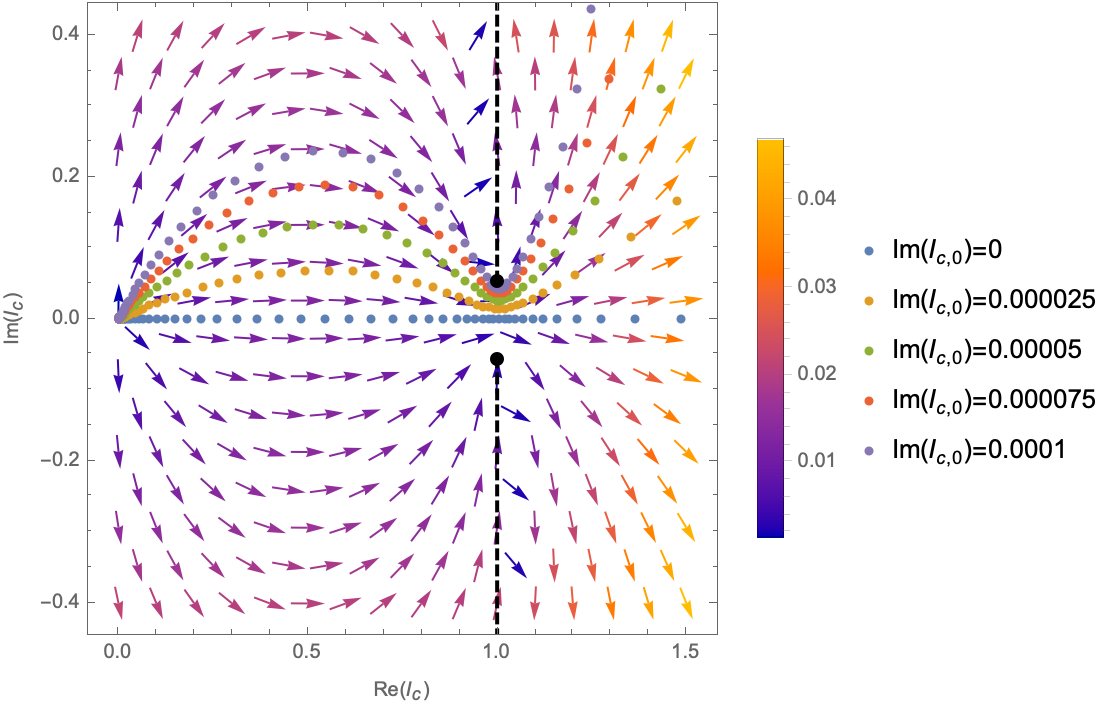} \hspace{0.5cm}
\includegraphics[width=7.2cm]{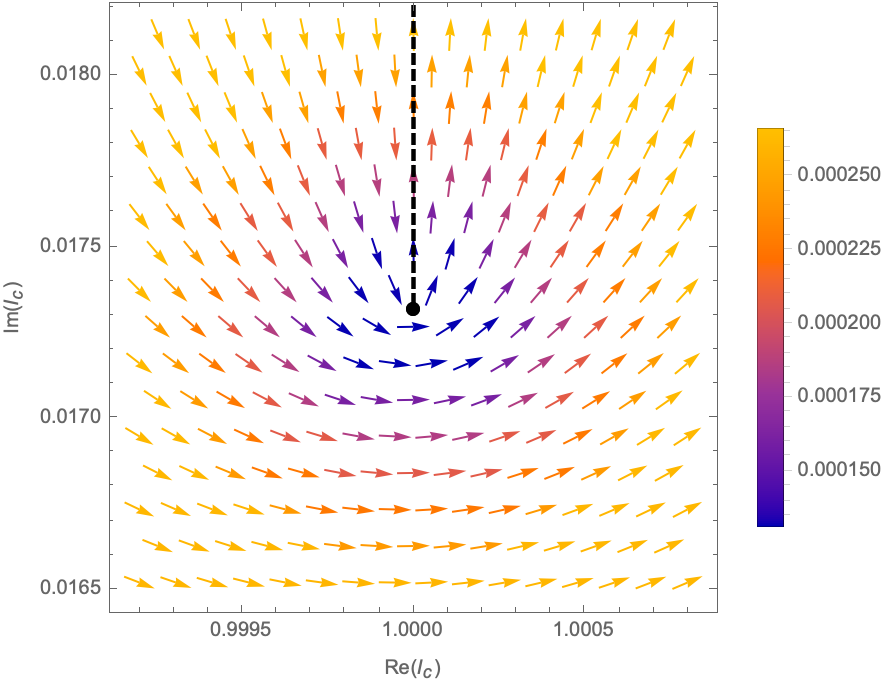} 
\caption{Schematic flow diagrams representing the $\beta$-function (\ref{eq:beta1}) with $A=1$, $\lambda=0.05$, $\delta=-0.003$ and $p=1/2$: the two-component vectors are given by $(\text{Re}(-\beta(I_{c})),\text{Im}(-\beta(I_{c})))$, with $I_{c}\in\mathbb{C}$ with the overall length represented by the colour-coding. Left panel: trajectories of the flow in the complex plane with initial conditions $I_c(t=0)=I_{c,0}$, with $\text{Im}(I_{c,0})$ specified in the figure. Right panel: close-up on the complex fixed point $I_{c,+}$. The dashed line represents a branch cut, which needs to be chosen such that it does not intersect the real axis.}
\label{Fig:ComplexFlow}
\end{figure} 

\begin{itemize}
\item For $|\delta|<\delta_{\text{max}}=\frac{p^2}{1+2p}$, the beta-function has a local maximum and $\Ic$ enters into a regime of near linear growth characterised by
\begin{align}
\frac{d\Ic}{dt}(t)\sim \text{const.}
\end{align}
In the context of epidemics, the linear growth phase can be associated to an endemic phase of the disease, when the virus keeps diffusing within the population without an exponential growth in the number of new infected (this  corresponds to a situation with reproduction number $R_0 = 1$, which keeps the number of infectious cases constant). A connection of this regime with compartmental models of the SIR type has been presented in Section~\ref{Sect:AnSolLinear}.

\item In the CeRG, the linear growth is only an intermediate phase, which preludes to a new exponential increase in the number of infections. The duration depends on $|\delta|$, and can can be estimated as \cite{cacciapaglia2020evidence}
\begin{align}
\Delta t_{\rm  endemic} = - 2 \int_A^\infty  \frac{d \Ic}{\beta (\Ic)}\ .
\end{align}
This time is plotted for different values of $p$ as a function of $\delta$ in the right panel of \figref{Fig:Strollingtime}.

\item For $|\delta|\geq \delta_{\text{max}}$ the beta-function no longer has a local maximum and $\Ic$ keeps growing exponentially, without a linear growing phase.

\end{itemize}
In Fig.~\ref{Fig:ComplexFlow} we represent the dynamics encoded in Eq.\eqref{eq:beta1} as a flow in the complex space of $I_c$. We clearly see that the system starts from the unstable fixed point at $I_c = 0$, and moves towards the approximate one at $\text{Re} I_c \approx A$, where the evolution slows down. This is represented by the closeness of the data-points, which are calculated at equal intervals of time. We also show flows in the complex plane, which are unrealistic as $I_c$ remains a real number when describing a pandemic. Anyhow, all the solutions feature a slowing down of the infection growth near the complex fixed points, which reproduced the endemic phase of linear growth.

The endemic linear-growing phase, therefore, is the prelude of a new wave of the epidemic diffusion. The CeRG approach can describe this endemic phase and the beginning of the next wave, however the number of infections would continue to grow indefinitely. In the following section we will further extend the approach to take into account the multi-wave pattern.

\subsection{Modelling multi-wave patterns}

Pandemics like the 1918 Spanish flu \cite{1918influenza} and COVID-19 have shown the appearance of multiple consecutive waves of exponential increase in the number of infections. In the case of COVID-19, the data support the fact that an endemic linearly-growing phase is always present  in between two consecutive waves \cite{cacciapaglia2020evidence}.
The CeRG model can be extended to take into account this structure, in a way that reproduces nicely the current data \cite{cacciapaglia2020multiwave}.

The multi-wave beta function, for an epidemic with $w$ consecutive waves, can be written as:
\begin{align}
- \beta_{\rm multi-waves} (\Ic)  = \rgg  \Ic \; \prod_{\rho=1}^w \left[\left( 1-\zeta_\rho\, \frac{\Ic}{A} \right)^2 - \delta_\rho \right]^{p_\rho}\,,
 \label{eq:beta2wave}
\end{align}
with $\zeta_\rho \leq 1$, $|\delta_\rho|\ll 1$ and $p_\rho >0$ for $\rho\in\{1,\ldots,w\}$. The normalisation $A$ can be fixed to match the first wave, so that 
\begin{equation}
0 < \zeta_w < \dots < \zeta_2 < \zeta_1 = 1\,.
\end{equation}
Besides the repulsive fixed point at $\Ic^\ast = 0$, the equation has a series of complex fixed points ruled by the parameters $\delta_\rho$. Without loss of generality, we can fix $\delta_w = 0$ so that the disease is extinguished after the last wave, and the total number of infections during the whole epidemic is given by $\lim_{t\to\infty} \Ic(t)=A/\zeta_w$. This description, however, only works for $\alpha (t) \propto \Ic (t)$, for which the value of the various fixed points are well separated \cite{cacciapaglia2020multiwave}, but not for $\alpha (t) \propto \ln \Ic (t)$. 

\begin{figure}
\begin{center}
\includegraphics[width=7.5cm]{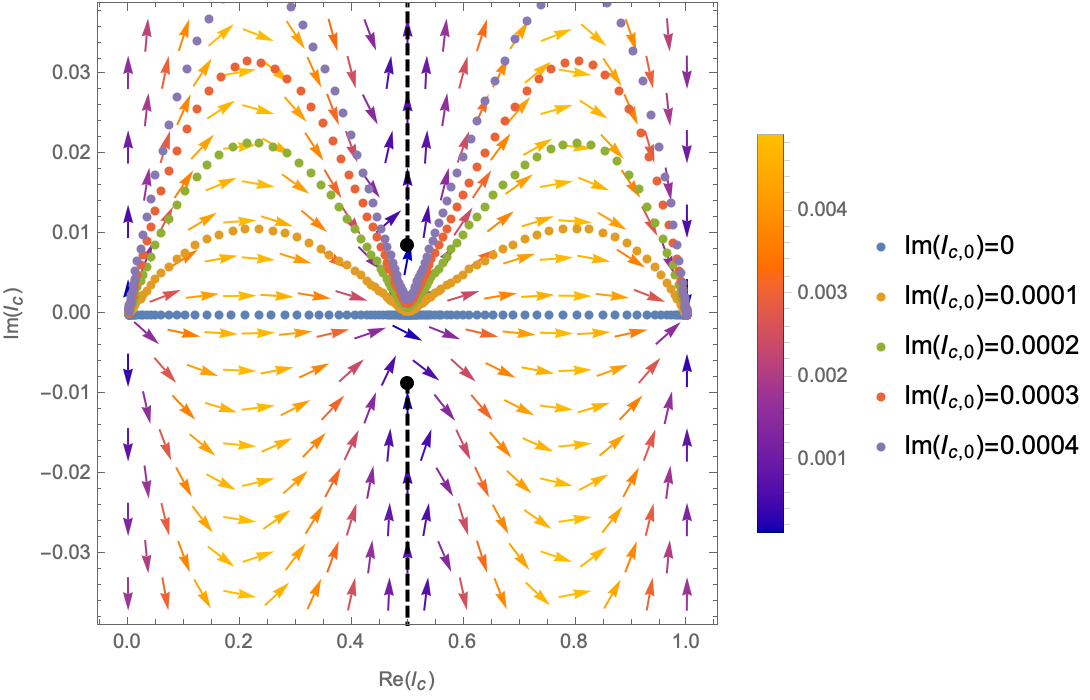} \hspace{1cm}\includegraphics[width=7.5cm]{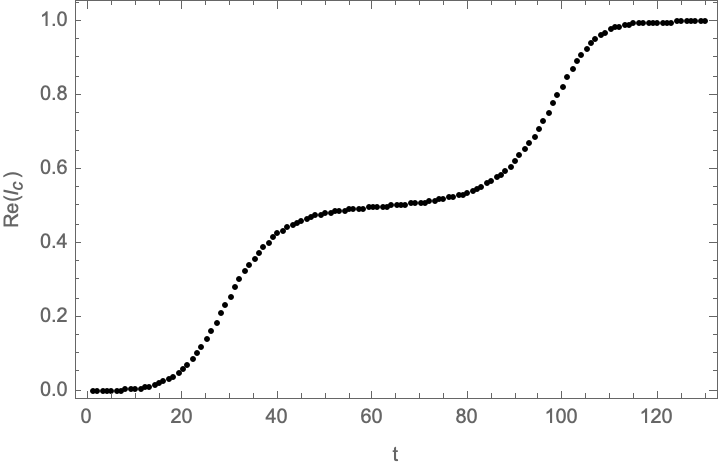}
\end{center}
\caption{Left panel: Schematic flow diagram representing the $\beta$-function (\ref{eq:beta2wave}) and trajectories for different initial conditions with $A=1/2$, $\zeta_1=1$, $\zeta_2=1/2$, $\lambda=0.05$, $\delta_1=-0.003$, $p_1=1/2$ and $p_2=1$. The trajectories of the flow start from the initial conditions $I_c(t=0)=I_{c,0}$, with $\text{Im}(I_{c,0})$ specified in the figure. Right panel: Cumulative number of infected $I_c$ for the trajectory with $\text{Im}(I_{c,0})=0$, \emph{i.e.} the light blue dots in the left panel.}
\label{Fig:MultiFlow}
\end{figure} 

In Fig.~\ref{Fig:MultiFlow} we show the flow in the complex plane for Eq.\eqref{eq:beta2wave} with two waves ($\omega = 2$). After leaving the unstable fixed point at $I_c = 0$, the system slows down near the  complex fixed points, hence generating the linear endemic phase like in the CeRG approach, before entering a second wave. The latter ends at the final attractive fixed point. In the right panel we show the time evolution of $I_c (t)$ for this fictitious example, clearly showing two exponential episodes. As for the CeRG, the time delay between the two waves is controlled by the number of new cases in the endemic phase, i.e. by the parameter $\delta$ in Eq.\eqref{eq:beta2wave}. Hence, this model highlights the importance of imposing some measures to limit the circulation of the virus after the end of an epidemic wave in order to tame and control the emergence of the next one. Note, finally, that this formalism can also be used for studying the diffusion in between different regions, by adding a coupling  term like the second term in Eq.\eqref{eq:deltaI}.


\section{COVID-19}\label{Sect:CovidSpecific}
The approaches that we have discussed in the previous sections are applicable to a large variety of infectious diseases. The main differences are in certain key parameters, such as the method of transmission of the pathogen, the incubation time, the infection and removal (mortality) rate, \emph{etc.}. They influence the resulting time evolution of the epidemic and lead, for example, to a different total duration of the epidemic, total number of infected and fatalities, \emph{etc}.  In this section, as a study case, we present data for the cumulative number of infected individuals in various countries during the COVID-19 pandemic, which started at the end of 2019 and is still ravaging the world. Since large scale testing is at the heart of many countries strategies to combat this pandemic, there is a large amount of publicly available data documenting the spread of the SARS-CoV-2 across the globe. Here we use data from public repositories \cite{Worldometer} for the time period of 15/02/2020 until 17/08/2021.

\begin{figure}[hp]
\begin{align}
&\includegraphics[width=0.46\textwidth]{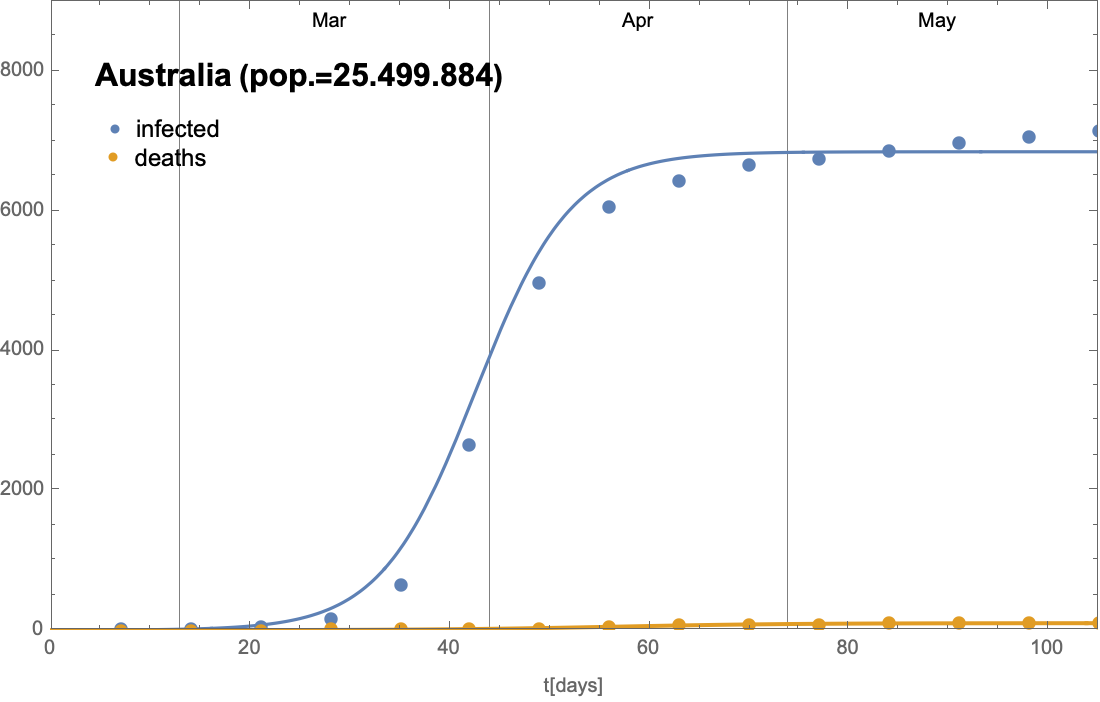} && \includegraphics[width=0.46\textwidth]{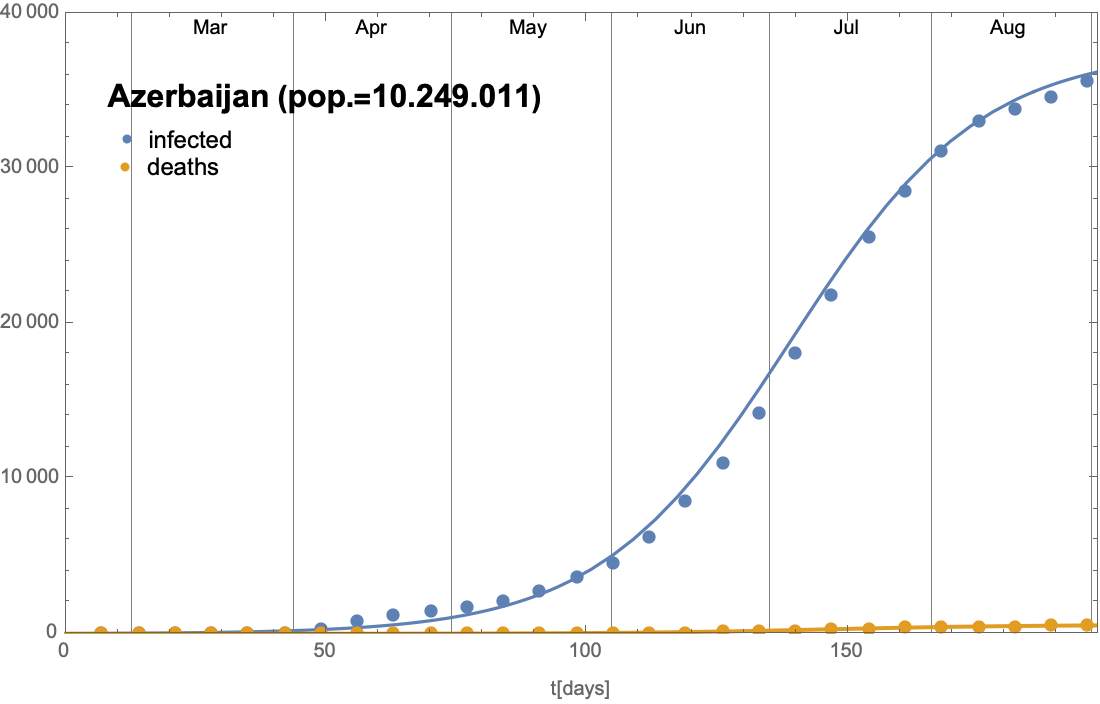}\nonumber\\
&\includegraphics[width=0.46\textwidth]{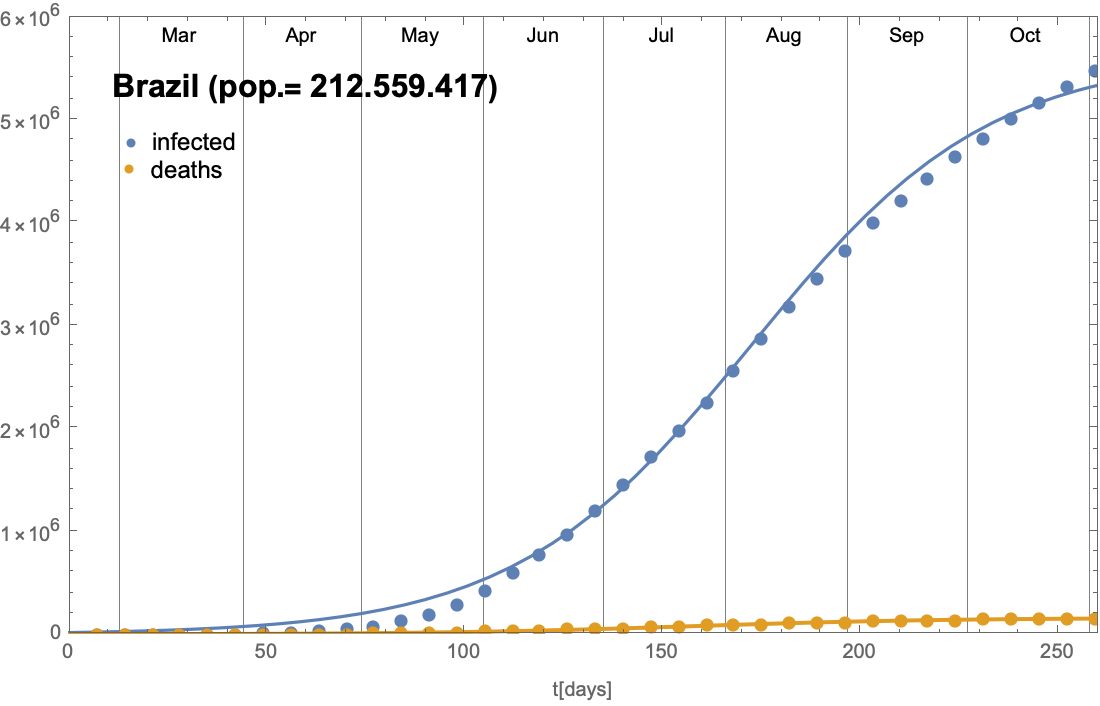} && \includegraphics[width=0.46\textwidth]{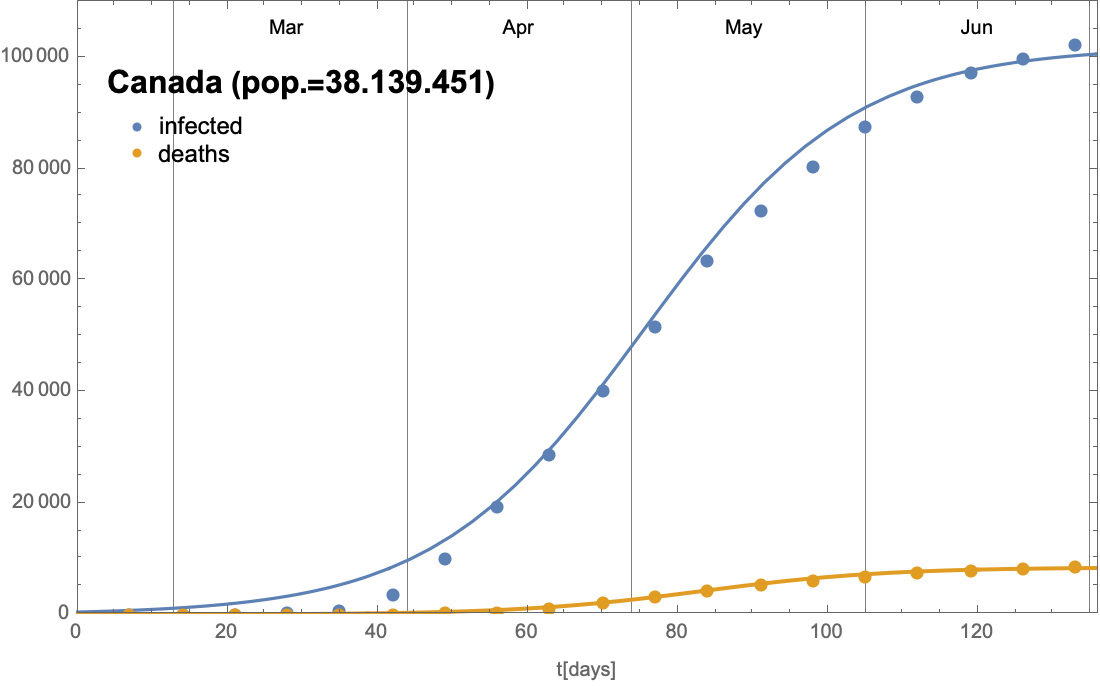}\nonumber\\
&\includegraphics[width=0.46\textwidth]{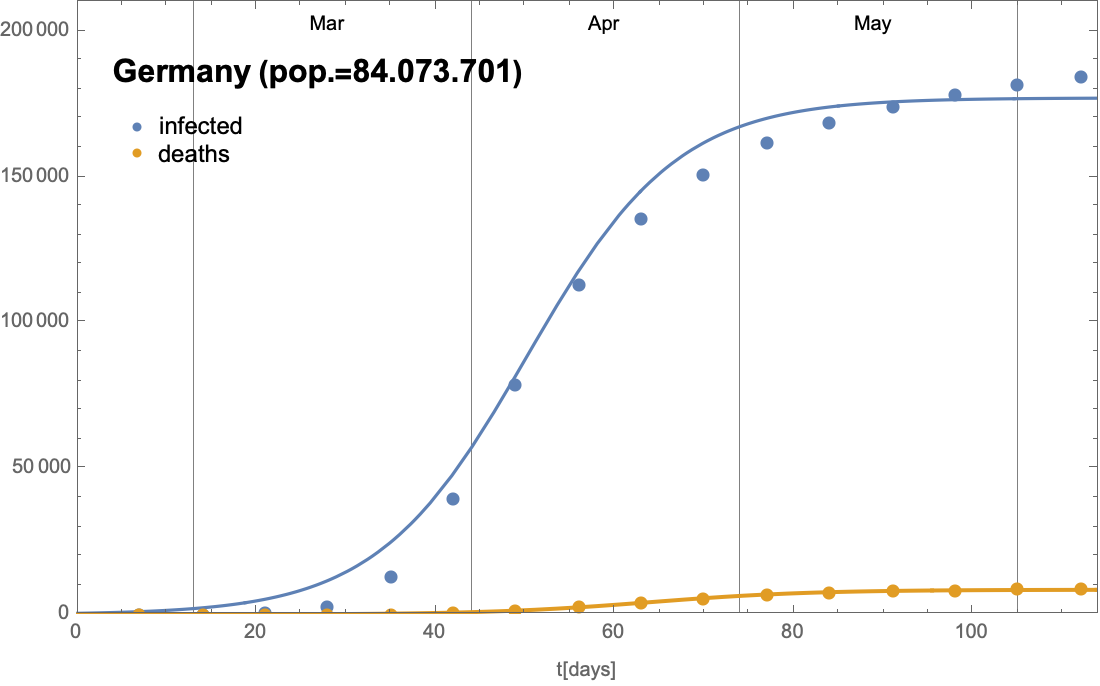} && \includegraphics[width=0.46\textwidth]{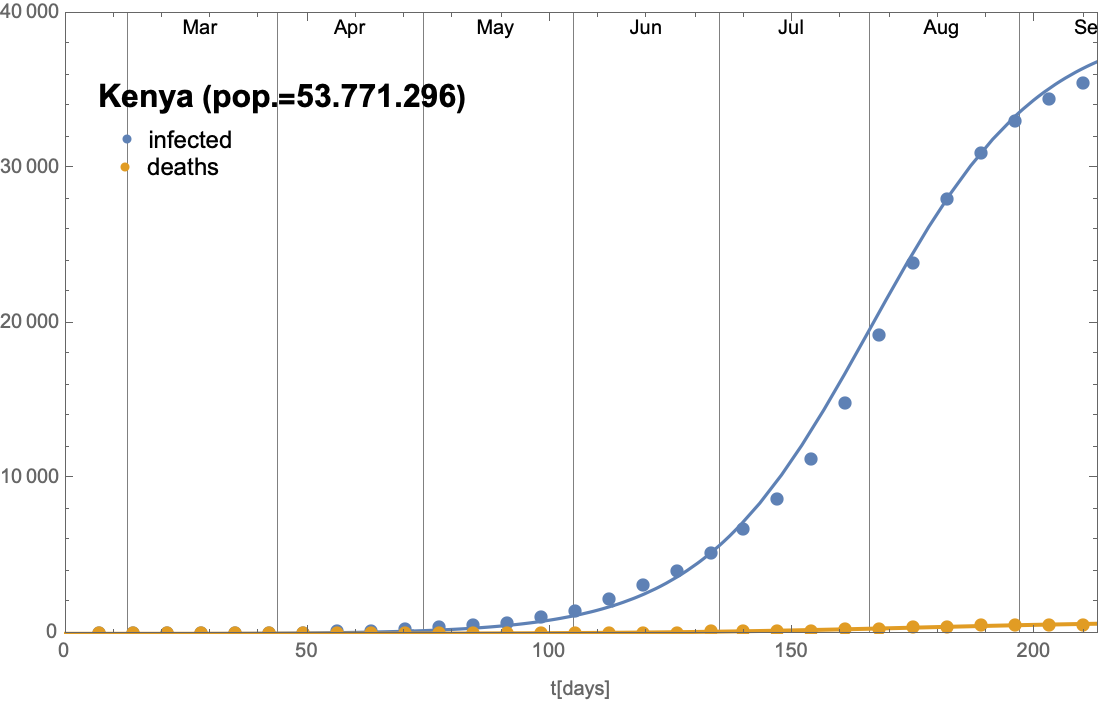}\nonumber\\
&\includegraphics[width=0.46\textwidth]{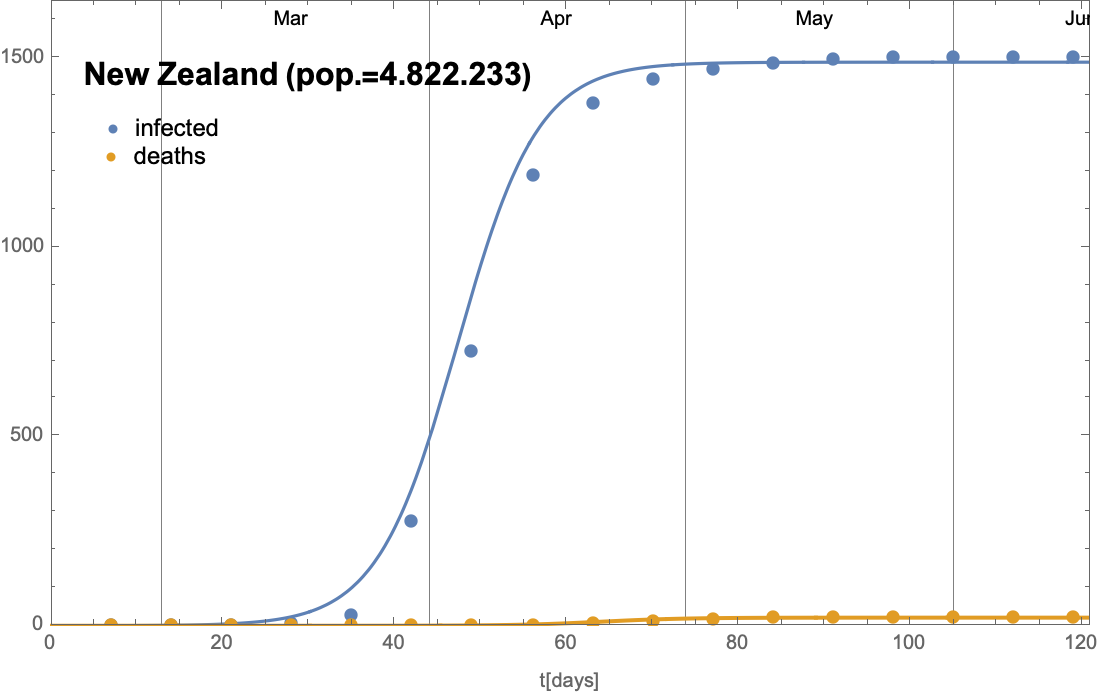} && \includegraphics[width=0.46\textwidth]{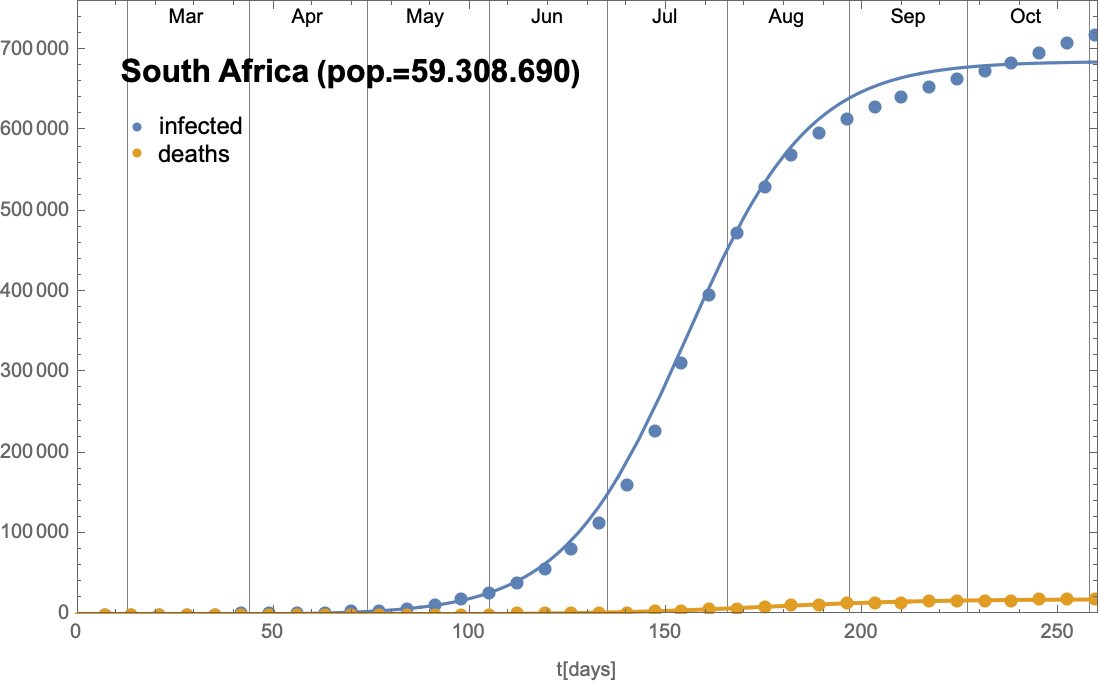}\nonumber
\end{align}
\caption{Cumulative number of individuals infected with SARS-Cov-2 and cumulative number of deaths during the first wave in countries across all continents. The dots represent the data reported at \cite{Worldometer} (averaged over a week) and the coloured lines fits with logistic functions of the form (\ref{LogisticFunctionGen}).}
\label{Fig:DataWave1}
\end{figure}

We use these data to highlight peculiarities of the time evolution of the spread of the disease, namely the previously mentioned distinct multi-wave structure of repeated phases of exponential growth in the number of infected individuals interspersed with phases of (quasi-)linear growth: Figure~\ref{Fig:DataWave1} shows examples of the first of such waves in countries taken from all around the globe. The plots show the cumulative number of infected individuals as well as the cumulative number of deaths. These plots provide examples of the epidemiological dynamics under very different conditions not only with regards to geographical (\emph{e.g.} size of the population, population density, level of urbanisation), climatic, economical (\emph{e.g.} the gross national product of each country), socio-cultural and political factors (\emph{e.g.} the level of medical care the population has access to), but also different strategies the countries have deployed to combat the epidemic. While this has lead to a different dynamics in each country (with regards for example to the total number of cases or the duration of the wave), the infection numbers all follow a similar shape. Indeed, as the solid lines in Fig.~\ref{Fig:DataWave1} shows, in each case the data can be fitted with a a logistic function of the form of Eq.\eqref{LogisticFunctionGen} and the differences only lie with the different numerical parameters, as reported in Table~\ref{tab:paramfit}.
\begin{table}[tb!]
\begin{center}
\begin{tabular}{|c|c|c|c|}\hline
{\bf Country} & $A$ & $\lambda$ & $B$\\\hline\hline
&&&\\[-8pt]
Australia & $6854\pm 24$ & $0.2095\pm 0.0051$ & $7509\pm 1615$\\[2pt]\hline
&&&\\[-8pt]
Azerbaijan & $37703\pm 136$ & $0.0547\pm 0.0004$ & $2013\pm 110$\\[2pt]\hline
&&&\\[-8pt]
Brazil & $5624661\pm 26343$ & $0.0333\pm 0.0003$ & $314\pm 14$\\[2pt]\hline
&&&\\[-8pt]
Canada & $101987\pm 544$ & $0.0716\pm 0.0012$ & $222\pm 18$\\[2pt]\hline
&&&\\[-8pt]
Germany & $177112\pm 910$ & $0.1192\pm 0.003$ & $399\pm 58$\\[2pt]\hline
&&&\\[-8pt]
Kenya & $39469\pm 236$ & $0.0571\pm 0.0006$ & $13234\pm 1225$\\[2pt]\hline
&&&\\[-8pt]
New Zealand & $1491\pm 3$ & $0.2133\pm 0.0032$ & $24095\pm 3636$\\[2pt]\hline
&&&\\[-8pt]
South Africa & $686350\pm 1586$ & $0.0637\pm 0.0007$ & $19449\pm 1910$\\[2pt]\hline
\end{tabular}
\end{center}
\caption{\label{tab:paramfit} Parameters of the logistic function in Eq.\eqref{LogisticFunctionGen} obtained by fitting the epidemiological curves (cumulative number of infected) shown in Fig.~\ref{Fig:DataWave1}.}
\end{table}
Furthermore, Fig.~\ref{Fig:DataWave1} also shows the cumulative number of deaths in each country, which can also be fitted with a logistic function of the same form. As discussed in Section~\ref{Sect:RGapproach}, the fact that epidemiological curves in very different regions of the world, under very different circumstances, can be fitted with a single class of functions is due to a self-similarity structure of the corresponding dynamics. Indeed, the underlying symmetry principle that organises the spread of the disease around (near) fixed points of flow equations is at the heart of the eRG approach.

\begin{figure}[hp]
\begin{align}
&\includegraphics[width=0.48\textwidth]{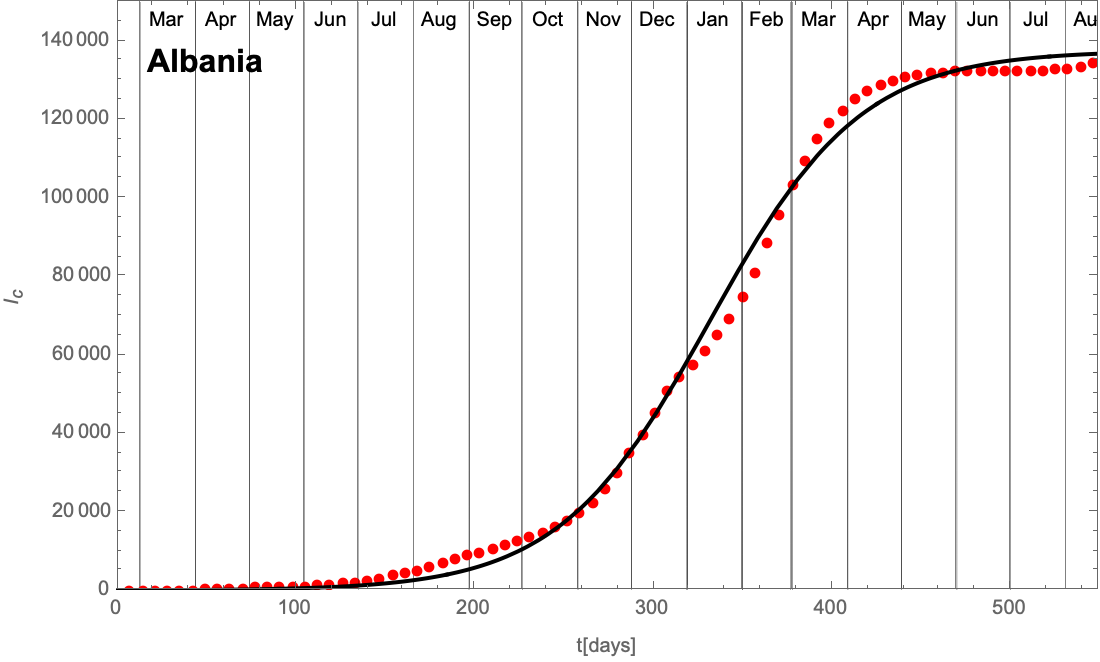} && \includegraphics[width=0.48\textwidth]{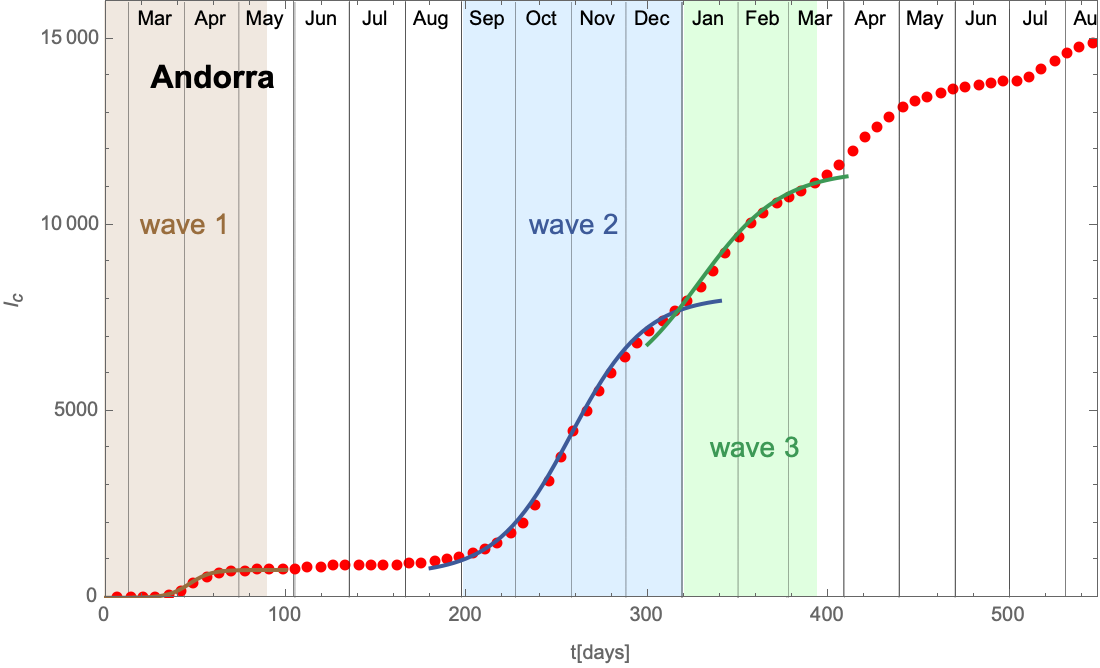}\nonumber\\
&\includegraphics[width=0.48\textwidth]{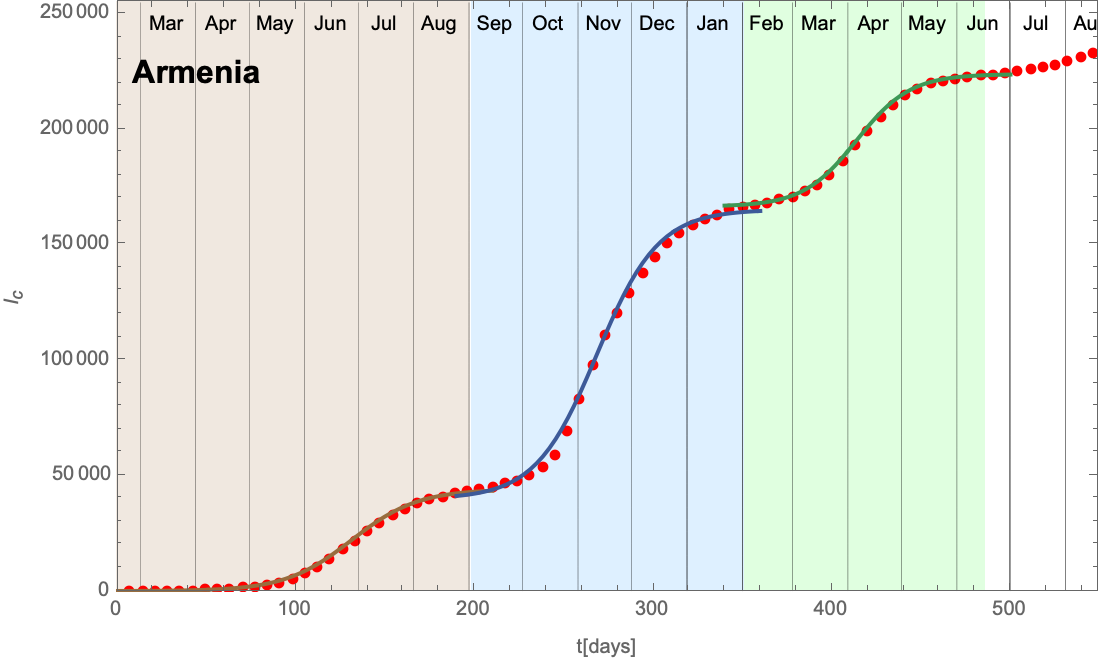} && \includegraphics[width=0.48\textwidth]{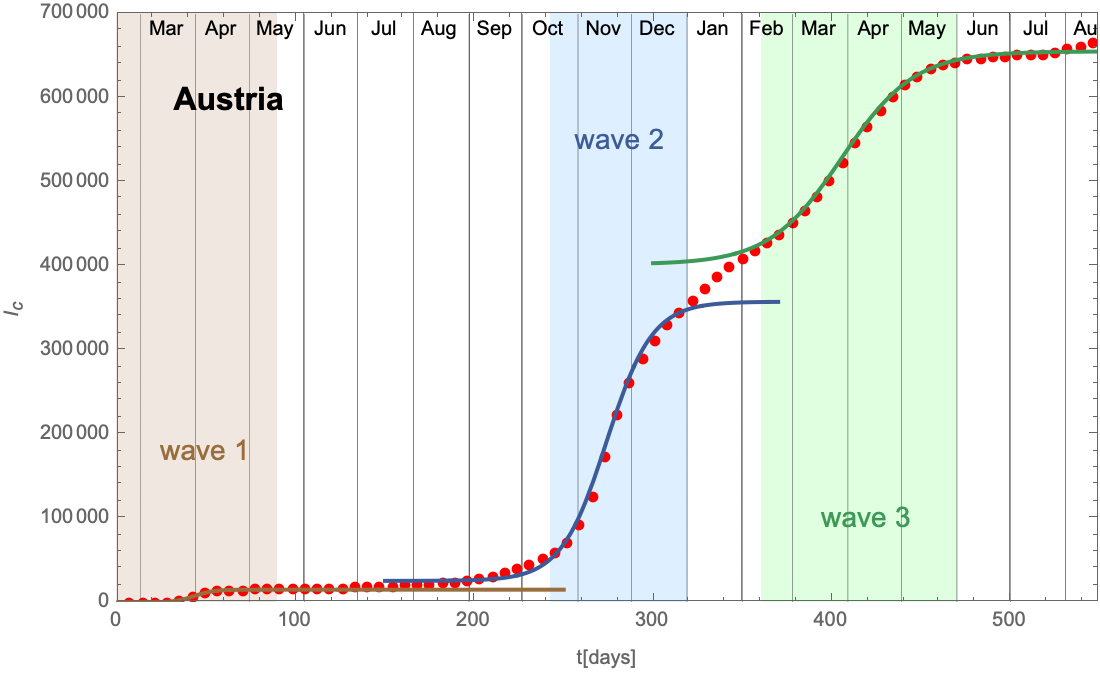}\nonumber\\
&\includegraphics[width=0.48\textwidth]{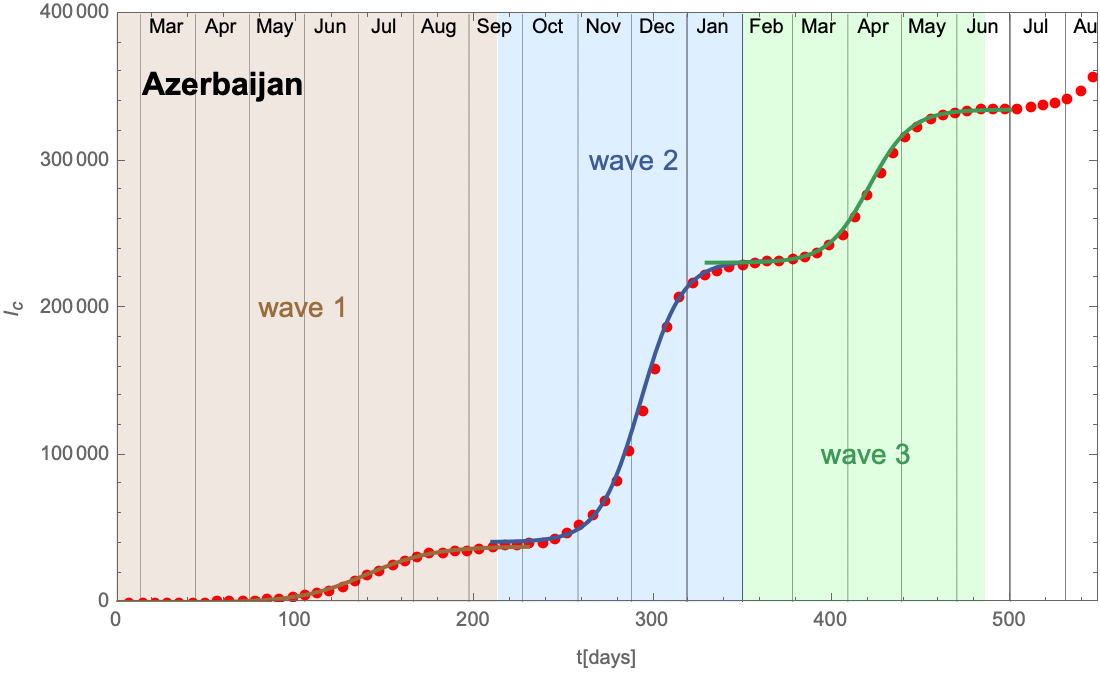} && \includegraphics[width=0.48\textwidth]{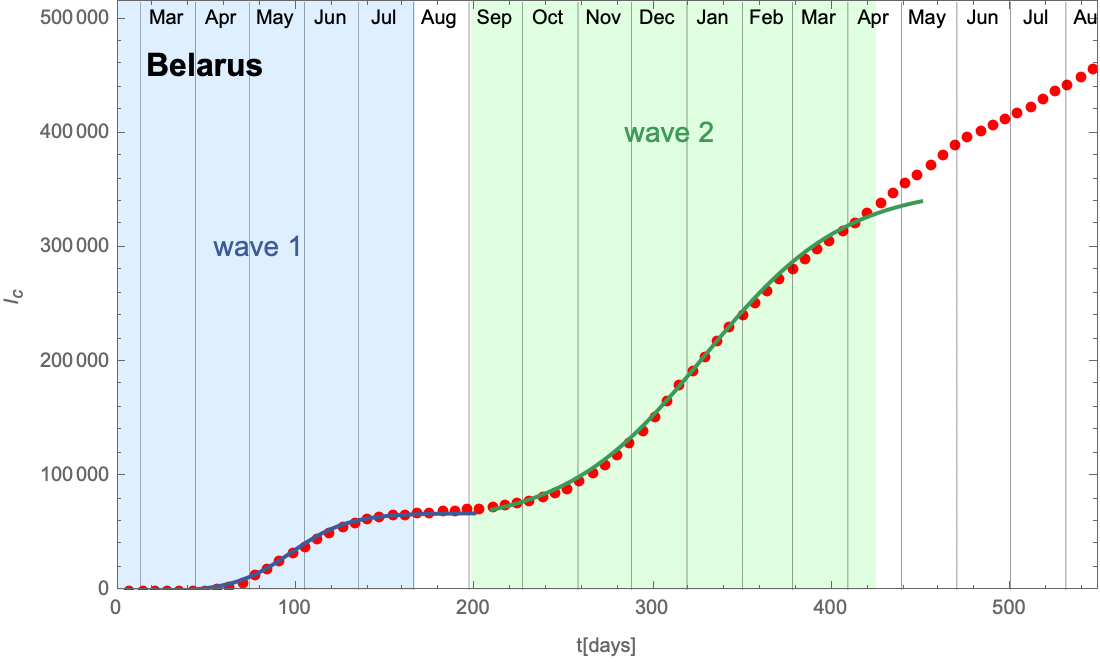}\nonumber\\
&\includegraphics[width=0.48\textwidth]{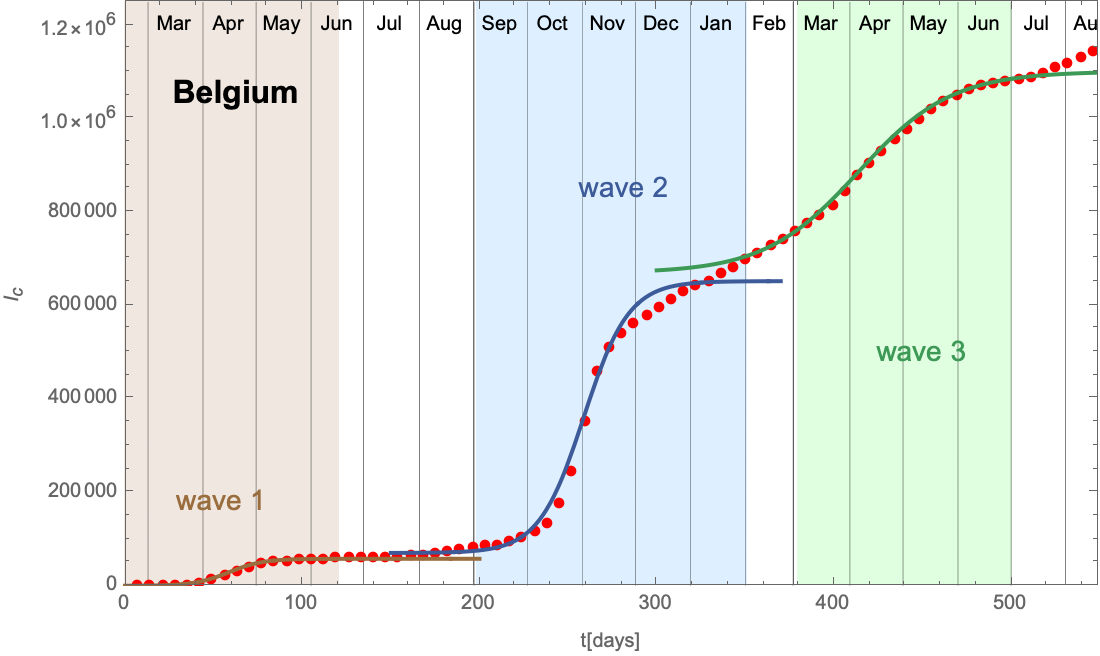} && \includegraphics[width=0.48\textwidth]{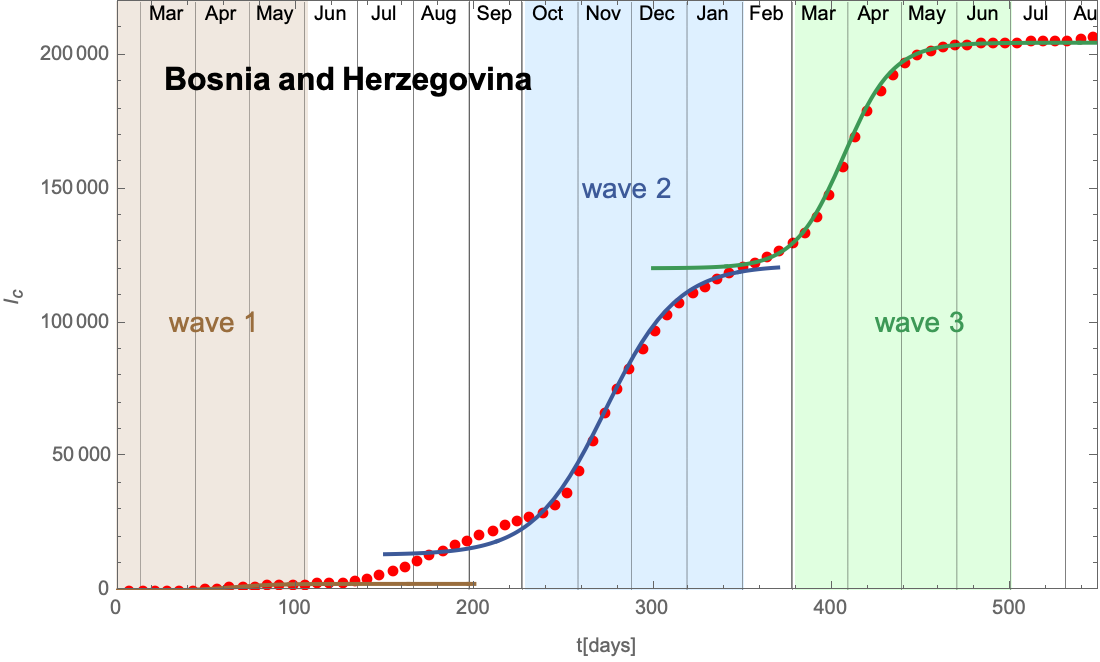}\nonumber
\end{align}
\caption{Cumulative number of individuals infected with SARS-CoV-2 from 15/02/2020 until 17/08/2021 in different countries of Europe. The red dots represent the data reported at \cite{Worldometer} and the coloured lines fits with logistic functions of the form (\ref{LogisticFunctionGen}). The coloured regions indicate the time frame over which the data were fitted for a single wave.}
\label{Fig:DataEurop1}
\end{figure}

\begin{figure}[hp]
\begin{align}
&\includegraphics[width=0.48\textwidth]{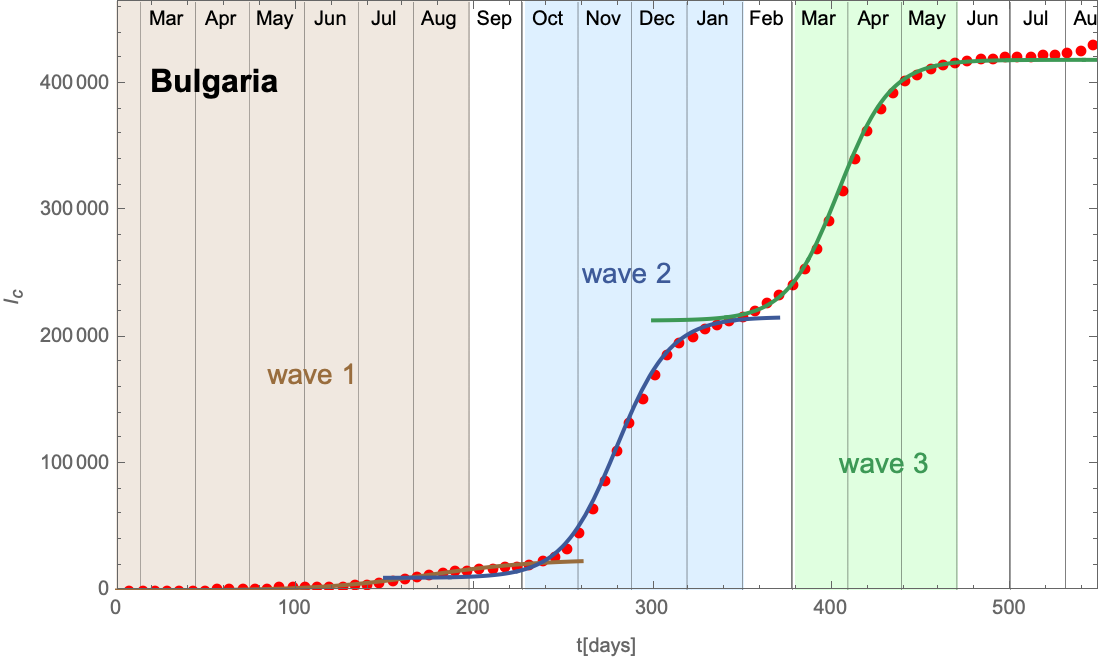} && \includegraphics[width=0.48\textwidth]{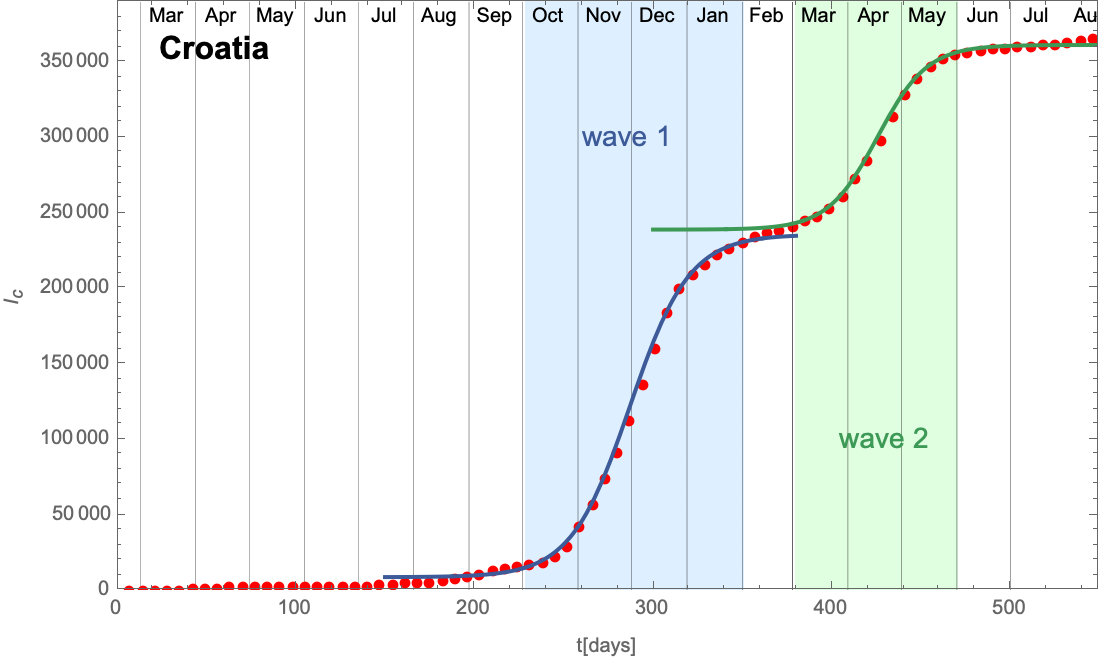}\nonumber\\
&\includegraphics[width=0.48\textwidth]{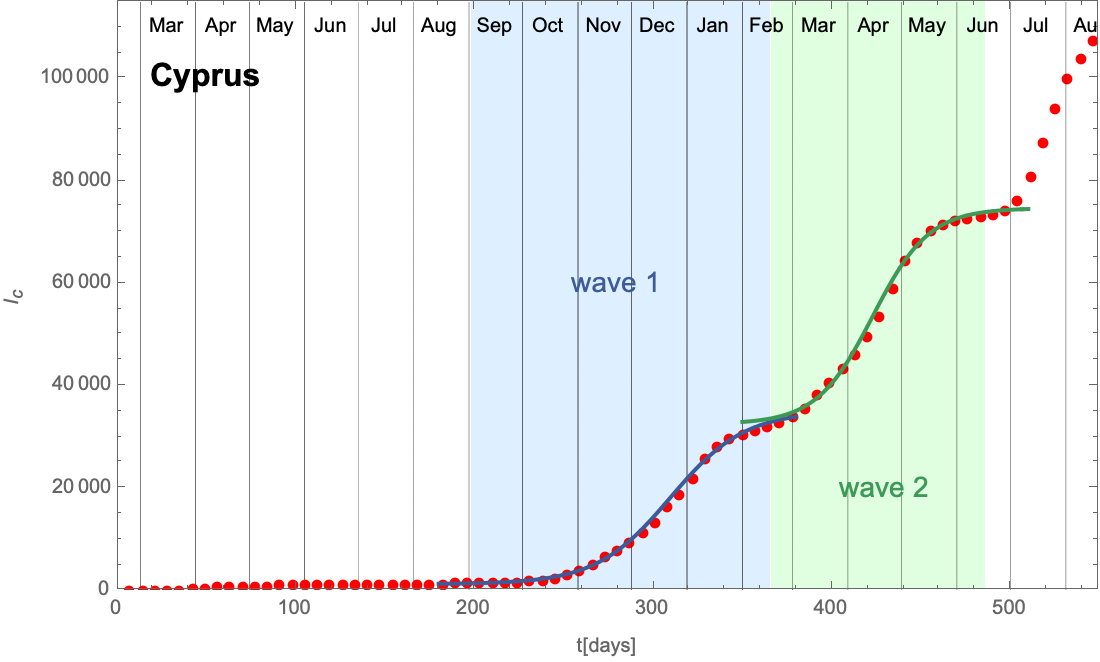} && \includegraphics[width=0.48\textwidth]{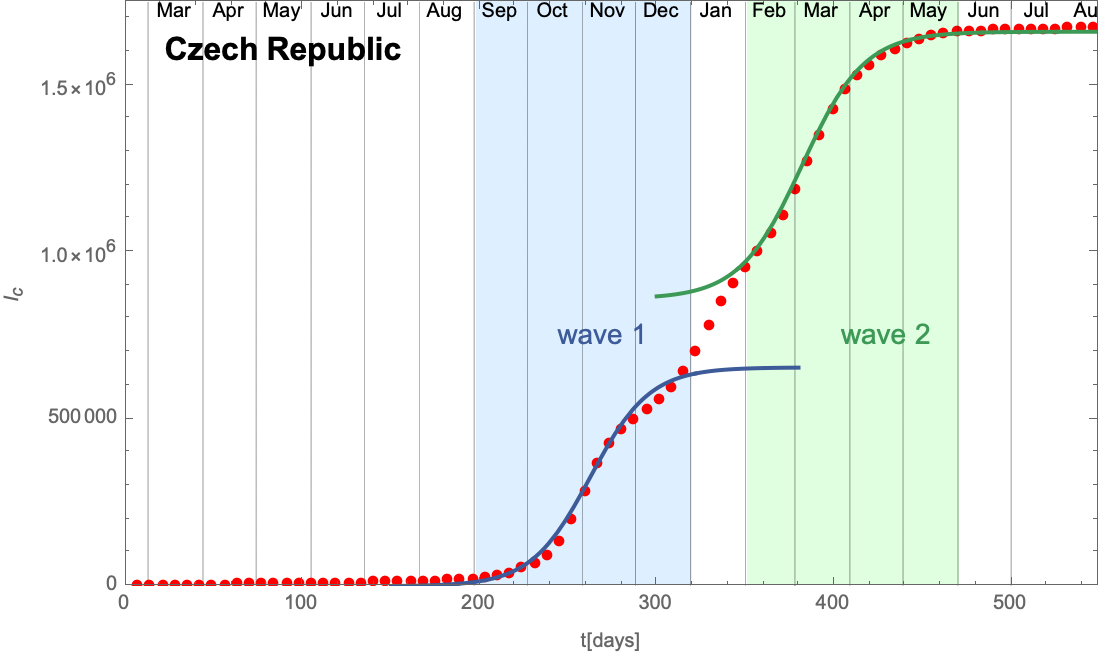}\nonumber\\
&\includegraphics[width=0.48\textwidth]{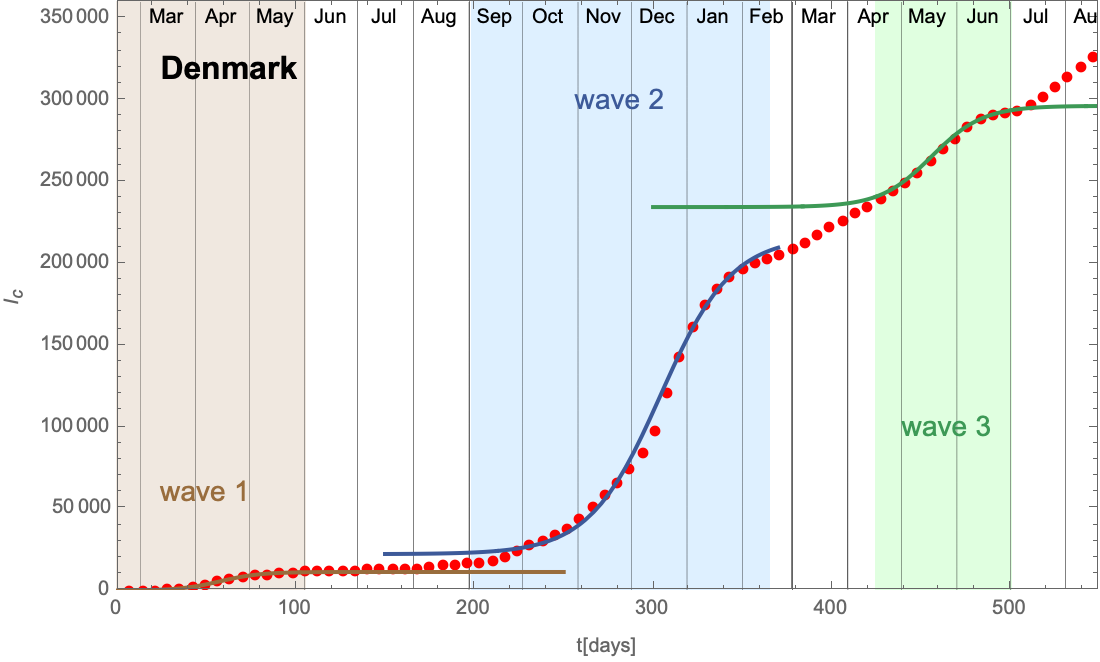} && \includegraphics[width=0.48\textwidth]{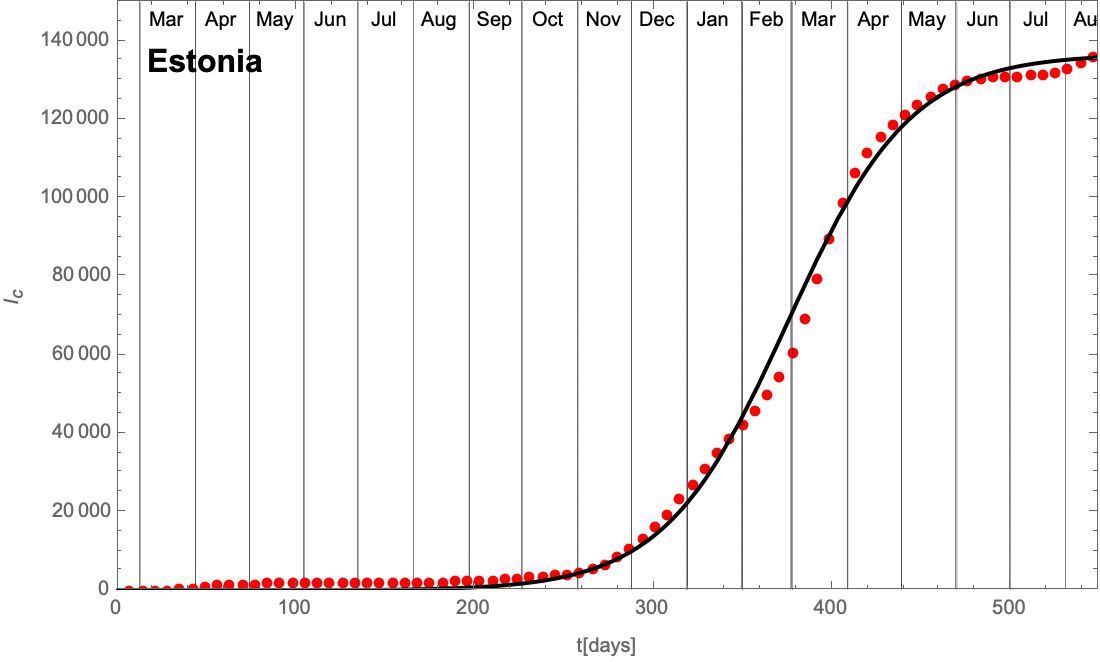}\nonumber\\
&\includegraphics[width=0.48\textwidth]{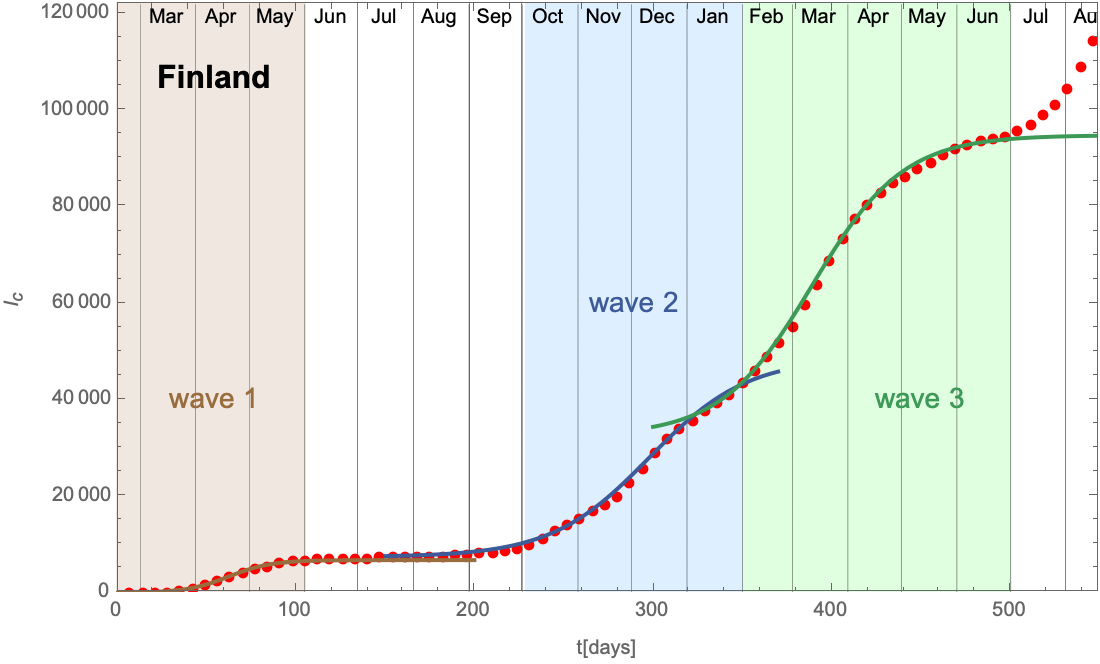} && \includegraphics[width=0.48\textwidth]{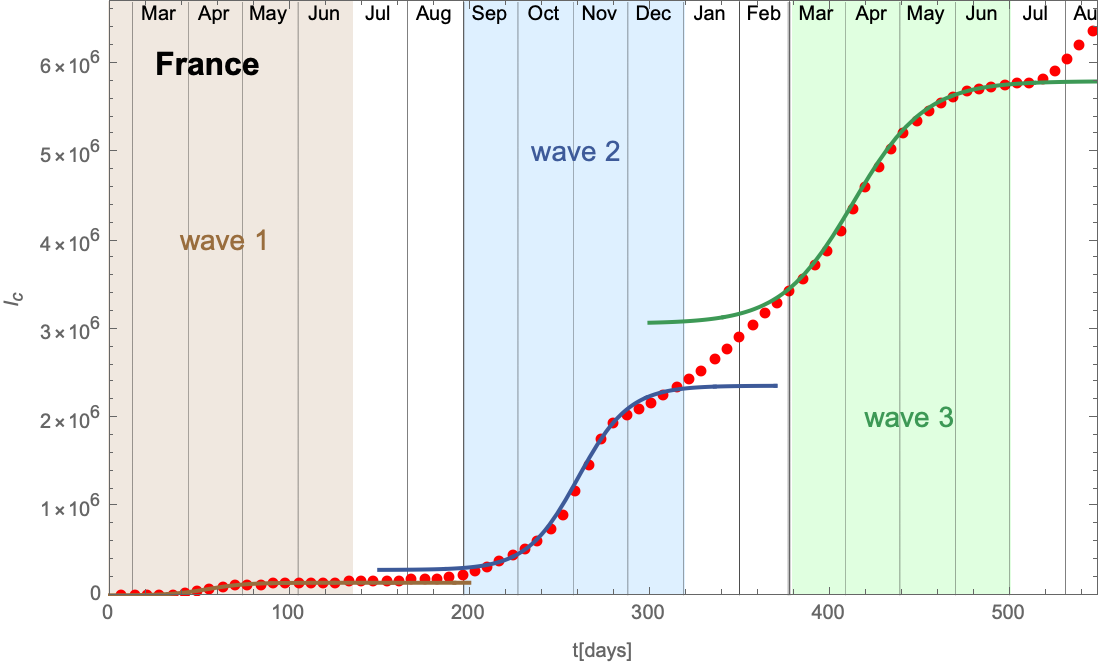}\nonumber
\end{align}
\caption{Cumulative number of individuals infected with SARS-CoV-2 (contd.)}
\label{Fig:DataEurop2}
\end{figure}

\begin{figure}[hp]
\begin{align}
&\includegraphics[width=0.48\textwidth]{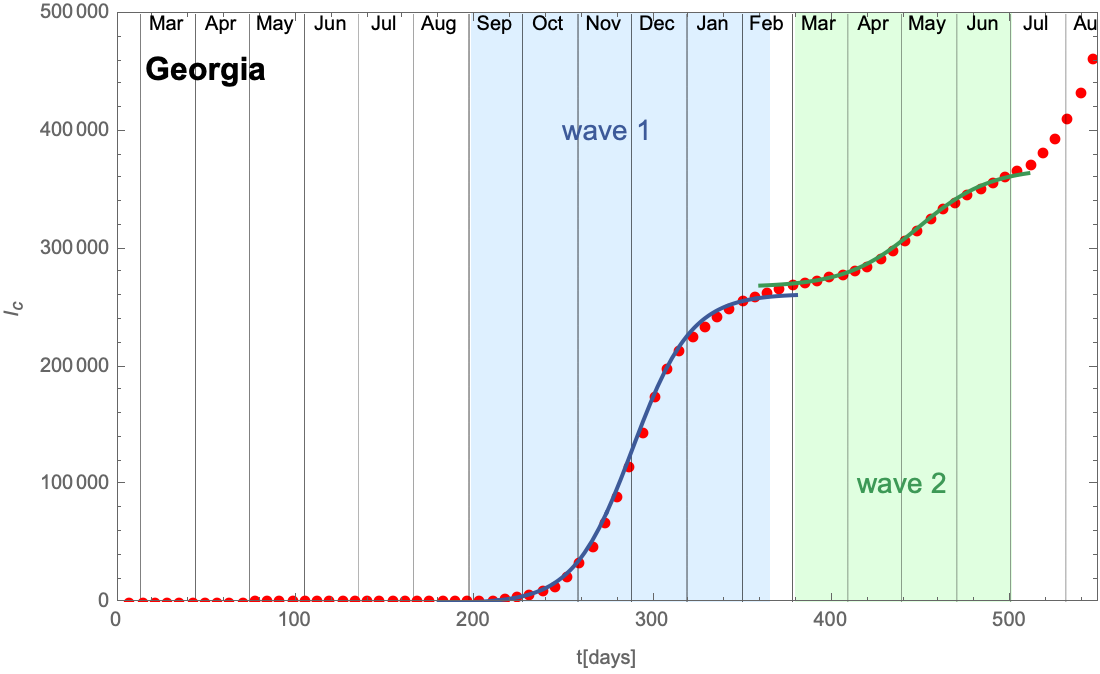} && \includegraphics[width=0.48\textwidth]{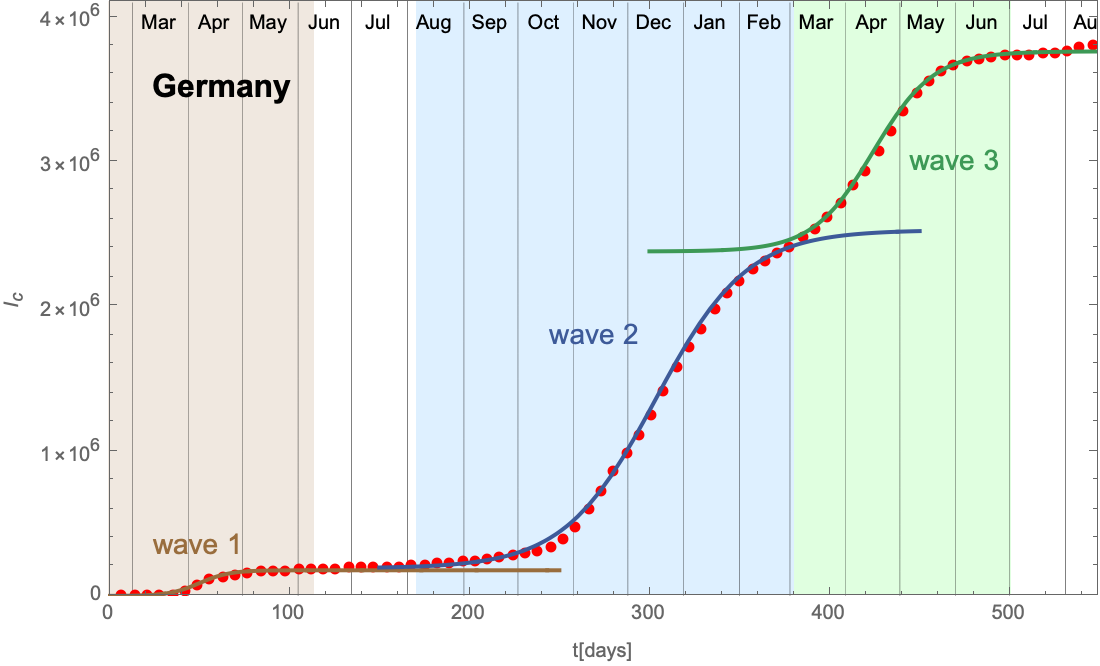}\nonumber\\
&\includegraphics[width=0.48\textwidth]{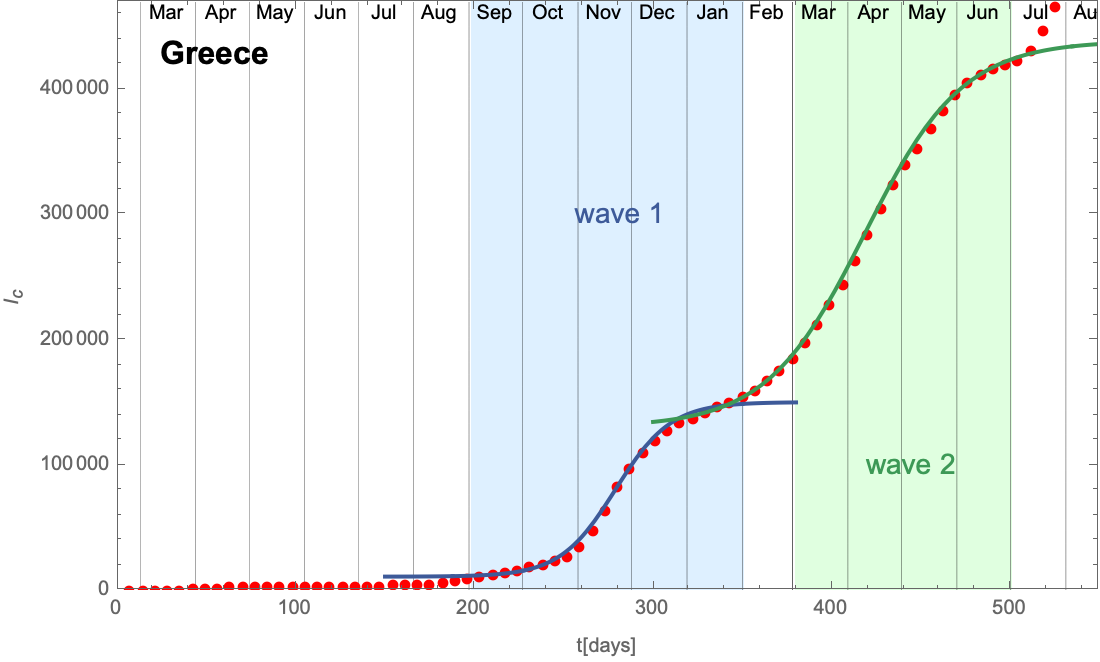} && \includegraphics[width=0.48\textwidth]{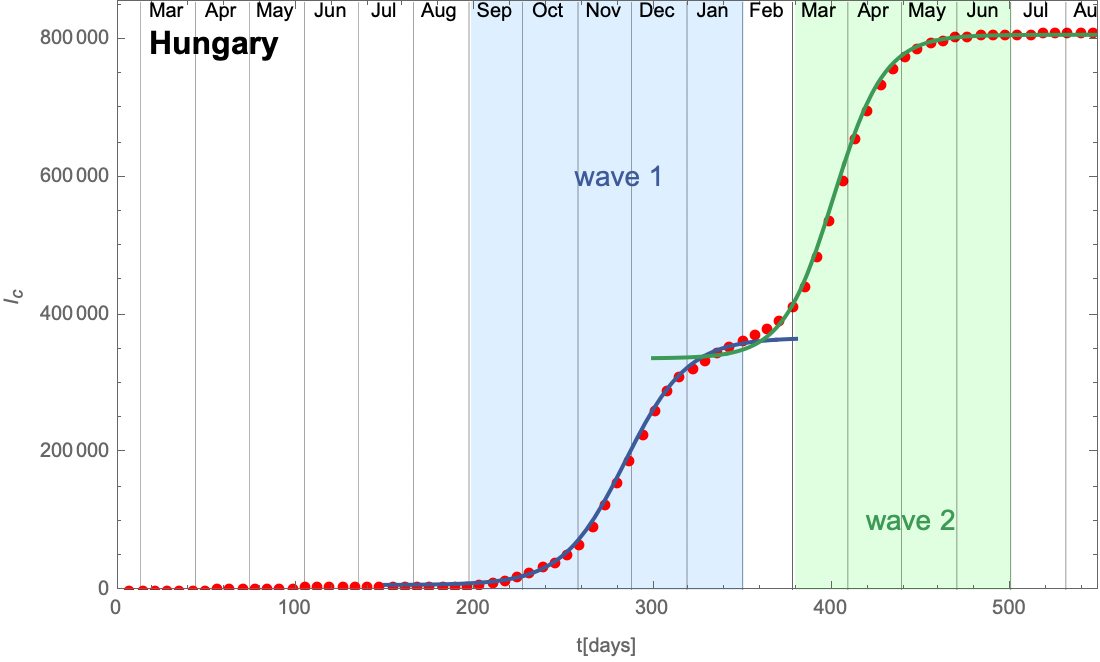}\nonumber\\
&\includegraphics[width=0.48\textwidth]{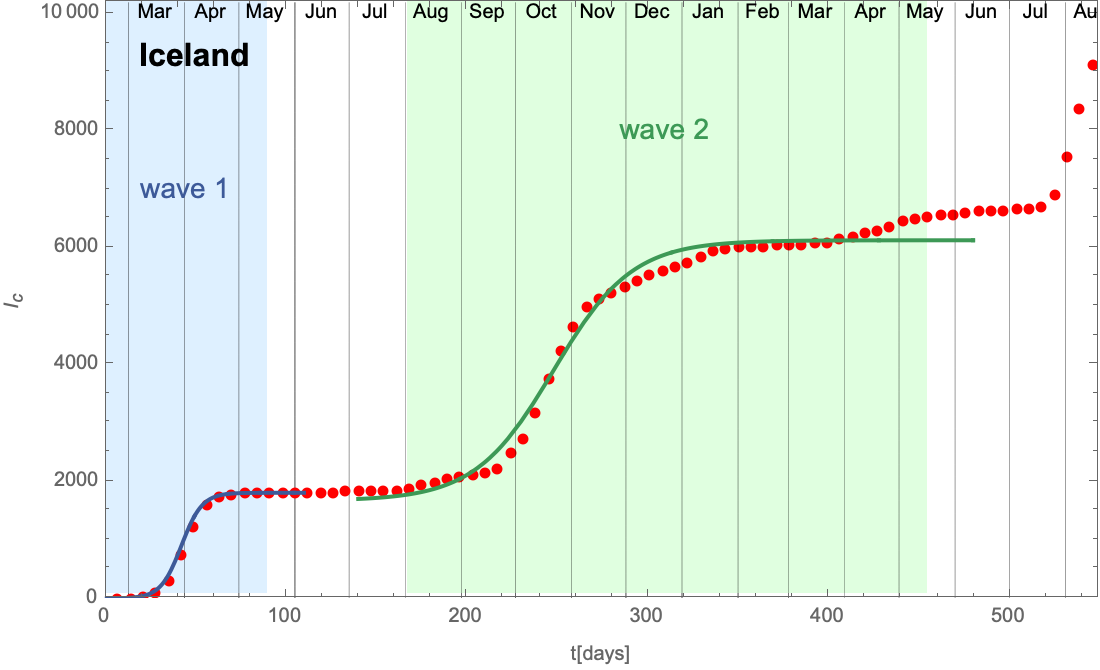} && \includegraphics[width=0.48\textwidth]{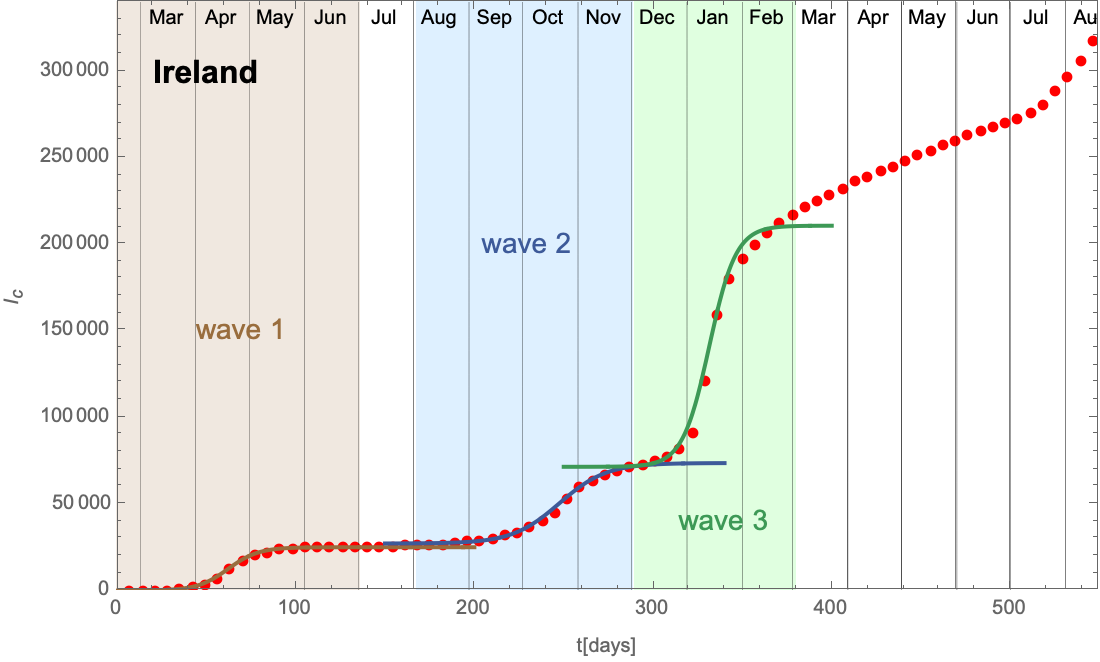}\nonumber\\
&\includegraphics[width=0.48\textwidth]{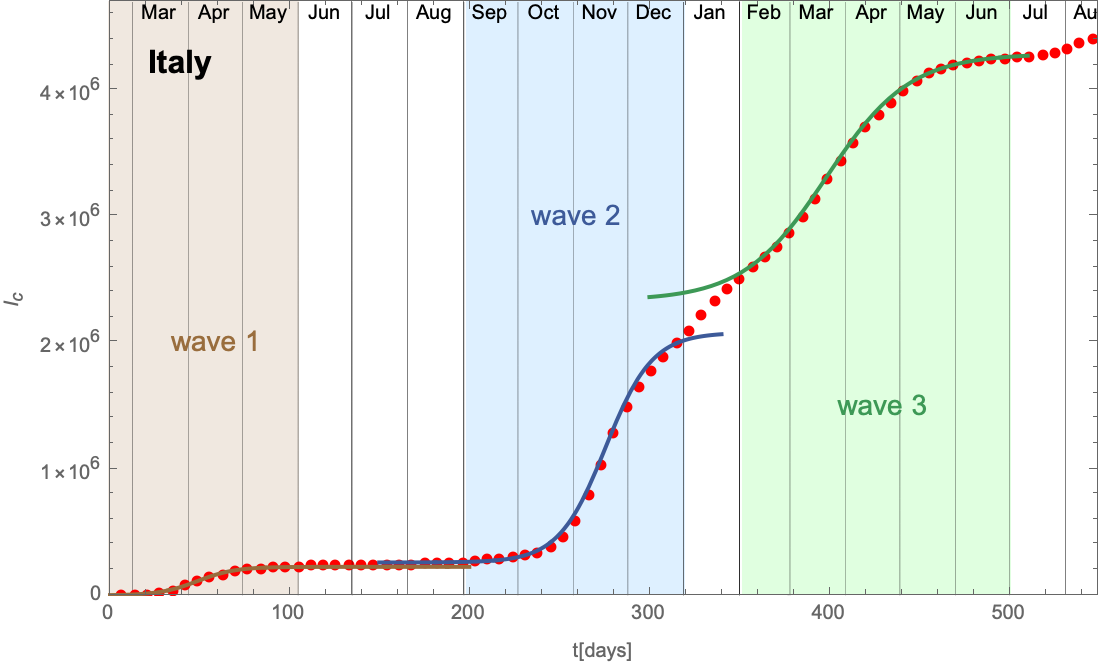} && \includegraphics[width=0.48\textwidth]{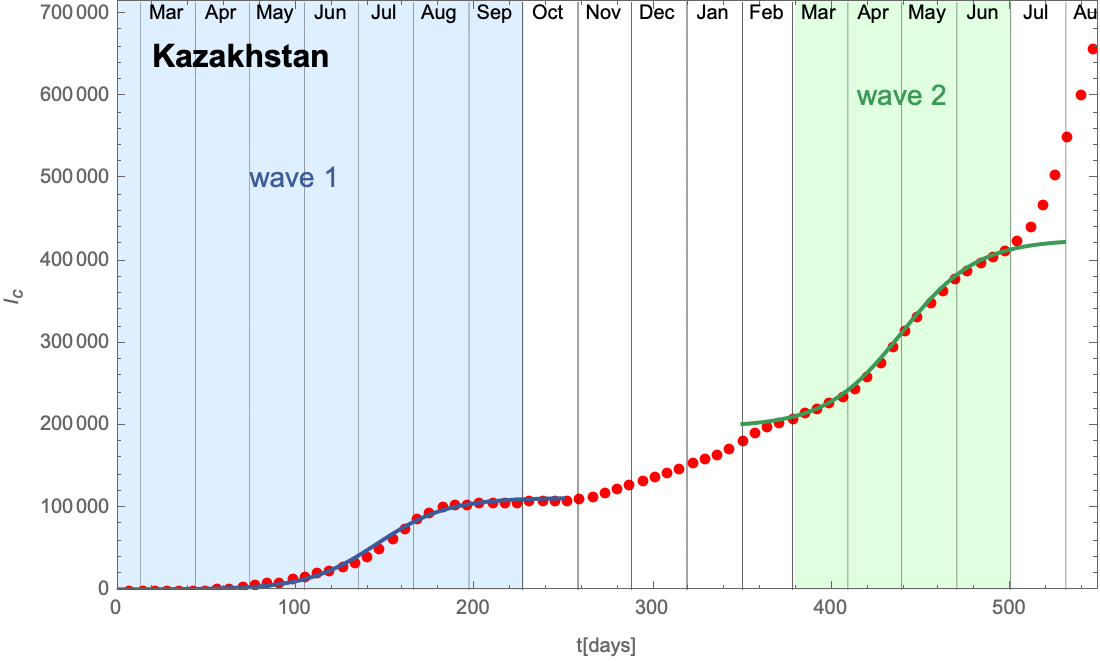}\nonumber
\end{align}
\caption{Cumulative number of individuals infected with SARS-CoV-2 (contd.)}
\label{Fig:DataEurop3}
\end{figure}

\begin{figure}[hp]
\begin{align}
&\includegraphics[width=0.48\textwidth]{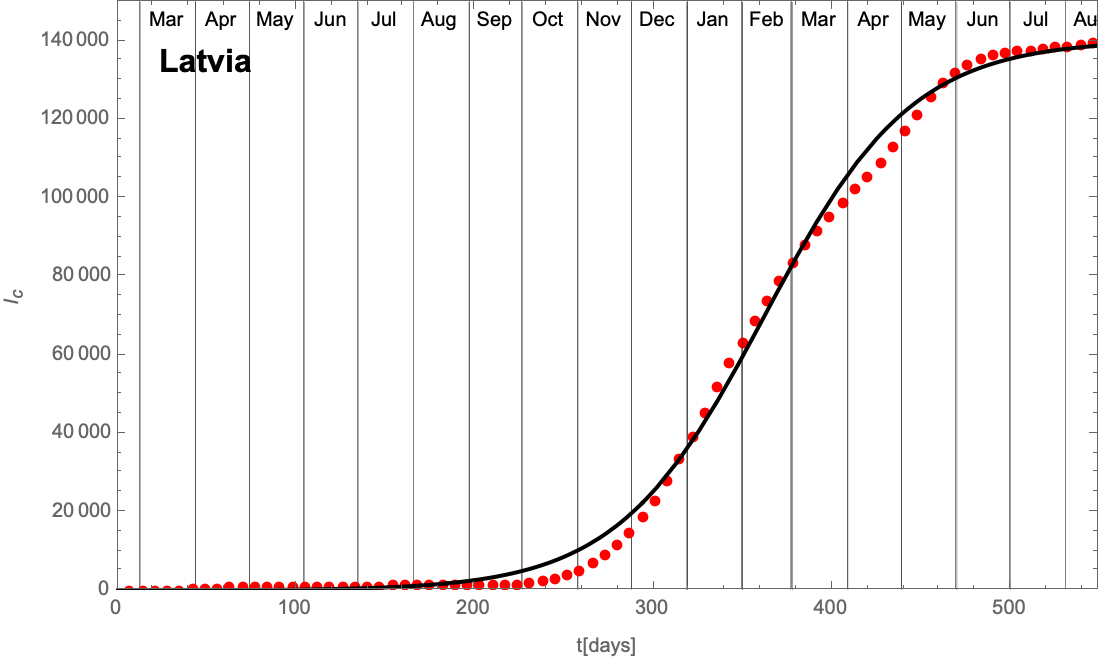} && \includegraphics[width=0.48\textwidth]{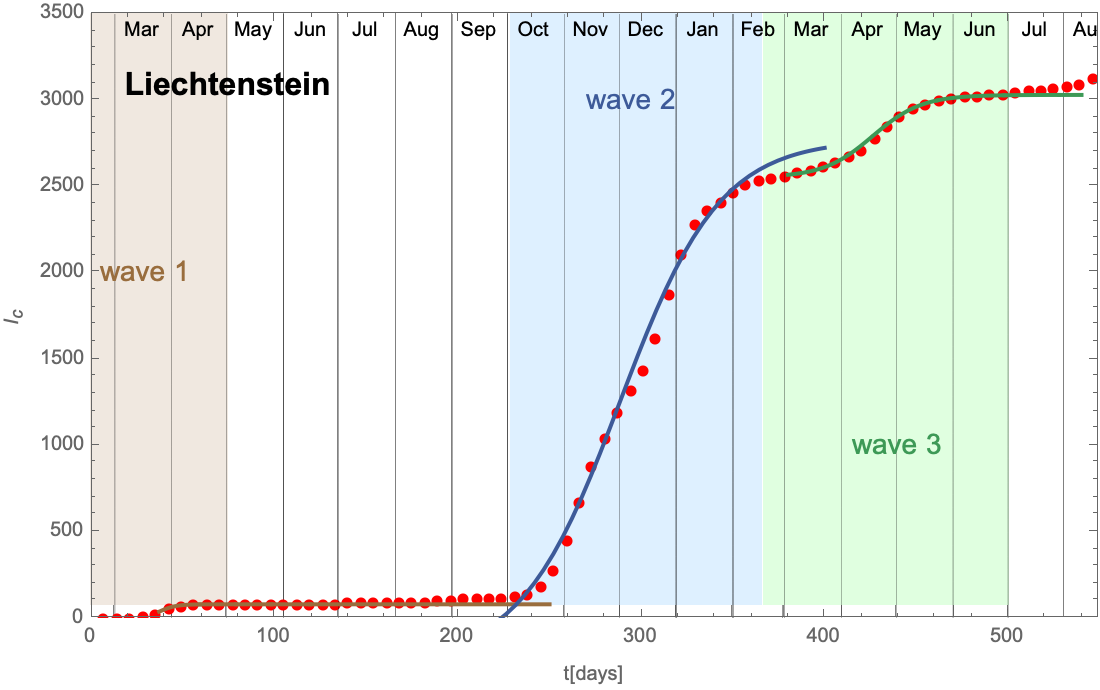}\nonumber\\
&\includegraphics[width=0.48\textwidth]{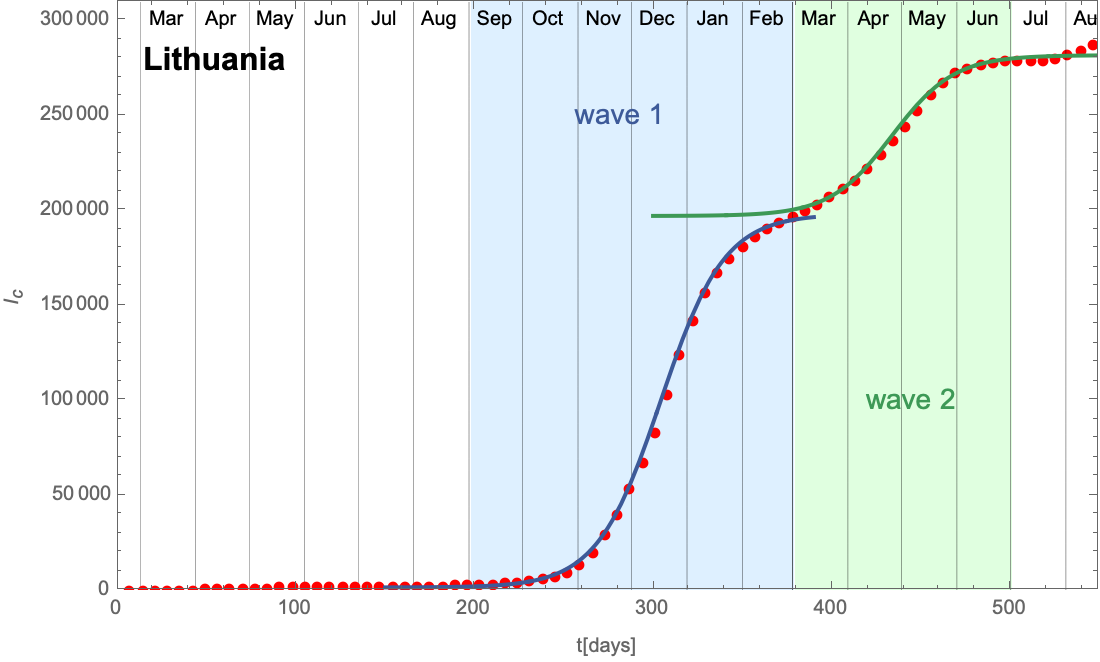} && \includegraphics[width=0.48\textwidth]{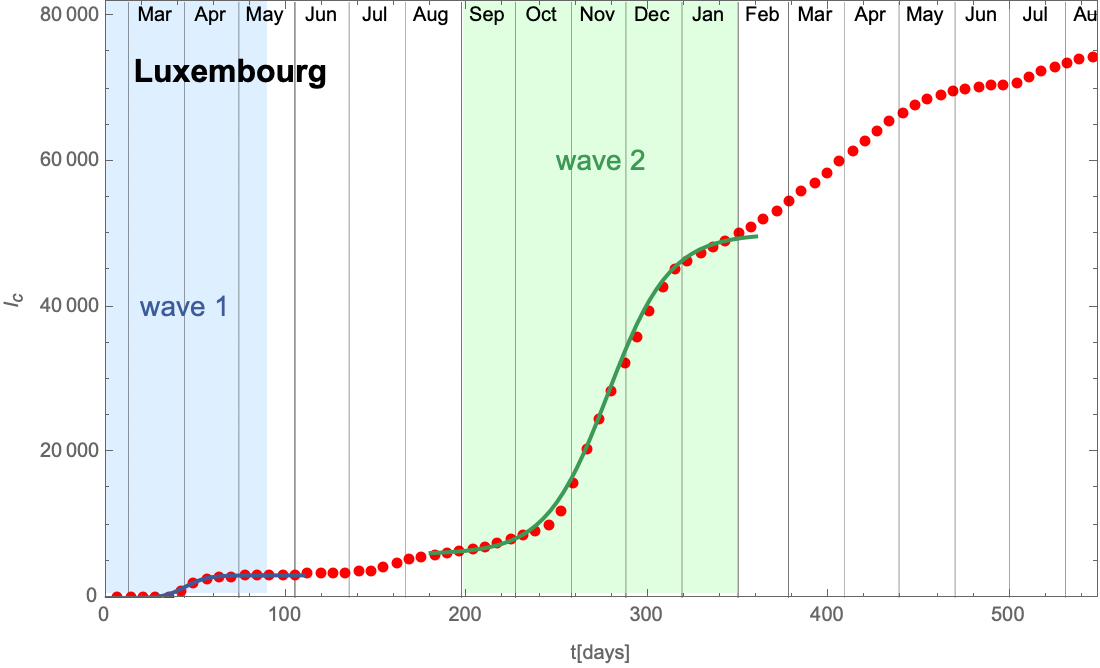}\nonumber\\
&\includegraphics[width=0.48\textwidth]{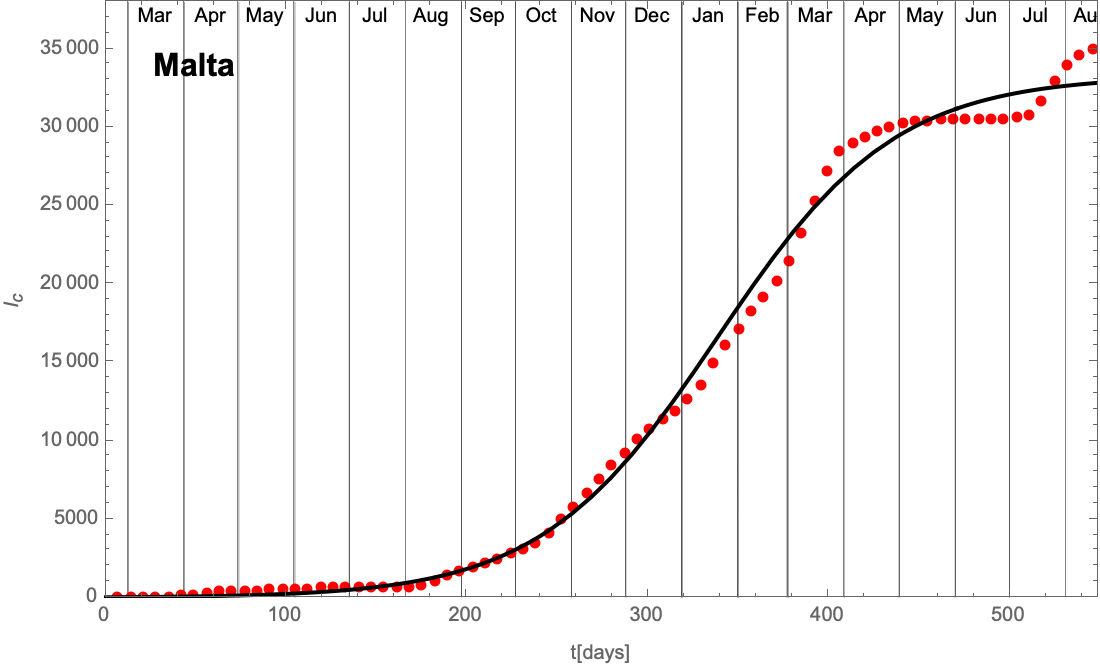} && \includegraphics[width=0.48\textwidth]{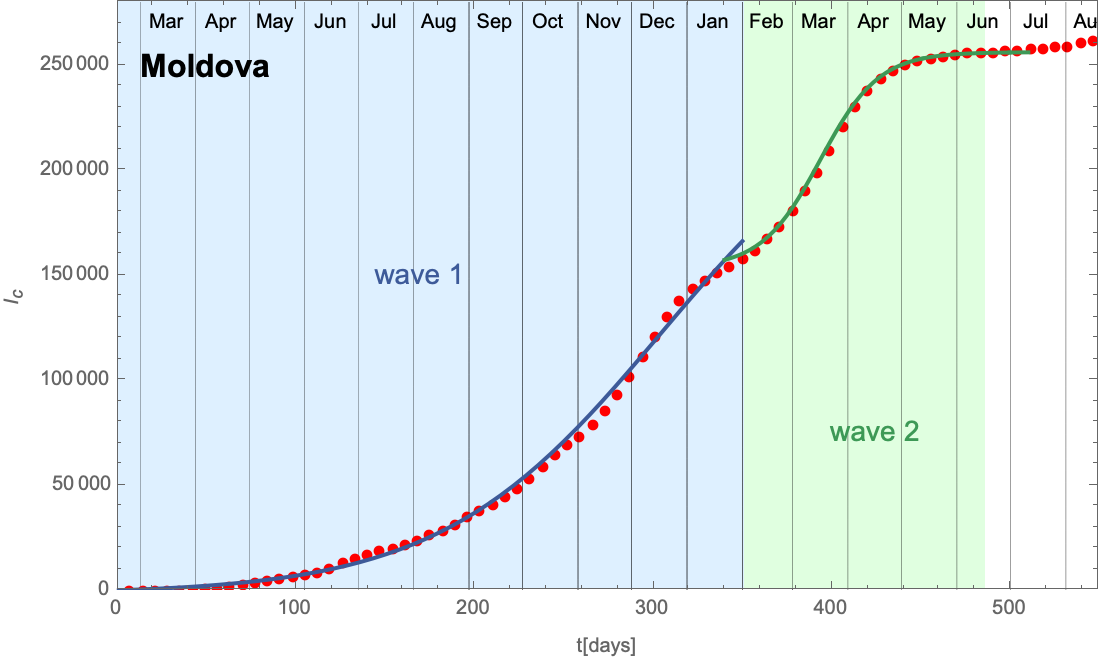}\nonumber\\
&\includegraphics[width=0.48\textwidth]{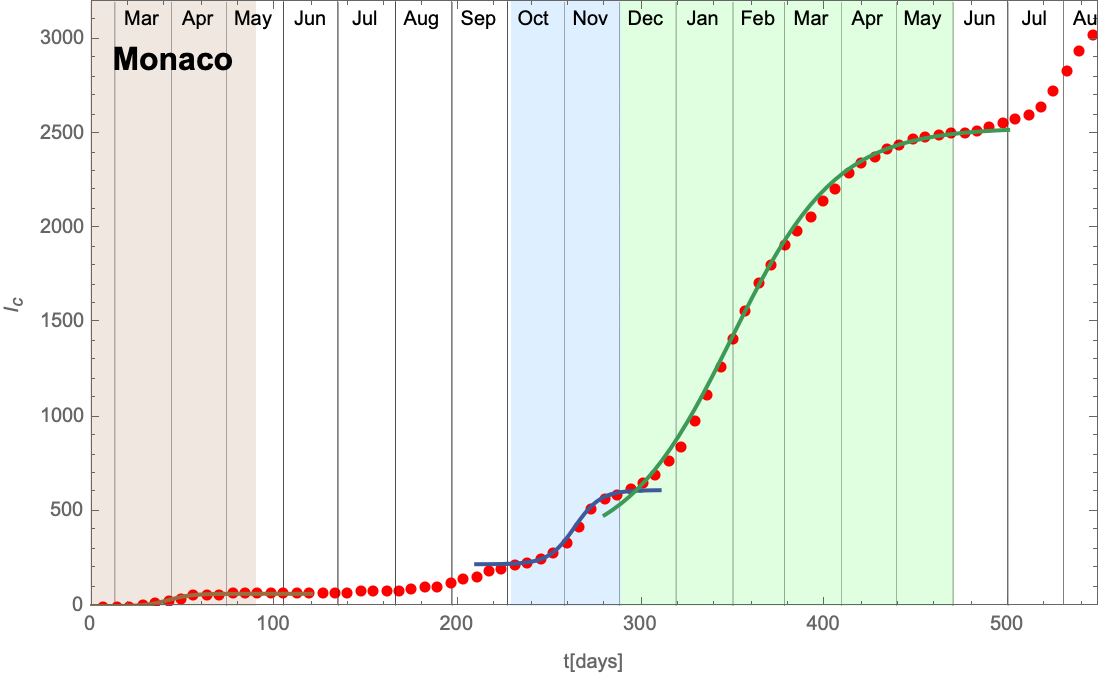} && \includegraphics[width=0.48\textwidth]{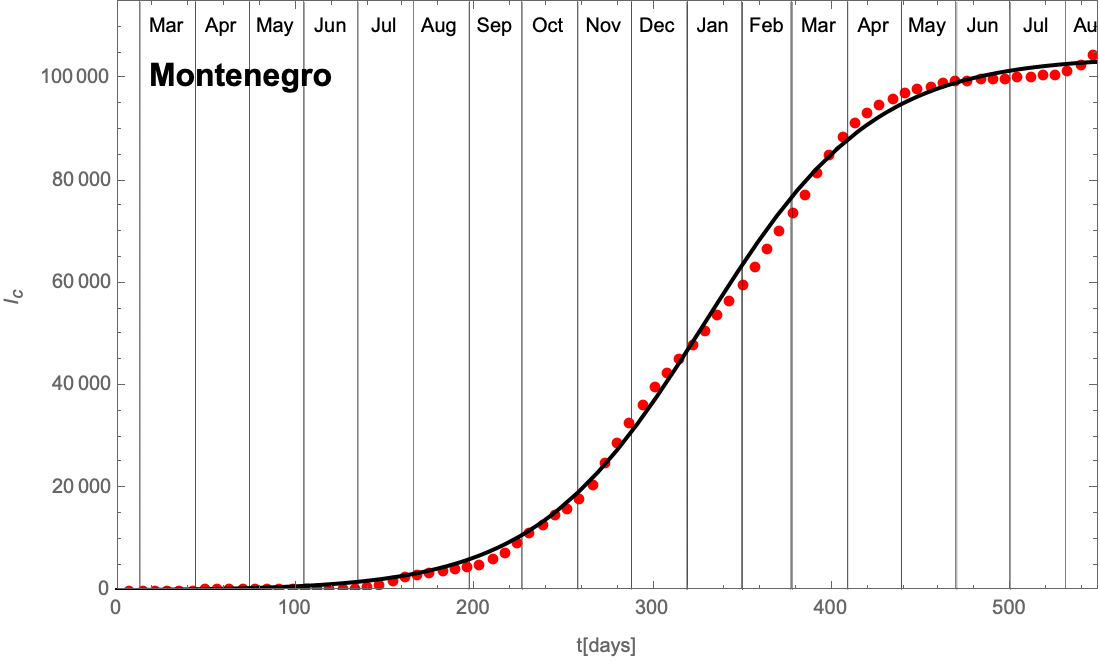}\nonumber
\end{align}
\caption{Cumulative number of individuals infected with SARS-CoV-2 (contd.)}
\label{Fig:DataEurop4}
\end{figure}

After the first wave of COVID-19, most countries have entered into an endemic phase, where the cumulative number of infected individuals has grown linearly, followed by further waves. As an example, Figs~\ref{Fig:DataEurop1}--\ref{Fig:DataEurop4} show the infection numbers in 48 European countries for the entire duration of the pandemic so far. We have highlighted in each of them individual waves and have fitted them with a logistic function as a solution of the eRG approach that we have reviewed in Section~\ref{Sect:RGapproach}.\footnote{For ease of visibility, we have focused on the larger such epidemiological episodes.} As can be seen from the quality of the fit, although these functions only have three parameters $(A,B,\lambda)$, they capture correctly the cumulative number of cases despite the fact that the data (even for different waves within the same country) represent very different epidemiological situations:
\begin{itemize}
\item the countries show large geographical, climatic as well as socio-cultural differences;
\item the waves occur during different seasons under different meteorological conditions; 
\item during each wave the governments of these countries have imposed different non-pharmaceutical interventions to reduce the spread of the virus;
\item since the beginning of 2021 all countries have started vaccination campaigns which have lead to a rate of roughly 60\% of all adults across Europe being fully vaccinated by summer 2021;
\item since the beginning of the pandemic, the SARS-CoV-2 virus has mutated multiple times and several different variants (with different infection and mortality rates as well as different efficacy for the vaccines) have dominated certain periods of the epidemiological dynamics.
\end{itemize}
As it is visible from the plots in Figs~\ref{Fig:DataEurop1}--\ref{Fig:DataEurop4}, despite all of these differences, the cumulative number of infected can still be organised by a self-similarity principle, which is characterised by logistic functions. 

Finally, in Figs~\ref{Fig:DataEurop1}--\ref{Fig:DataEurop4} we have restricted ourselves to fit waves that occurred before the summer of 2021. Many countries, however, show in late summer/early fall of 2021 once more a tendency of growing infection numbers, which (despite the vaccination efforts), may indicate the onset of new waves.

\begin{figure}[hp]
\begin{align}
&\includegraphics[width=0.48\textwidth]{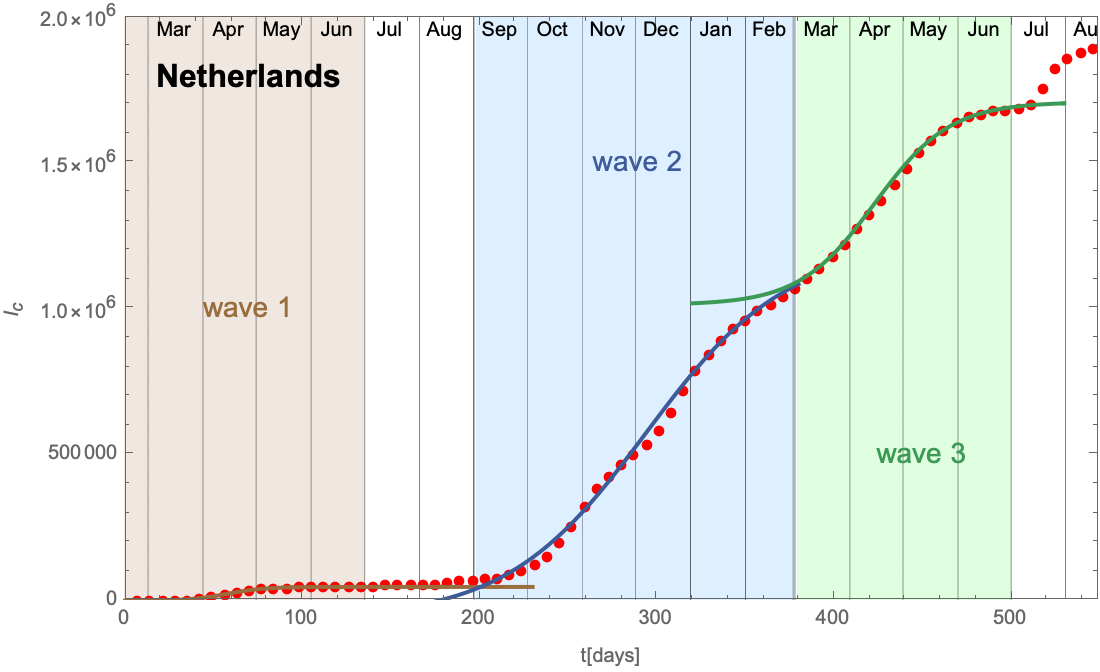} && \includegraphics[width=0.48\textwidth]{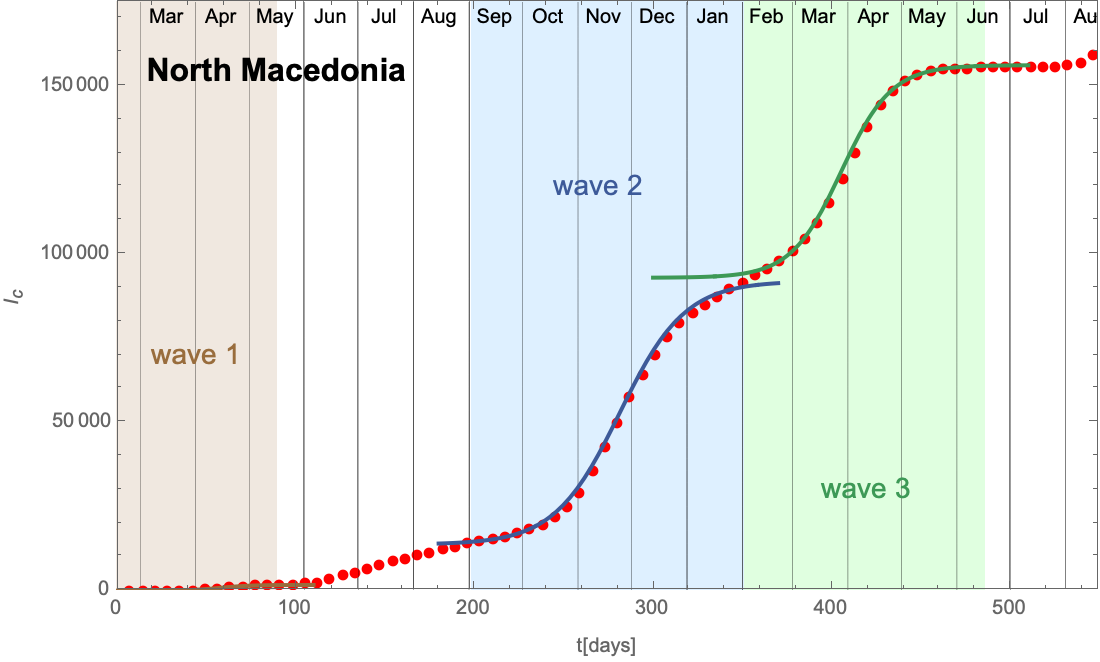}\nonumber\\
&\includegraphics[width=0.48\textwidth]{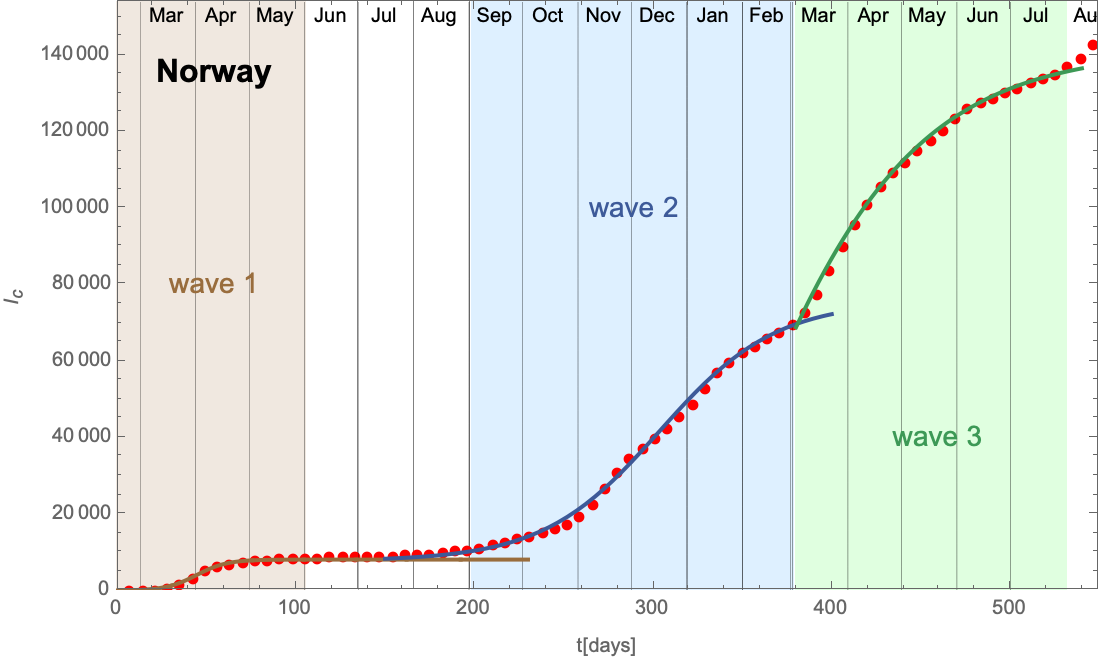} && \includegraphics[width=0.48\textwidth]{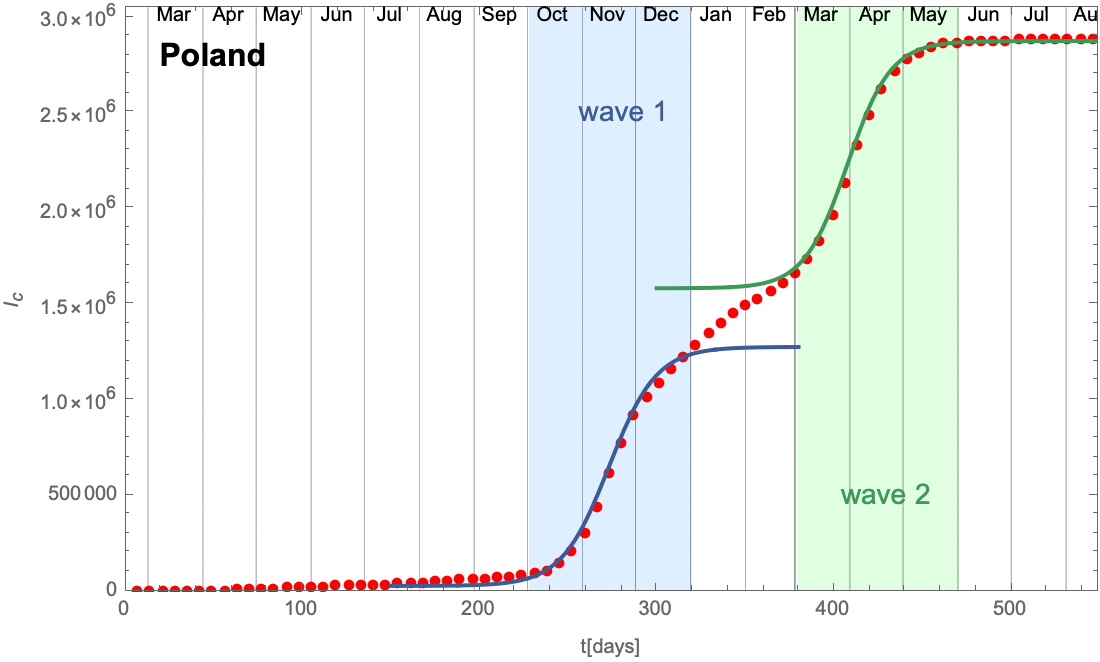}\nonumber\\
&\includegraphics[width=0.48\textwidth]{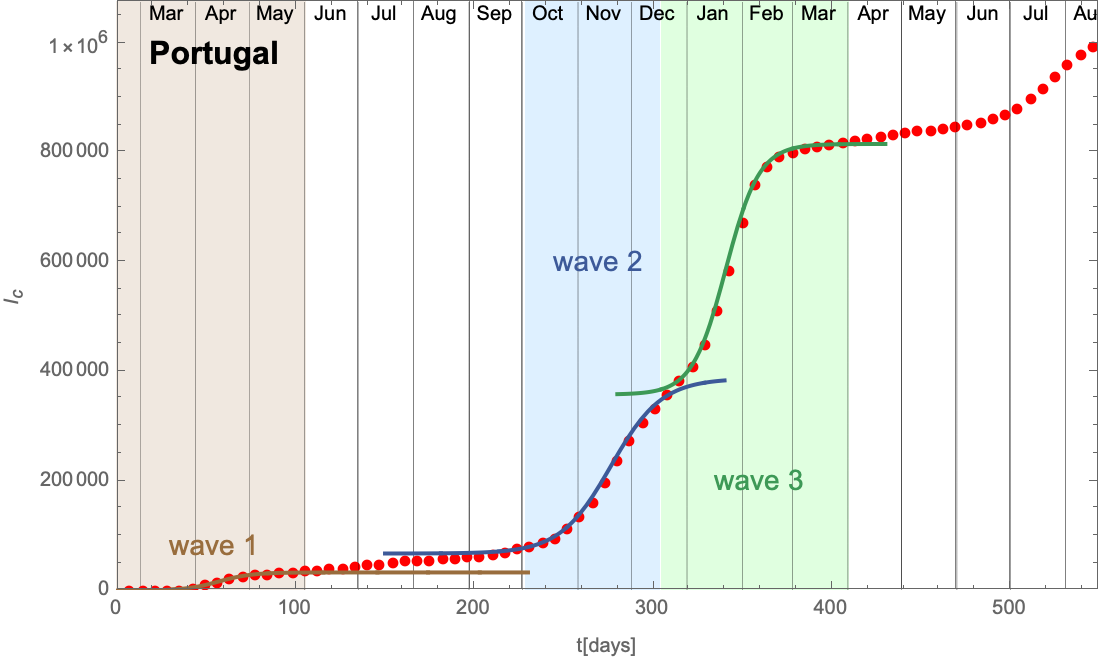} && \includegraphics[width=0.48\textwidth]{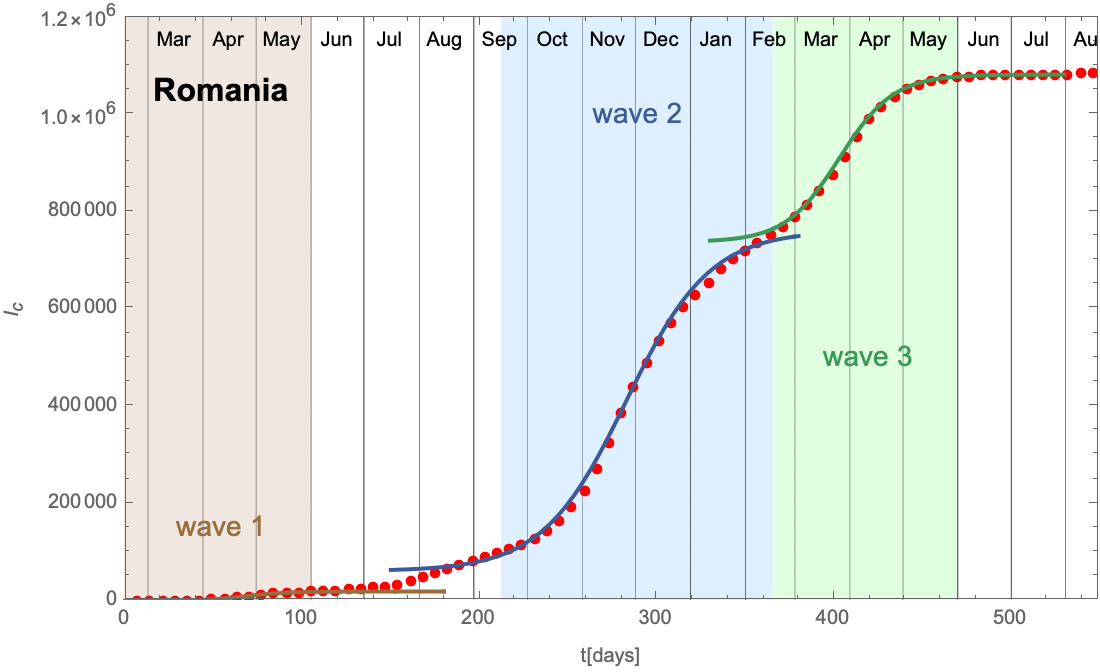}\nonumber\\
&\includegraphics[width=0.48\textwidth]{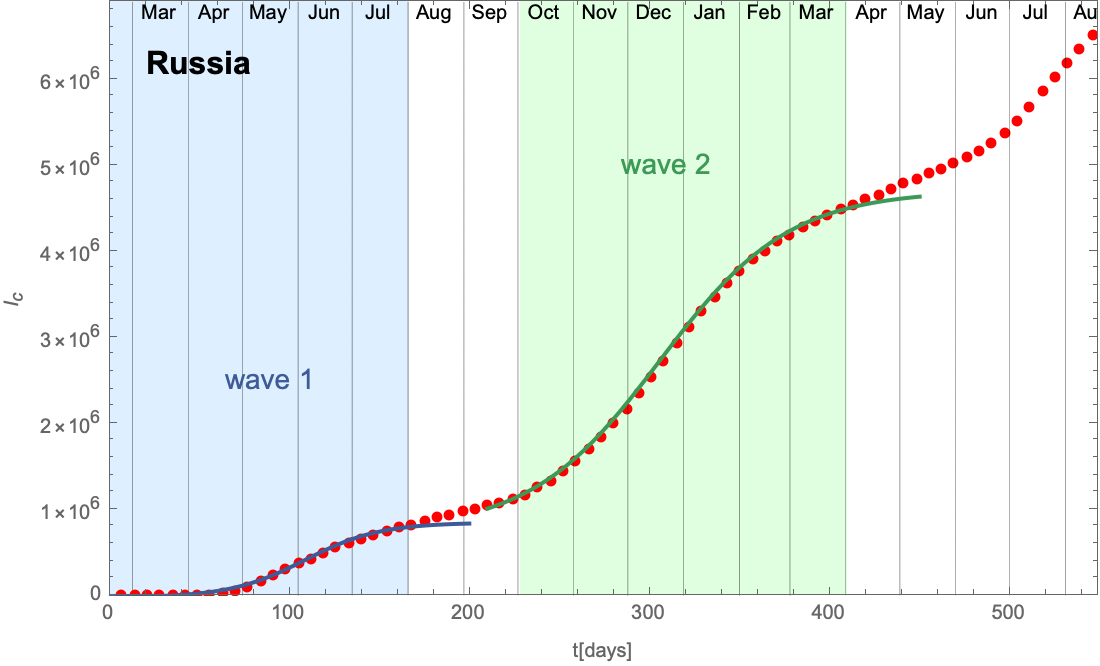} && \includegraphics[width=0.48\textwidth]{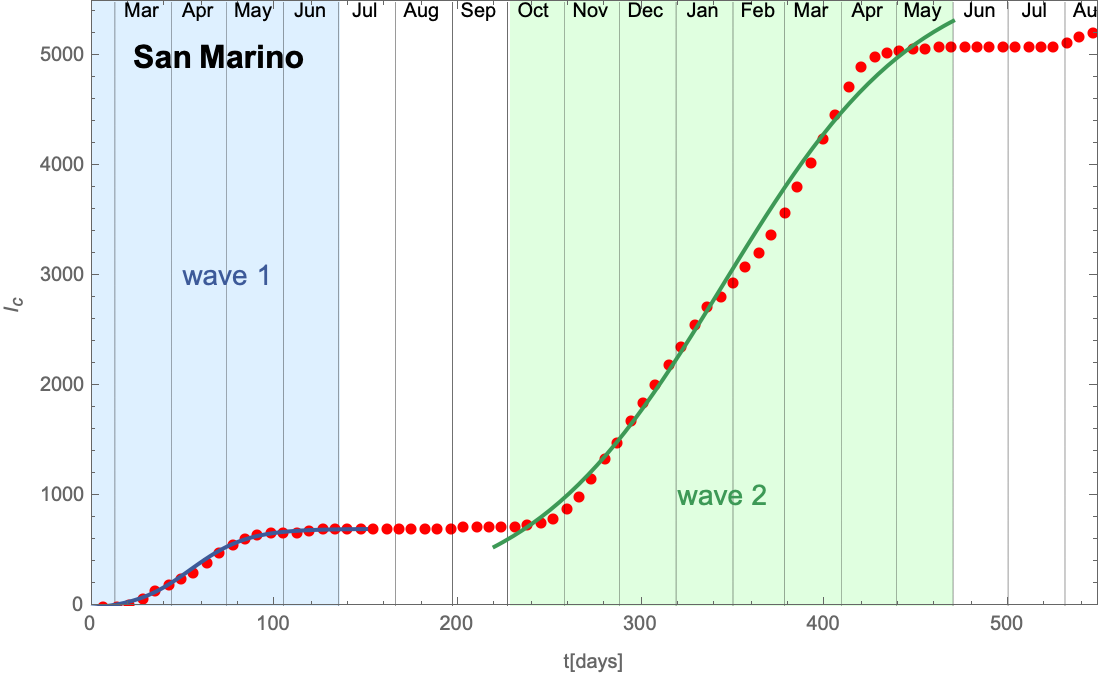}\nonumber
\end{align}
\caption{Cumulative number of individuals infected with SARS-CoV-2 (contd.)}
\label{Fig:DataEurop5}
\end{figure}

\begin{figure}[hp]
\begin{align}
&\includegraphics[width=0.48\textwidth]{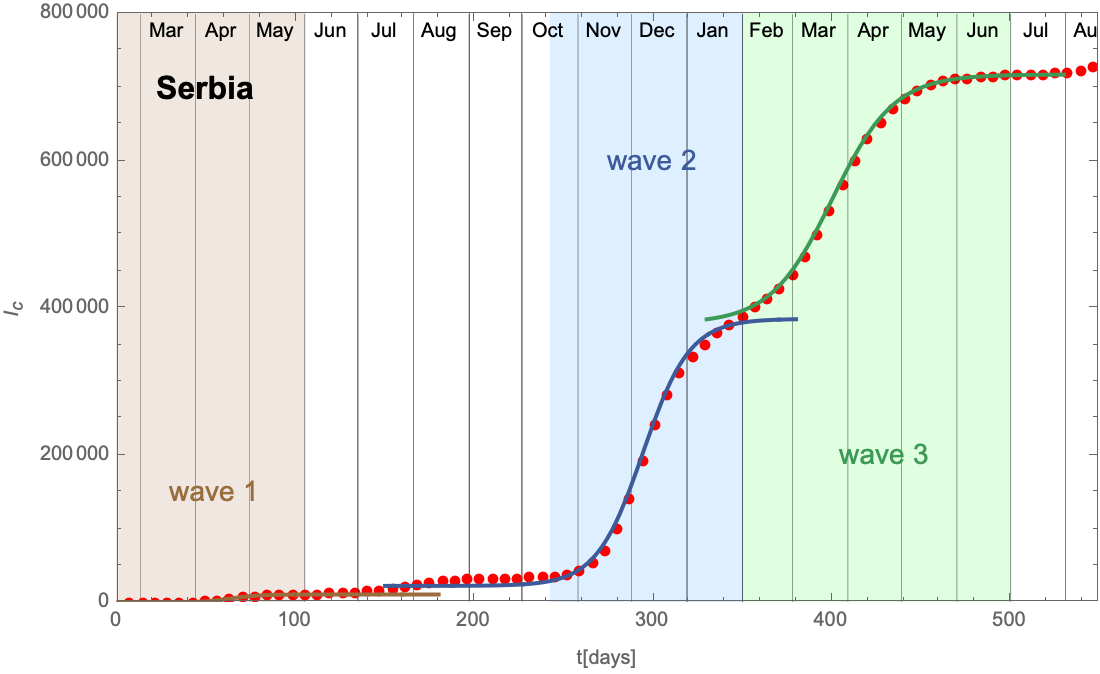} && \includegraphics[width=0.48\textwidth]{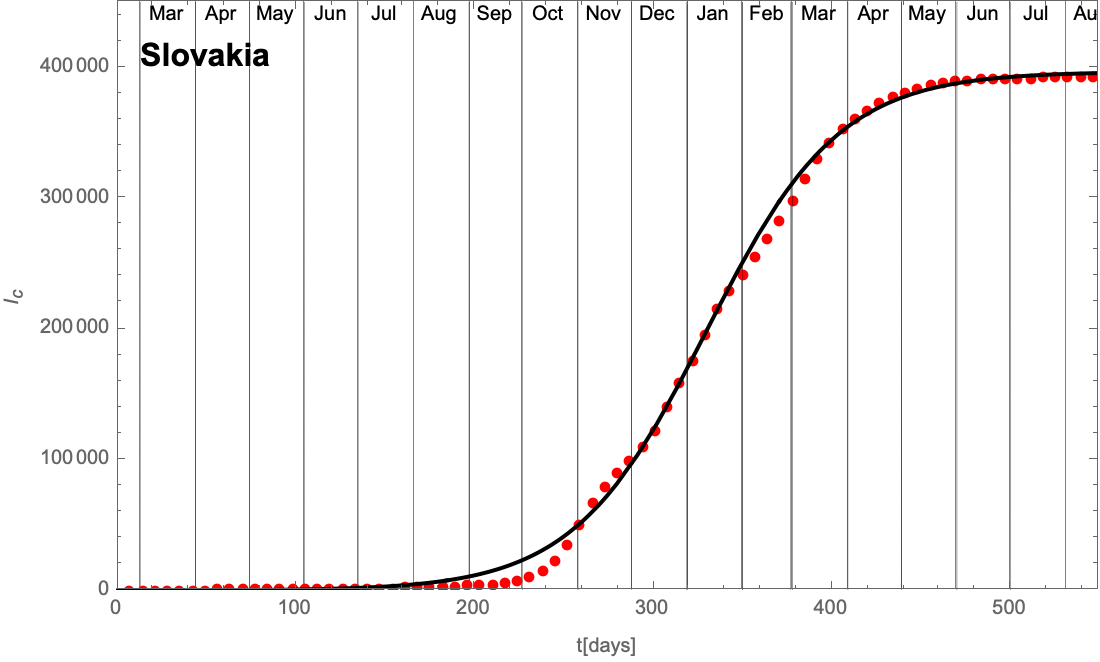}\nonumber\\
&\includegraphics[width=0.48\textwidth]{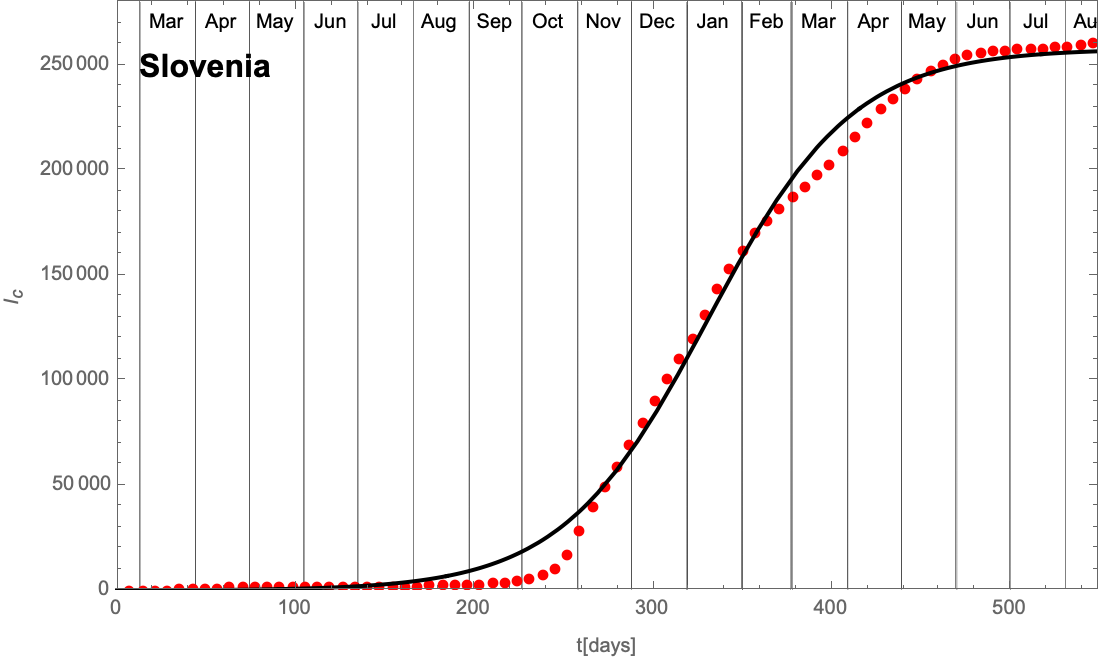} && \includegraphics[width=0.48\textwidth]{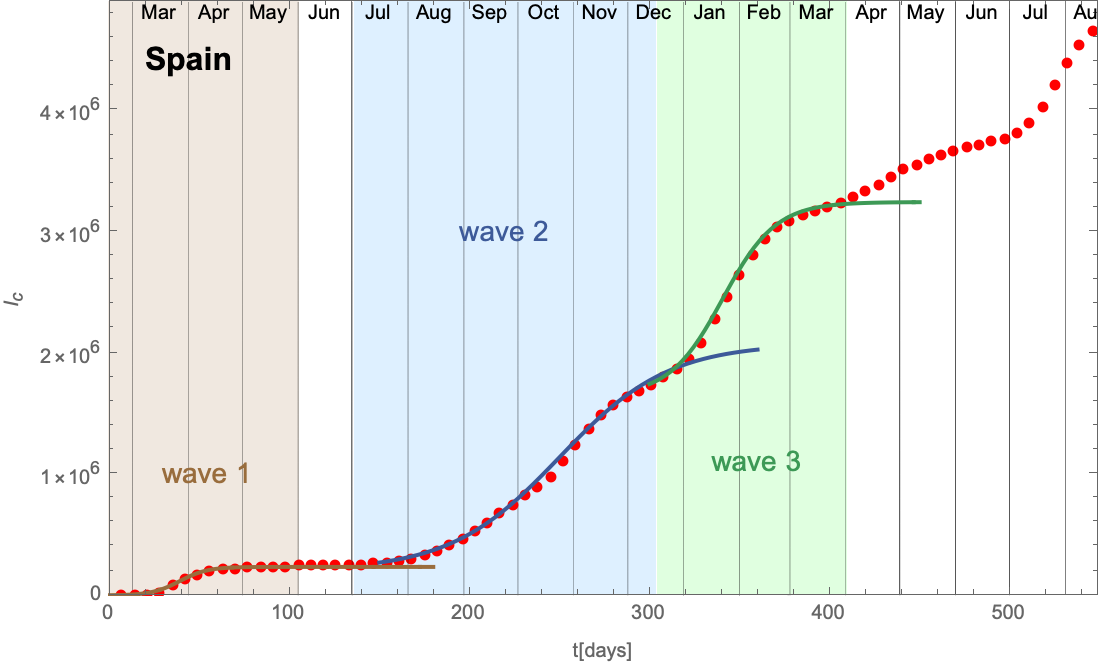}\nonumber\\
&\includegraphics[width=0.48\textwidth]{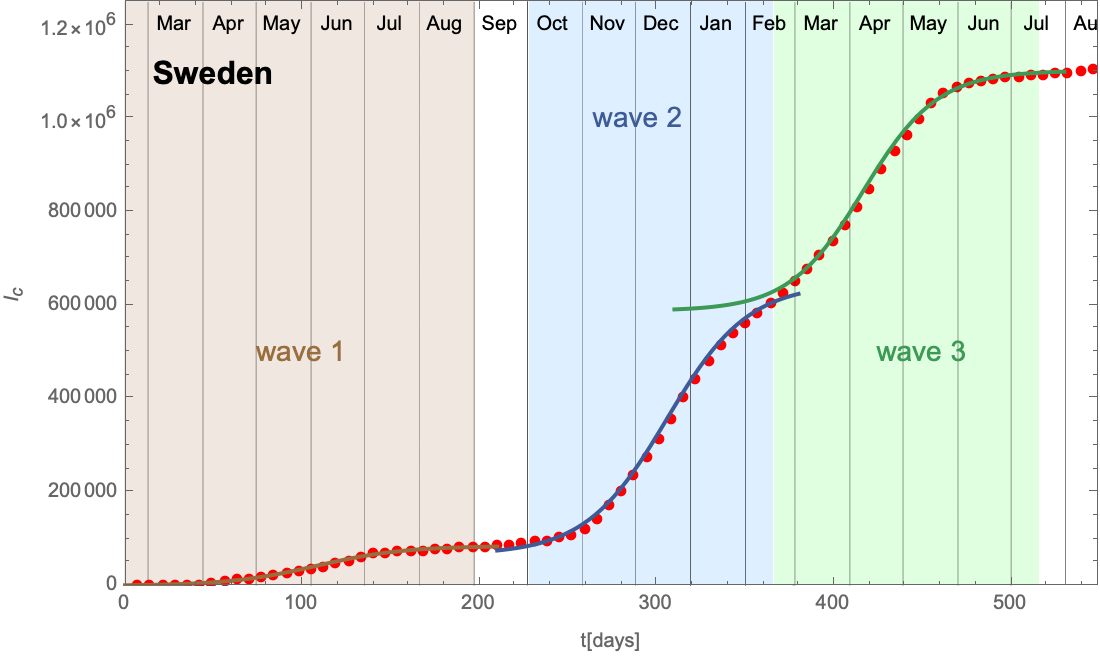} && \includegraphics[width=0.48\textwidth]{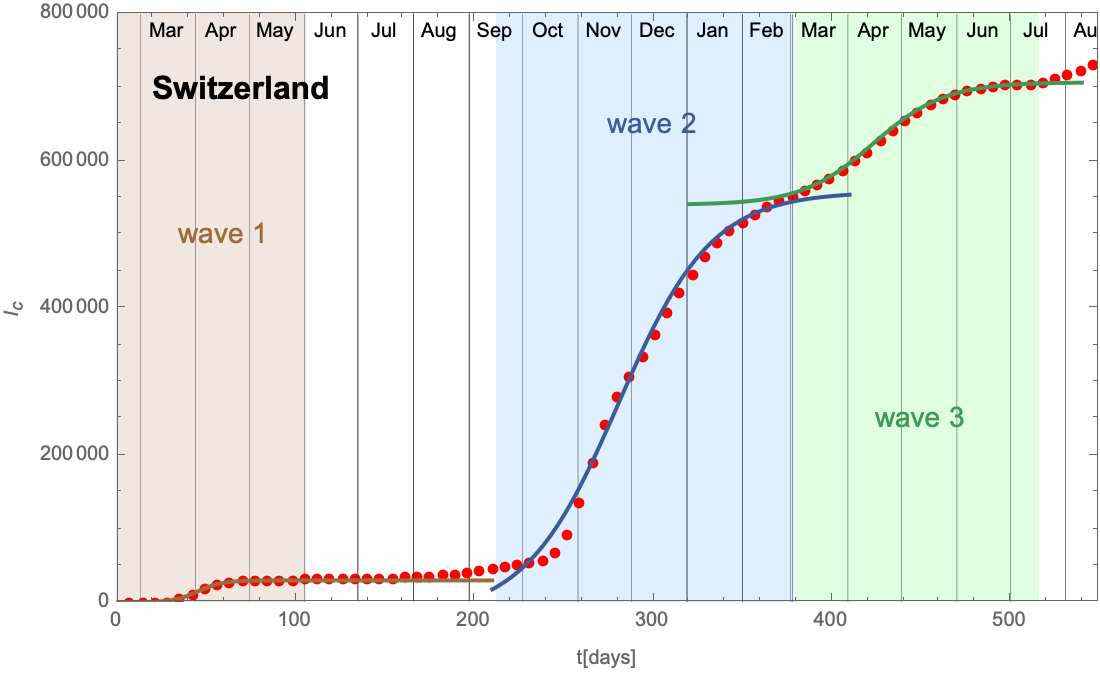}\nonumber\\
&\includegraphics[width=0.48\textwidth]{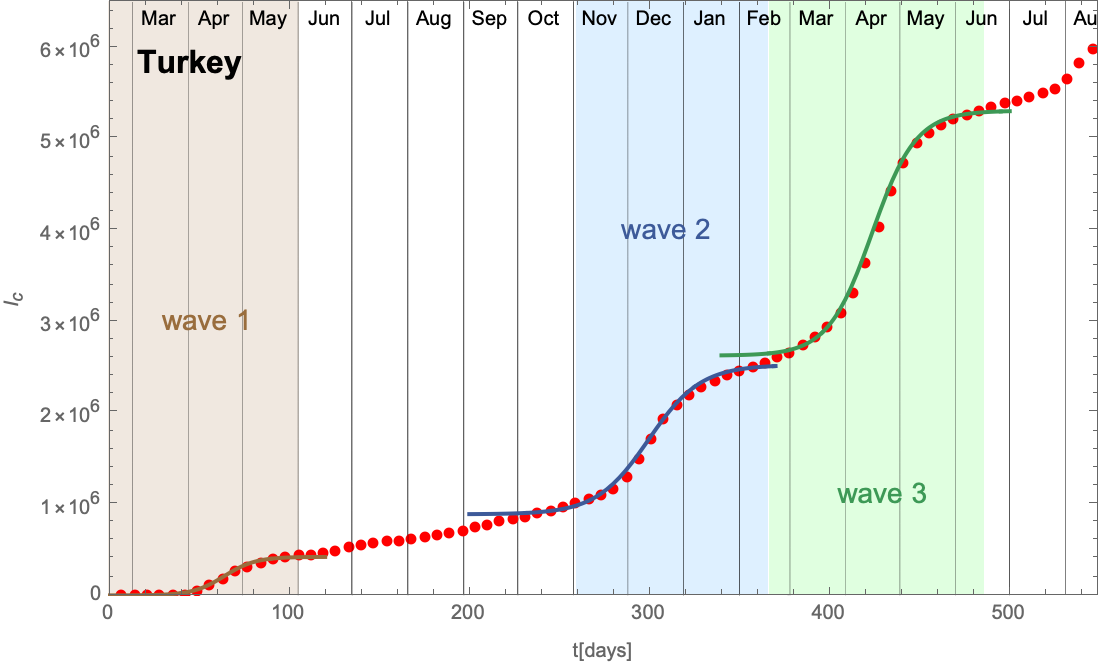} && \includegraphics[width=0.48\textwidth]{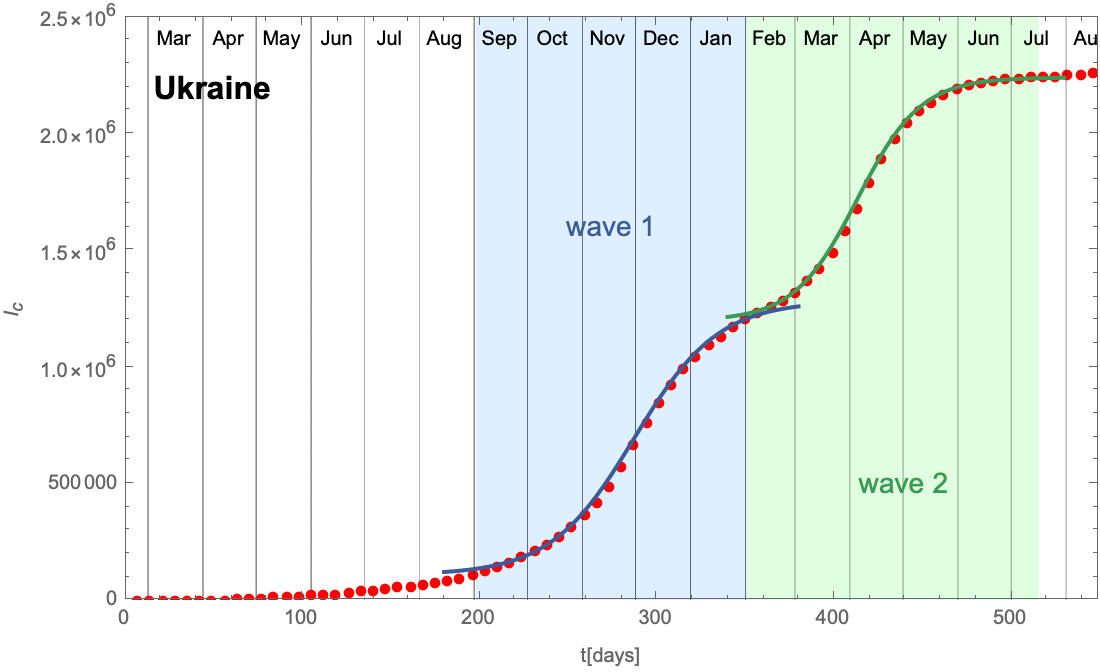}\nonumber\\
\end{align}
\caption{Cumulative number of individuals infected with SARS-CoV-2 (contd.)}
\label{Fig:DataEurop6}
\end{figure}


\section{Outlook and Conclusions}

The study of the time evolution of infectious diseases is a long standing subject: the impact of pandemics on human society cannot be overstated (as the recent devastating case of COVID-19 has highlighted). Consequently, over the course of more than a century, numerous approaches and mathematical models have been proposed with the aim to predict the spread of diseases among a population, devise tools to estimate their biological, social and economical impact and develop strategies to mitigate the harm done to society as a whole. In this report we give a review of this endeavour that is inspired by theoretical physics, in particular the study of phase transitions and critical phenomena, encompassed by the framework of field theory. Indeed, we organise mathematical models ranging from `microscopic' models, in which the spread of the disease is modelled at the individual level, to `effective' models, in which these microscopic interactions have been `summed up' and replaced by the description of the time evolution of suitable macroscopic degrees of freedom. We give concrete examples in each case and show how they are related to one another. We also show how to extend the models to account for observed phenomena, like multi-wave dynamics and the emergence of time-dependent symmetries such as approximate time-dilation invariance.

We start with lattice and percolation models in Section~\ref{Sect:Percolation}. These are among the most `microscopic' models and allow to simulate the spread of a disease at the level of individuals, therefore permitting to easily incorporate biological and social peculiarities related to the transfer of the disease from an infected individual to a susceptible one. Typically at great computational cost, these models provide insight into how these details influence the time evolution of the disease at larger scales and can highlight emerging patterns and symmetries. Indeed, via numerical analyses of simple models, we show in Section~\ref{Sect:Percolation} the emergence of critical behaviour: as a function of some key parameters, the system undergoes a phase transition from a state where only a small fraction of the population gets infected over time to a state where a significant portion of individuals is affected. Near the critical point, this behaviour can be cast into a field theoretical description for which we review an action formalism.

We next argue that mean field and averaging procedures of percolation models naturally lead to compartmental models. The latter are among the oldest descriptions of epidemiological processes (the SIR-model dating back almost a century) and are ubiquitous in the modern study of infectious diseases. As we review in Section~\ref{Sect:CompartmentalModels}, following our classification of approaches, compartmental models are effective descriptions: rather than describing the spread of a disease among individuals of the population, they comprise (first order) differential equations that yield (among others) the total number of infectious individuals in the population. The microscopic details of the spread of the disease have been `averaged' and enter into the details of the equations. The seeming loss of control over the microscopic details of the infectious dynamics comes at the benefit of a more `global' description of the disease (and typically a reduced computational cost). In Section~\ref{Sect:CompartmentalModels}, we provide an in-depth review of SIR-like compartmental models that, from a theoretical vantage point, elucidates their mechanics and dynamics. We analyse, review and extend the models to take into account single-wave dynamics, multi-wave patterns and even superspreaders, thus highlighting the flexibility of the approach as a whole. Finally, we also discuss that these models can be re-organised in a fashion to make efficient use of time-scaling symmetries of the epidemiological dynamics and which emphasises the role of fixed points.

In Section~\ref{Sect:RGapproach} we develop these ideas further and discuss the epidemic Renormalisation Group framework, which is in fact organised around the symmetry principle of time-scale invariance of the diffusion solutions. Using intuition from particle physics, the epidemiological process is described through flow equations (called beta-functions), which govern the trajectories of the system that connect different fixed points. The latter correspond to stationary solutions of the dynamics, in which either no disease is present in the first place or it has been completely eliminated. By invoking an even richer structure of fixed points, an extended eRG approach allows to model multi-wave pandemics.

The approaches and models outlined in this review can be adapted to a large range of different situations and cases: in Section~\ref{Sect:CovidSpecific} we have presented results related to the COVID-19 pandemic. We highlight how the multi-wave dynamics, as well as the impact of non-pharmaceutical interventions, vaccines and the geographical mobility of the (a portion of the) population can be modelled by the approaches outlined in the previous sections.

In order to keep the discussion as simple as possible and to focus on the underlying `physics', we have illustrated the ideas in this review by rather simple models. The latter can be much more refined and, for example, take into account other aspects and phenomena related to the spread of diseases. These range, for example, from developing strategies for protecting the population by implementing efficient vaccination campaigns and concrete strategies on the use of non-pharmaceutical interventions (such as lockdowns, social distancing measures and travel restrictions), to gauging the impact of mutations and adaptation mechanisms of pathogens \cite{Wang2021.01.15.426911,McCallum2021.01.14.426475}. We have refrained from working out the latter in detail, but instead refer the reader to more specialised literature. 

Furthermore, due to their obvious applications, the tools developed in this review have been exclusively focused on the description of infectious diseases among a human population. While many of them have in fact been inspired by other systems (notably chemical reactions), they can be applied to other fields as well with equal ease and success: apart from the immediate applicability to other species (\emph{e.g.} the spread of diseases among livestock), the ideas underlying the concrete models discussed here can be applied to a much larger range of problems. In fact, similar problems to the ones tackled here can be found in other complex systems as well, ranging from applications in network systems (\emph{e.g.} the spread of computer malware in a decentralised system) to human behaviour \cite{10.1371/journal.pone.0233410,10.3389/fpubh.2020.583181} as well as social engineering and media science (\emph{i.e.} the spread of ideas and information in a network/society).

\section*{References}

\bibliography{biblio}

\end{document}